\documentclass{article}

\usepackage{comment}
\usepackage{PRIMEarxiv}
\usepackage{subcaption}
\usepackage{color, colortbl}
\usepackage[utf8]{inputenc} %
\usepackage[T1]{fontenc}    %
\usepackage{hyperref}       %
\usepackage{url}            %
\usepackage{booktabs}       %
\usepackage{amsfonts}       %
\usepackage{nicefrac}       %
\usepackage{microtype}      %
\usepackage{lipsum}
\usepackage{fancyhdr}       %
\usepackage{graphicx}       %
\graphicspath{{figures/}}     %
\usepackage[dvipsnames]{xcolor}
\usepackage{placeins}
\usepackage{soul,color}

\graphicspath{{figures/}}     %
\title{Global mapping of structures and properties of crystal materials
\thanks{\textit{\underline{Citation}}: 
\textbf{Li et al.. Global mapping of crystal materials.... DOI:000000/11111.}} 
}

\author{
  Qinyang Li\\
 Department of Computer Science and Engineering\\
  University of South Carolina\\
  Columbia, SC 29201 \\
  \And
  Rongzhi Dong, Nihang Fu, Sadman Sadeed Omee, Lai Wei\\
 Department of Computer Science and Engineering\\
  University of South Carolina\\
  Columbia, SC 29201 \\ 
   \And
 Jianjun Hu *\\
 Department of Computer Science and Engineering\\
  University of South Carolina\\
  Columbia, SC 29201 \\
  \texttt{jianjunh@cse.sc.edu} \\
}

\begin{document}
\maketitle

\begin{abstract}

Understanding materials' composition-structure–function relationships is of critical importance for the design and discovery of novel functional materials. While most such studies focus on individual materials, we conducted a global mapping study of all known materials deposited in the Material Project database to investigate their distributions in the space of a set of seven compositional, structural, physical, and neural latent descriptors. These two-dimensional materials maps along with their density maps allow us to illustrate the distribution of the patterns and clusters of different shapes, which indicates the propensity of these materials and the tinkering history of existing materials. We then overlap the material properties such as composition prototypes and piezoelectric properties over the background materials maps to study the relationships of how material compositions and structures affect their physical properties. We also use these maps to study the spatial distributions of properties of known inorganic materials, in particular those of local vicinities in structural space such as structural density and functional diversity. These maps provide a uniquely comprehensive overview of materials and space and thus reveal previously undescribed fundamental properties. Our methodology can be easily extended by other researchers to generate their own global material maps with different background maps and overlap properties for both distribution understanding and cluster-based new material discovery. The source code for feature generation and generated maps are available at \url{https://github.com/usccolumbia/matglobalmapping}

\end{abstract}


\section{Introduction}

Analyzing and understanding the structure-property relationship is a fundamental scientific challenge in physical sciences, from materials to organic compounds and proteins. Although it has been widely studied in the field of protein structure research \cite{hou2005global}, only a few studies have been done so far in material sciences \cite{suzuki2022self}. With a global mapping of all known material structures into a 2D or 3D space, it is possible to navigate the design space, identify structural and functional clusters, and even discover new functional materials within the vicinity of known functional materials. For over a century, materials scientists have accumulated a large number of crystal structures through tedious experimental processes, leading to modern crystal structure databases such as Inorganic Crystal Structure Database (ICSD) \cite{zagorac2019recent,allmann2007introduction,belsky2002new}, which contains more than 200,000 structures since 1913 with about 7,000 structures added each year, and more open databases such as the Materials Project \cite{jain2013commentary}, which contains more than 124,515 inorganic materials along with their properties. However, the composition-structure-property relationship of inorganic materials still remains elusive. Just as a navigator on the sea, it is always desirable to have a map that gives an overview of what materials we have explored, where we are right now, and what direction we should go forward. 

Despite the fact that both crystal materials and protein structure spaces are mostly well discovered in terms of their building blocks, structure prototypes, and topology, protein sequences structures, and functions have been investigated more thoroughly using a wide range of artificial intelligence and modern deep learning techniques as shown in AlphaFold \cite{jumper2021highly} for protein structure prediction. Protein scientists have widely studied the mapping of protein function and structure space for function predictions. Protein mapping is first introduced by Holm and Sander \cite{holm1996mapping}, which aims to organize known protein shapes, identify new types of protein architectures, and discover evolutionary relationships or new gene functions.  Global mapping of protein structures was further studied by Kim, et al \cite{hou2005global,hou2003global,choi2006evolution}, in which they plotted the protein structure space using the pairwise structural similarity scores calculated for all non-redundant protein structures. They used the DALI algorithm and calculate the similarity score based on the matrix distance geometry. Then the protein structures were visualized in both 2D and 3D space. 
They found that similar structures clustered together in their protein map and the overall distribution of structural classes of this map followed closely that of the map of the protein fold space. They further found that proteins sharing similar molecular functions were also found to co-localize in the protein structure space map. In \cite{osadchy2011maps}, Osadchy and Kolodny used the FragBag descriptor vector to represent a protein structure, in which a library of 400 12-mer fragments are used to check for each entry in the vector what is the number of times it is the best approximation of any of the 12-mer fragments in the backbone of the protein. 

A map of the inorganic ternary metal nitrides has also been proposed as a matrix representation by sun et al \cite{sun2019map}, colored to represent the thermodynamic stability of the ternary nitride with the lowest formation energy. Their large stability map of the inorganic ternary metal nitrides clusters the ternary nitrides into chemical families with different stability and metastability, which proposed hundreds of promising new ternary nitride spaces for experimental investigation. In Zhu et al. \cite{zhu2021charting}, the authors predicted the thermal conductivity of all known inorganic materials in the ICSD and charted the structural chemistry of lattice thermal conductivity into extended van-Arkel triangles. The analysis of the maps allows them to discover candidate materials with ZT exceeding 1.0. 

Similar to global protein mapping analysis, the crystal structures can be directly transformed into a structure space representation with their topological information. In material sciences, the structural-property relationship is well studied with a scope usually limited to small groups of crystal structures such as Metal-Organic Frameworks (MOF) structures \cite{allendorf2015crystal}. Global mapping has been used in materials structure and function analysis. In our previous work \cite{dan2020generative}, a t-distributed stochastic neighbor embedding (t-SNE) based global mapping is used to map newly discovered hypothetical materials by their generative adversarial network models as regards existing materials. In one of the recent studies \cite{suzuki2022self}, they were able to conduct the structural-property mapping using a custom descriptor. By combining a CGCNN \cite{xie2018crystal} learned structure representation and XRD-based periodicity representation using contrastive learning, they were able to reflect the relationship on the global scale utilizing the t-SNE \cite{van2008visualizing} map. Using this method, they found that structures of similar properties are clustered together in the local area. In \cite{aykol2019network}, Aykol et al. applied a network analysis of the synthesizability of all inorganic materials, and they found a scale-free network constructed by combining the convex free-energy surface of inorganic materials computed by high-throughput density functional theory. However, there is no systematic study of the global mapping of known inorganic materials with diverse materials representations.

There are many different structural descriptors for representing a crystal material \cite{musil2021physics}. Commonly used representations include atom-centered symmetry functions\cite{behler2011atom}, Coulomb matrices\cite{neese2003improvement}, and the smooth overlap of atomic positions (SOAP) %
\cite{willatt2019atom}. Materials can also be represented by their composition features, the XRD spectrum of their structures, and pair-wise distance distributions, each characterizing different aspects of their structure and physics. These standalone descriptors can also be combined to infer more expressive representation. For example, the XRD descriptor and structure descriptor generated by graph neural networks have been combined using contrastive learning \cite{suzuki2022self} for better classifying and visualizing different groups of sub-classes from the whole dataset.

In this work, we develop an approach for global mapping of known crystal materials with structure densities for uncovering distribution patterns using a set of seven structural descriptors and composition representation. This includes structural information such as atomic coordinates, pair-wise atomic distance, physical information such as XRD, topological information, and latent representation learned from deep graph neural networks \cite{jiang2021topological, zhang2019unsupervised,dan2020generative,omee2022scalable}. For each descriptor, we transform the structures with varying numbers of atoms into fixed-dimension vectors so that we can apply dimension reduction to map the material representations into a 2D space. 
Our approach allows us to create highly detailed and accurate maps of the structure and properties of crystal materials. These maps can reveal patterns and trends in the relationship between the structure and properties of these materials and can help researchers understand how these properties are influenced by factors such as structure and chemical composition. 

Our contributions can be summarized as follows:

(1) We proposed and developed visualization approaches for global mapping of all inorganic materials or given material datasets using seven different structural, physical, and neural latent descriptors.

(2) We used global mapping methods to study the relationships between structure space and property space with different materials families including ABC$_3$ prototype materials and piezoelectric materials. We found different properties tend to show better patterns with different structural/physical/neural feature maps. We also found that the clusters of the global materials maps show the tinkering material discovery process over history. 

(3) We also proposed a measurement to evaluate the mapping quality of the materials family over the global material map. Our work disclosed novel insights in terms of structure-function/property relationships of materials. The availability of all the source code and processed data for visualization makes it easy for others to apply the same global mapping approach for studying similar material analysis.

\section{Method}
\label{sec:headings}

To characterize the compositional and structural properties of crystal materials using fixed dimensions for global mapping, we use a diverse set of compositional and structural information, including inter-atomic pairwise distance matrices, atomic coordinates of crystals, physical information such as X-ray diffraction patterns and elemental compositions, neighboring structural information such as persistent homology topology, and latent features derived from deep graph neural networks. These descriptors can map crystals into fixed-length vectors.

In order to visualize our high-dimensional data, we make use of a non-linear dimension reduction algorithm known as t-distributed stochastic neighbor embedding (t-SNE) \cite{cieslak2020t}. The purpose of the t-SNE technique is to convert data of high dimensions into samples with extremely low dimensions (usually 2d or 3d) for easy visualization. After applying t-SNE mapping, the pairwise neighborhood relationships among the data points remain mostly unchanged. Therefore, locations that are physically closer to each another in a higher dimension tend to stay physically near after being mapped to the space with lower dimensions, which can highlight the structural similarities of crystal materials in local distribution regions.

\subsection{Structural fingerprint representation}

We use two sets of structural fingerprints to represent the geometric aspects of crystal structures without considering the elemental differences of atoms.

\paragraph{Atomic sites coordinates.}
In a crystal, the atoms are arranged in a repeating pattern that forms the unit cell. The unit cell is the basic building block of a crystal, which can be thought of as a small box that encloses a portion of the crystal structure. The atoms within the unit cell are arranged in a specific pattern that is repeated throughout the 3D space. 
Fractional coordinates refer to the fractional location of an atomic site within a unit cell. Cartesian coordinates, on the other hand, describe the position of an atom as a set of three numbers that indicate the distance of the atom from a fixed reference point in three-dimensional space which consists of orthogonal basis vectors.
Using the zero-padding scheme, we are able to transform crystals with different numbers of atomic sites into fixed-length feature vectors for t-SNE analysis.

\begin{figure*}[ht]
  \centering
  \includegraphics[width=0.5\linewidth]{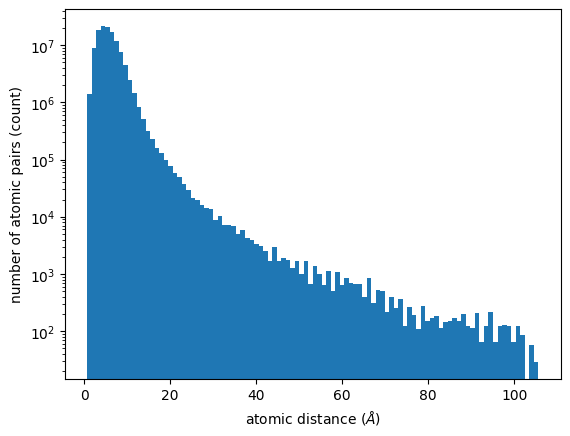}
  \caption{Histogram for all MP material atomic pairwise distances.
  }
  \label{fig:histogram}
\end{figure*}

\paragraph{Site pair-wise distance matrix.}
Many material properties such as formation energy and band gap are related to the pair-wise distances between atomic sites within the unit cell. 
To convert the pairwise distance matrices of varying sizes into a fixed-dimension representation, we use a histogram encoding scheme. First, we calculated pairwise atomic distances for each material of the entire MP data set and keep only the upper triangle part of the pairwise distance matrices. From the sampled 118,156,495 distances, we analyzed the frequency distribution (Figure \ref{fig:histogram}) of these distances and divide the distance range into 100 percentile ranks. Then for each crystal material, we count the frequencies of its atomic pairwise distances in each of the 100 intervals, leading to a vector of 100 dimensions to represent the crystal.

\paragraph{Topological representation of crystal structures.} We use the atom-specific persistent homology (ASPH), a persistent homology featurization approach to calculate 200-dimension descriptors for any given crystal structure \cite{jiang2021topological}. The ASPH can capture both pairwise and many-body interactions and reveal the topology-property relationship of a group of atoms at various scales. Standard persistent homology only offers global structural information without considering that crystal structures have a wide range of chemical compositions and structural complexity. The atom-specific persistent homology can better embed atom-wise chemical information into topological invariant.

\subsection{Physical fingerprint representation}
\paragraph{XRD representation.} 
Each crystal structure can be represented by the X-ray diffraction pattern based on Bragg’s law as explained in \cite{zhang2019unsupervised}. The X-ray diffraction of a periodic lattice is determined by the size and shape of the unit cell along with the position and identity of atoms on a given plane. The calculation of the diffraction pattern is performed at a fixed set of $2\theta$ values from 0 to 89.98 degrees with a step size of 0.1 degrees using the pymatgen package\cite{ong2013python}, which generates a 900-dimensional vector for each diffraction pattern. Gaussian smearing is then performed to normalize the integrated intensity of diffraction to a unitary value. These high-dimension vectors can then be fed to dimension reduction algorithms such as t-SNE or PCA for mapping to 2D or 3D dimension space for visualization.

\paragraph{Elemental composition fingerprint.}
Materials can also be represented by their elemental compositions. Using one-hot encoding, we are able to transform a formula into an 
87 x 32 matrix, which is then flattened into a vector of 2,784. The reason why we choose those parameters is that there are 87 unique elements within the MP data set and the number of atoms for each unique element does not exceed 32 within an individual crystal material. Details of one-hot encoding can be found in \cite{dan2020generative}.

\subsection{Neural latent fingerprint representation}

\paragraph{DeeperGATGNN latent space representation.} 
Recent studies have shown that deep graph neural networks (GNNs) have dominated performance in materials property prediction compared to traditional structural or topological descriptors \cite{fung2021benchmarking,omee2022scalable}. This implies that the latent feature learned by these graph neural networks may be used as a better representation for crystal material characterization. 

We use the DeeperGATGNN \cite{omee2022scalable}, a graph neural network with multiple graph-attention layers for latent feature extraction. This model is one of the state of the art GNN models for materials property prediction. In this network, each crystal is converted into a node-edge graph that uses multiple layers of message passing to process the information. The global attention mechanism of NN adds additional whole-crystal information and the crystal node’s location to the context vector. Compared to the standard of 4-9 graph convolution layers, this network architecture contains more than 30 layers without significant performance degradation. 
In this work, we train a DeeperGATGNN network model for formation energy prediction using a masked MP materials(36837) trainning set for 500 epoches with 20 graph-convolution layers. The feature vectors are extracted right before the last output layer. For global attention, the maximum number of neighbors is set to 12 with a maximum local radius of 8. We use the default parameters of the algorithm for training the GNN model for latent feature extraction.

\begin{table}[h]
\begin{center}
\caption{Fingerprint representation}
\label{latticepattern}
\begin{tabular}{lll}
\hline
\textbf{Structural}                 &\textbf{Physical}                  & \textbf{Deep learning}          \\ \hline
Cartesian \& Fractional coordinates & XRD                      & DeeperGATGNN                 \\ 
Pair-wise distance                  & Composition              &                         \\ 
Persistent Homology                 &                          &                        \\ \hline
\end{tabular}
\end{center}
\end{table}

\subsection{Datasets and feature engineering}

Each crystal structure is represented in a CIF file, which contains all basic information about this crystal. We obtain our raw structure and property dataset from the Materials Project database \cite{jain2020materials} using the Material Project APIs on June 30, 2022.
We represent the whole Materials Project's crystal materials using three different categories of spaces including structural spaces, physical spaces, and neural network embedding. In our data set, there are 136,071 unique crystal structures including 19,283 binary, 59,024 ternary, and 35,206 quaternary structures. 
Since there are 130k materials in the MP dataset, It is wise to divide them into sub-classes for t-SNE visualization. Here we used the number of different elements in crystals as our criterion to divide the whole dataset into five sub-classes, including the materials with 2 to 6 different elements. Crystal materials that have more than 6 elements are not well discovered and studied. For materials of the same subset, all materials feature vectors are computed as fixed-dimension vectors as described above. Then they are visualized using t-SNE analysis by mapping the high-dimension vectors into 2D space. 

\subsection{Creating functional mapping using crystal fingerprints}

\paragraph{Global local material map generation.} With seven different crystal representations, we conduct global mapping of all 136,071 materials in the Materials Project database as shown in Figure \ref{fig:globaldist}. In each sub-figure, each point represents a unique crystal structure, which is calculated using the t-SNE dimension reduction algorithm from raw fingerprints. Since t-SNE mapping has the characteristic to preserve the sample neighborhood relationship, the distance between any two points roughly represents the similarity of the structural/composition/physical features of the corresponding pair of crystal structures based on its descriptors. Since there are over 136,000 materials on every figure, there should be overlapping points. To address this issue, we used a density heat map coloring scheme to identify high-density areas. The density estimation method is the Gaussian kernel density estimation function from the scipy package \cite{bashtannyk2001bandwidth}. The red cluster indicates a lot of similar crystal structures that exist in this area. This is useful for the later subset study of global material distribution. For some piezoelectric properties, there are limited annotated materials available in the MP dataset. So we also generate local mapping for these materials over given prototype materials (such as ABC$_3$ or ABO$_3$) to better analyze their distribution. 

\paragraph{Mapping material properties over the global map.}
While the t-SNE maps of the seven descriptors show the locations of different materials in the 2D space, we can overlap the properties of a given materials family (a subset of the MP materials) by simply mapping the property values into different colors. We use a four-color scheme and each color with five different lightness levels, creating a 20-level coloring scheme to show the property distribution.

\section{Results}
\label{sec:Results}

\FloatBarrier

\subsection{Global mapping of crystal materials}

\begin{figure*}[thb]
    \centering

    \begin{subfigure}[b]{0.32\textwidth}
        \centering
        \includegraphics[width=1.0\textwidth]{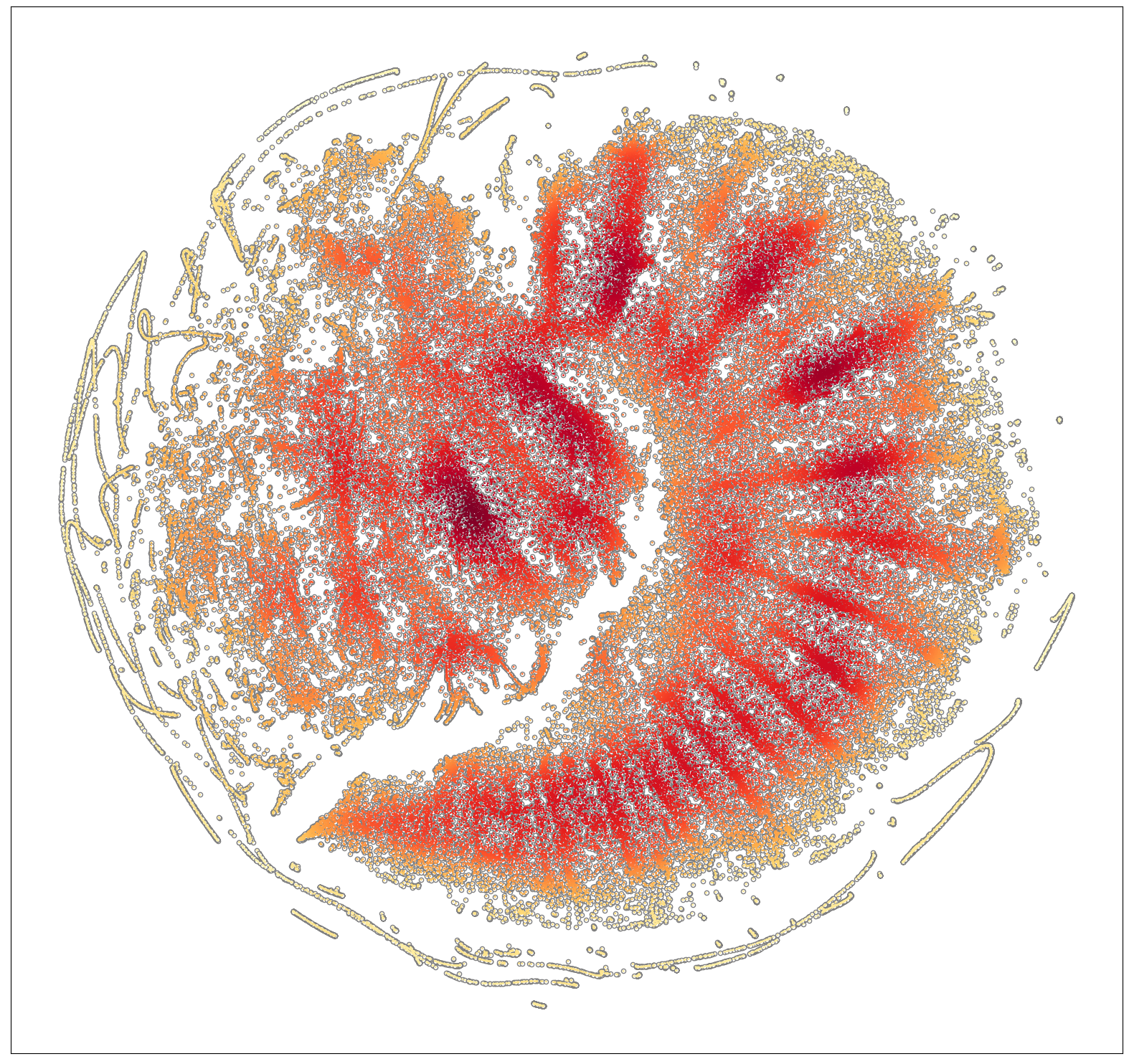}  
        \caption{}
        \label{fig:cod0g}
    \end{subfigure}%
    \hfill
    \begin{subfigure}[b]{0.32\textwidth}
        \centering
        \includegraphics[width=1.0\textwidth]{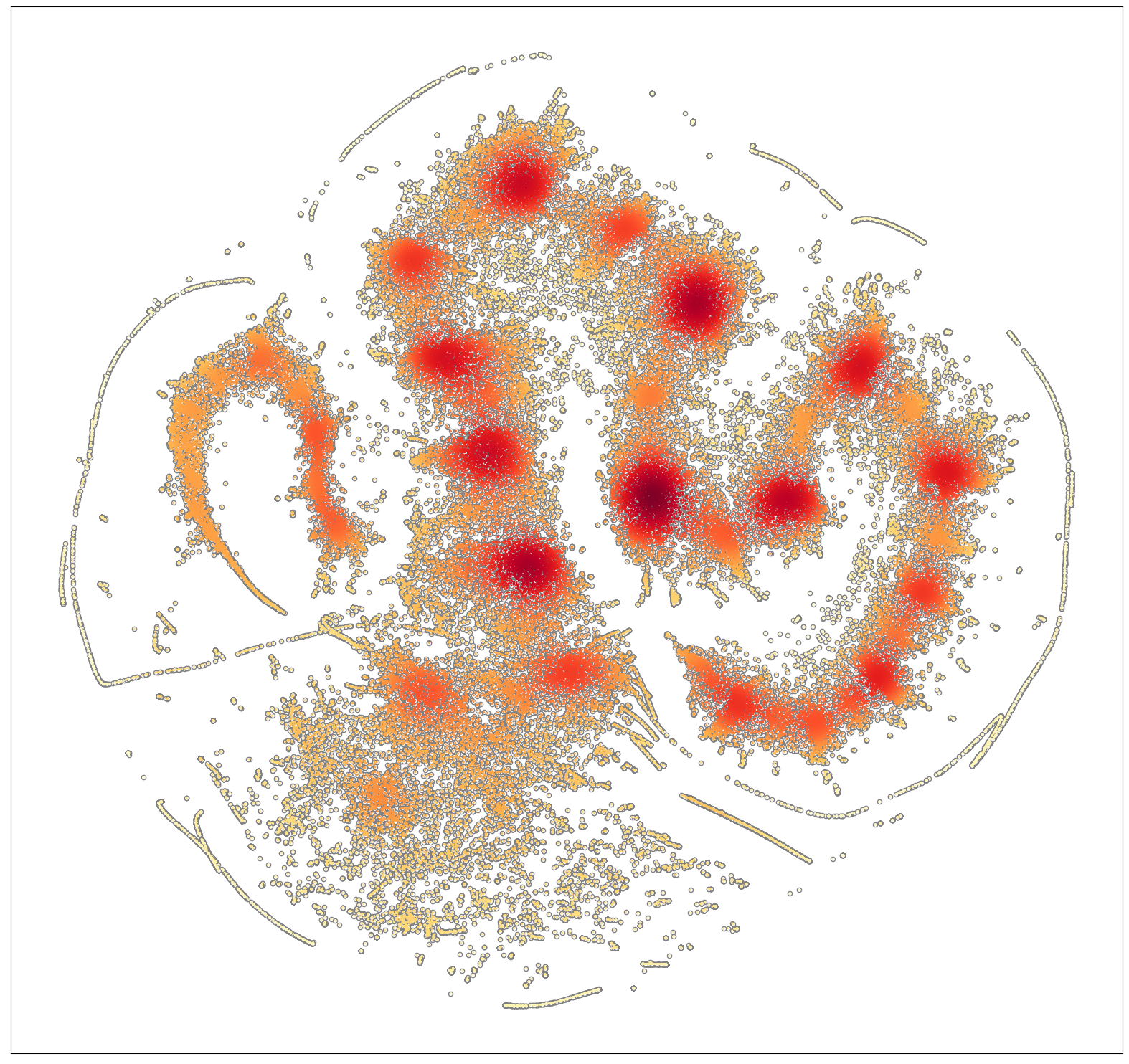}  
        \caption{}
        \label{fig:fod0g}
    \end{subfigure}%
    \hfill
        \begin{subfigure}[b]{0.32\textwidth}
        \centering
        \includegraphics[width=1.0\textwidth]{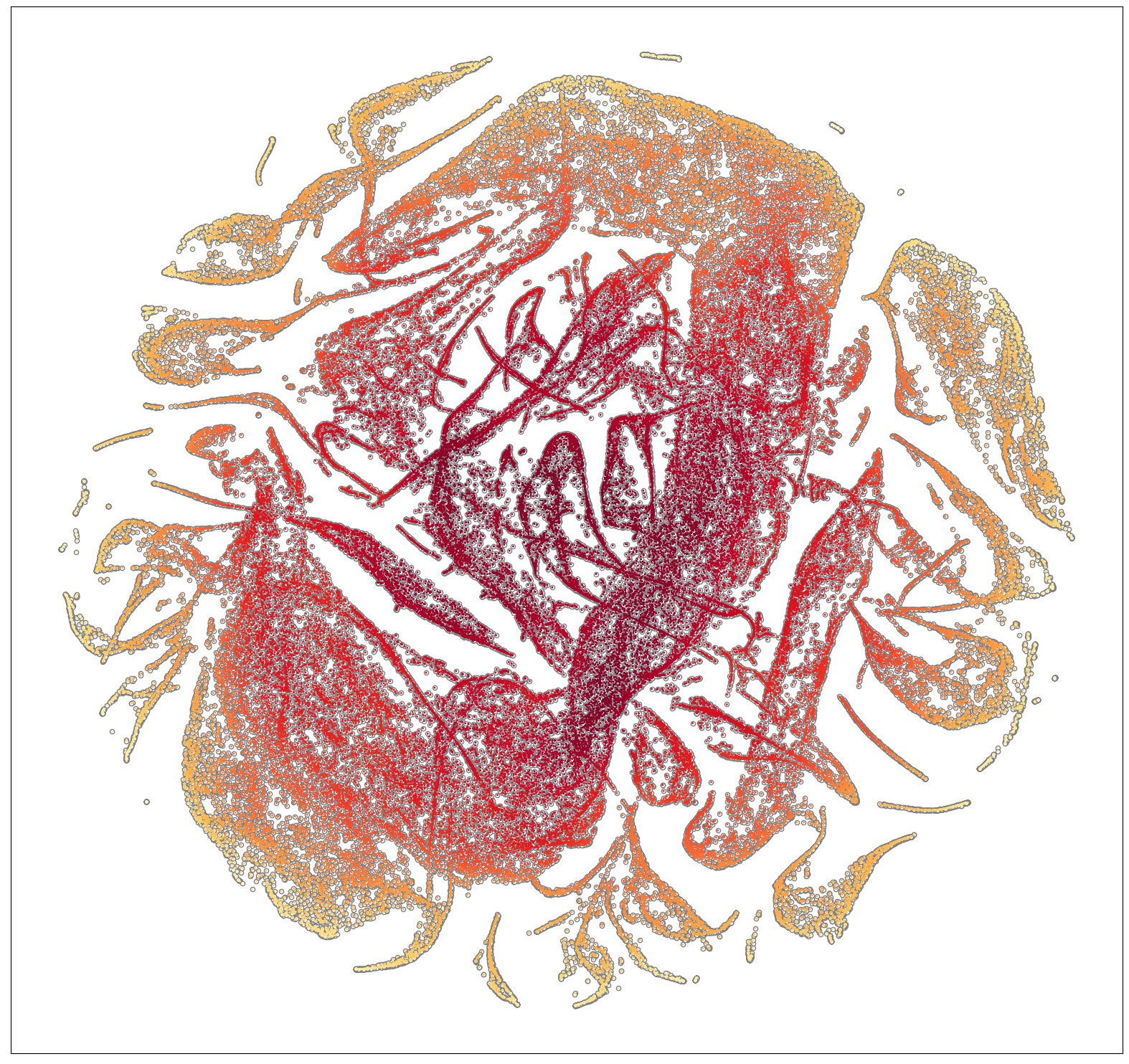}  
        \caption{}
        \label{fig:Topog}
    \end{subfigure}%
    \hfill

    \begin{subfigure}[b]{0.32\textwidth}
        \centering
        \includegraphics[width=1.0\textwidth]{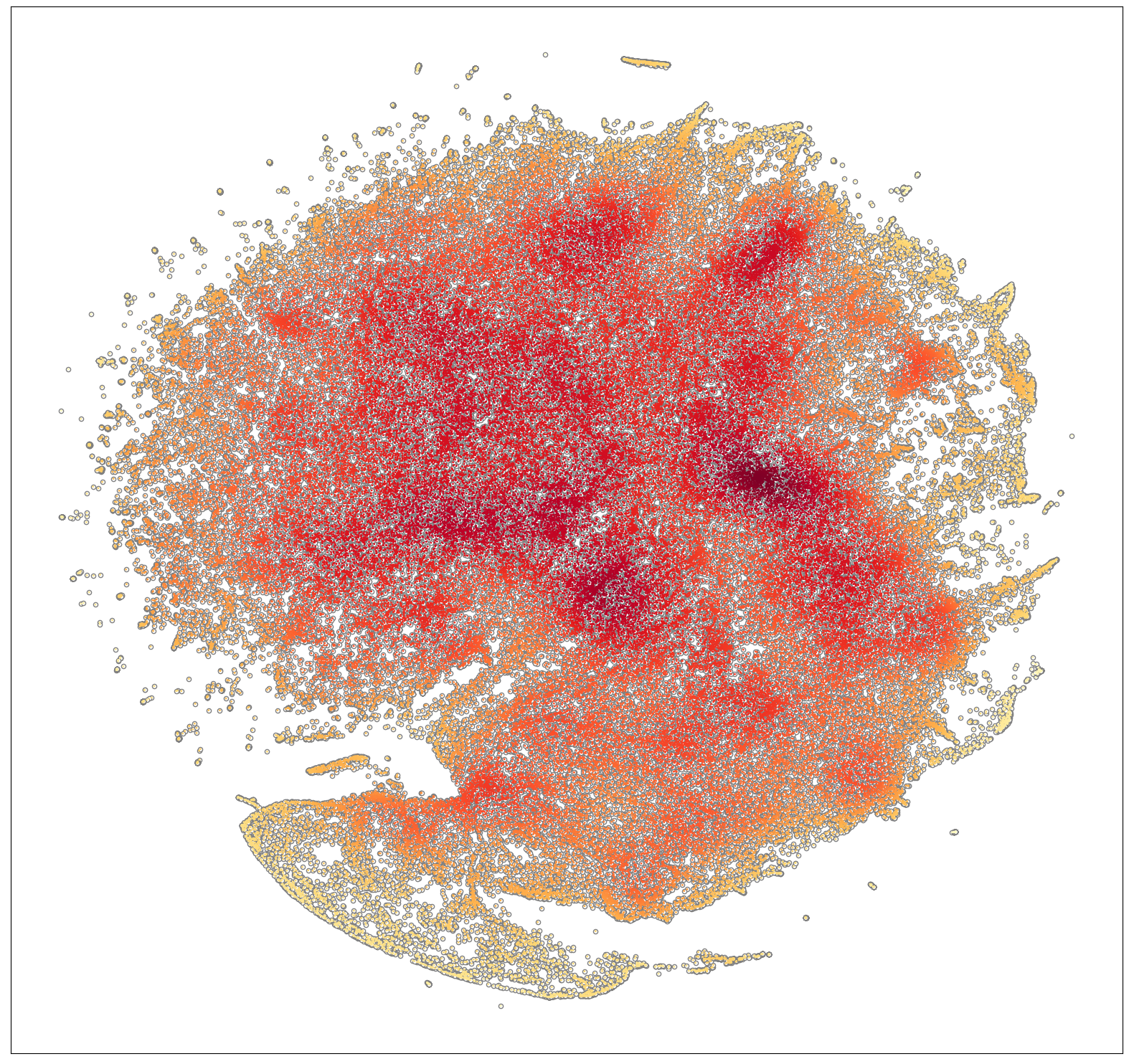}  
        \caption{}
        \label{fig:pwdm100g}
    \end{subfigure}%
    \hfill
    \begin{subfigure}[b]{0.32\textwidth}
        \centering
        \includegraphics[width=1.0\textwidth]{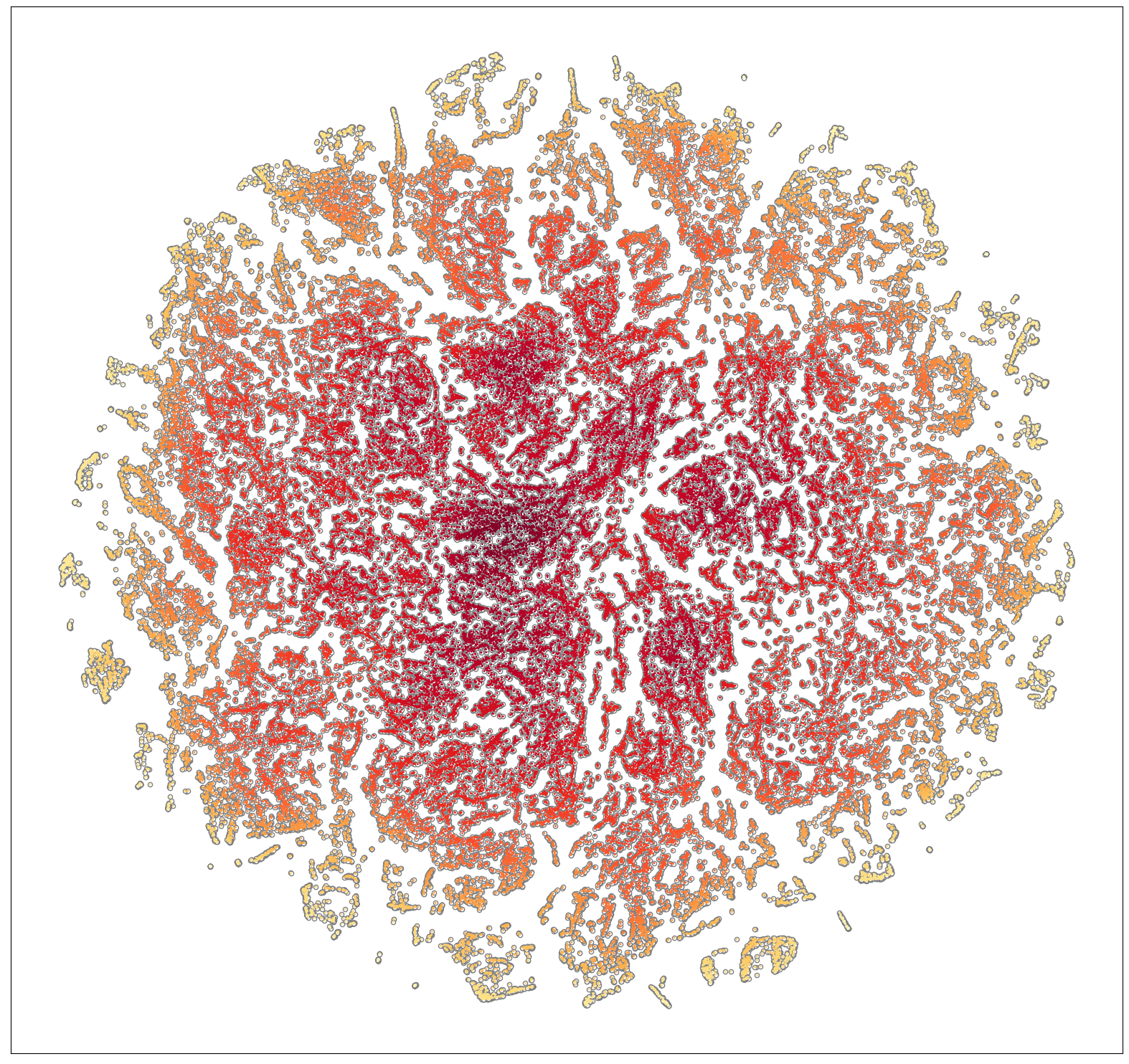}
        \caption{}
        \label{fig:GATGNNg}
    \end{subfigure}%
    \hfill
    \begin{subfigure}[b]{0.32\textwidth}
        \centering
        \includegraphics[width=1.0\textwidth]{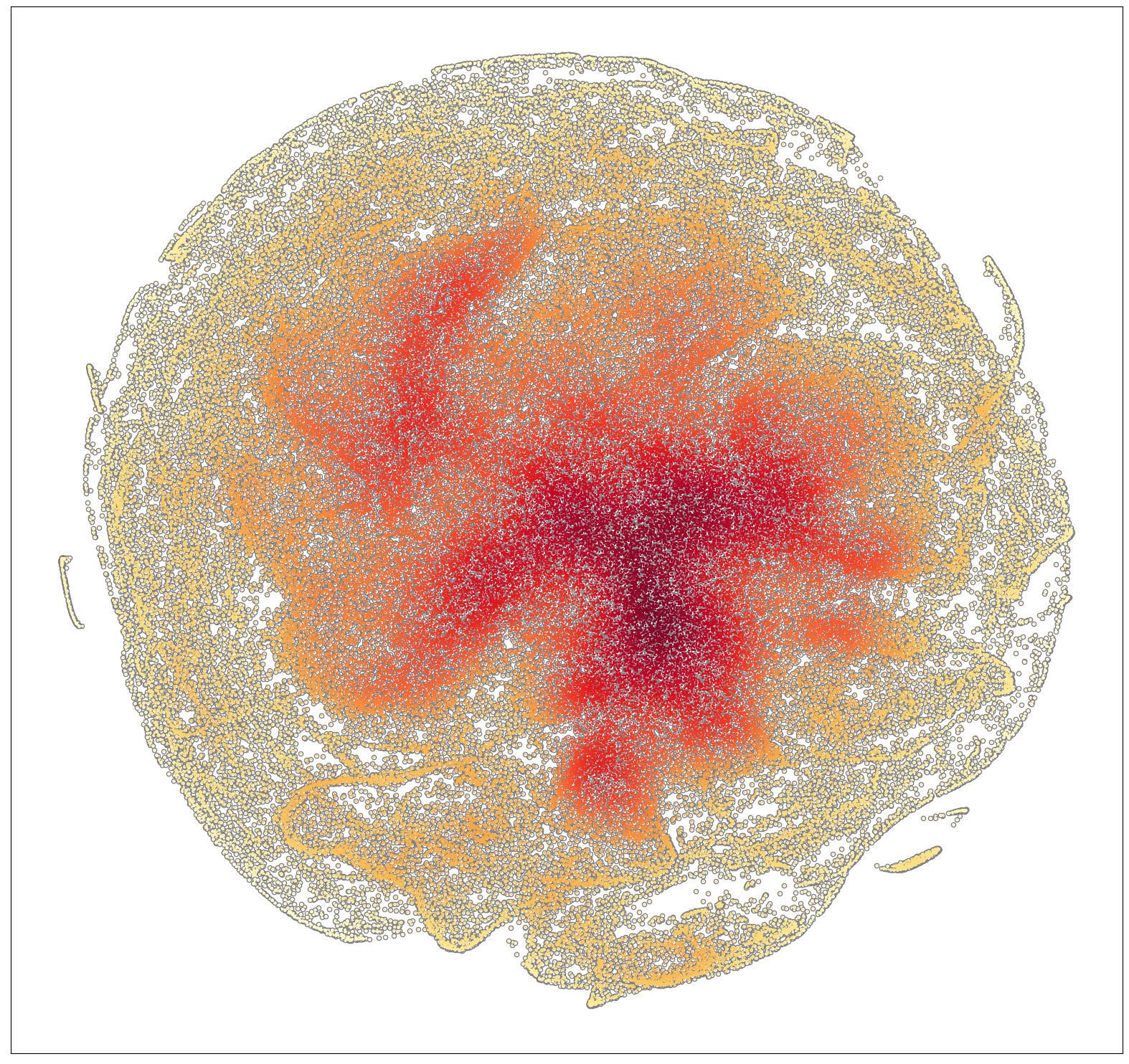}  
        \caption{}
        \label{fig:xrdg}
    \end{subfigure}%
    \hfill
    
    \begin{subfigure}[b]{0.32\textwidth}
        \centering
        \includegraphics[width=1.0\textwidth]{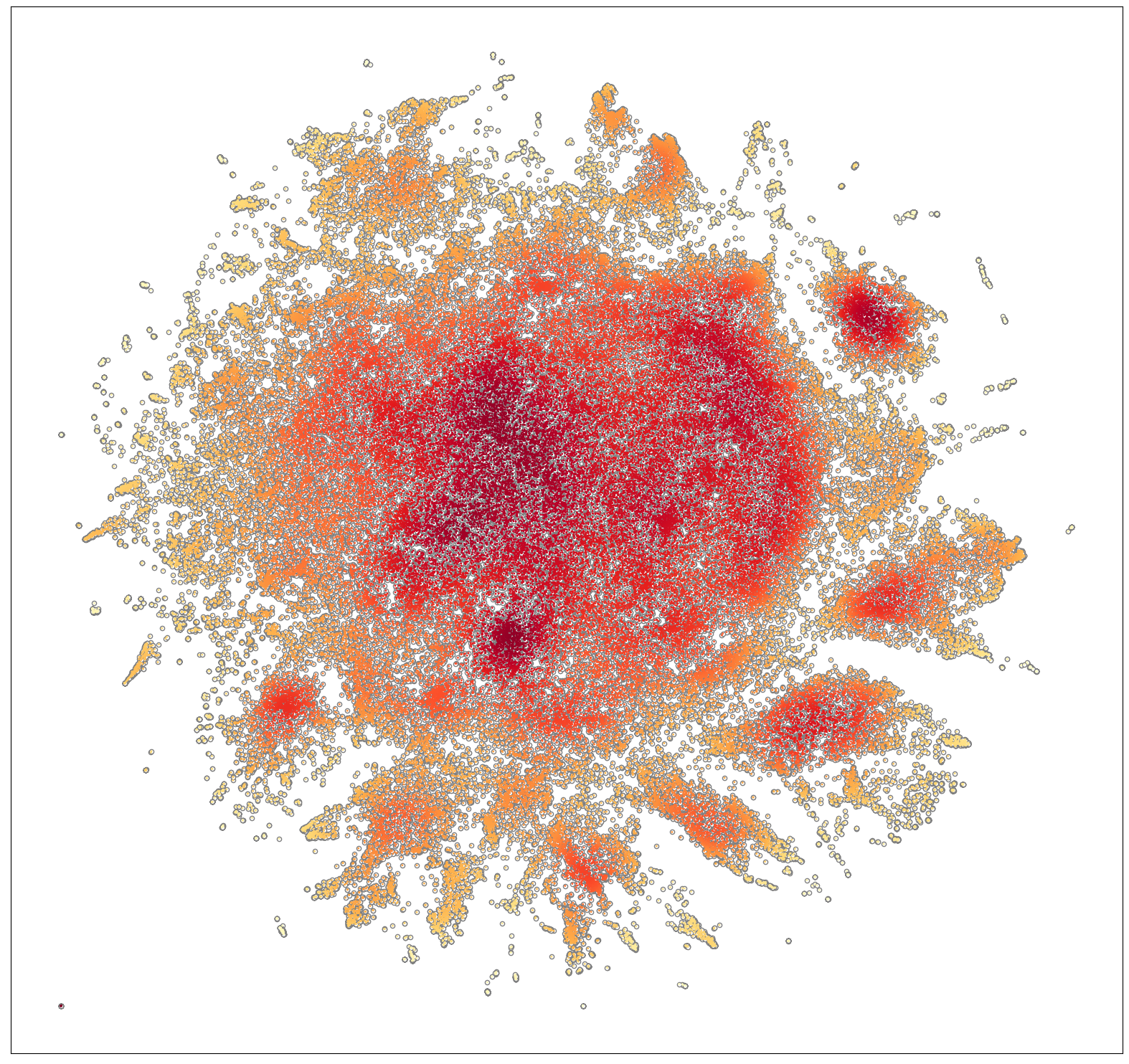} 
        \caption{}
        \label{fig:compg}
    \end{subfigure}%
    \hfill

\caption{Global density distributions of all MP materials using different descriptors.
(a) Cartesian Coordinates 0-padding 
(b) Fractional Coordinates 0-padding 
(c) Topology 
(d) Atomic pairwise distance  
(e) DeeperGATGNN latent feature  
(f) XRD feature 
(g) Composition 
}     
\label{fig:globaldist}
\end{figure*}

Figure \ref{fig:globaldist}(a) shows the global mapping of MP materials in the space of Cartesian coordinates with zero-padding. 
We can see a set of distinct shallow clusters (seen as the red density clusters), which reflect the fact that MP materials form many different material families. Some of these families are derived due to the tinkering discovery process of materials over history. The clusters of similar materials due to tinkering (e.g. element substitutions) are also shown as the string-shaped clusters formed in almost every mapping method. Figure \ref{fig:globaldist}(b) shows the global map of MP materials in the space of fractional coordinates after t-SNE dimension reduction. We find that it also contains distinct clusters similar to Figure \ref{fig:globaldist}(a), except that the density red clusters are more compact round ones. This is probably because the fraction coordinates remove the variation of unit cell lengths and angles. It also contains chain-like clusters. 

Figure \ref{fig:globaldist}(c) shows the MP material distribution in the persistence homology feature space. The map has many connected island-shaped clusters while there are no clear density clusters as seen in Figure \ref{fig:globaldist}(a) and (b). 
Figure \ref{fig:globaldist}(d) shows the global map of MP materials in the space of atomic pairwise distances histogram features. Compared to the previous three maps(a), (b), and (c), this map is much more evenly distributed and there are few distinct density clusters. There are also few island-shaped clusters as seen in other maps.

While Figure \ref{fig:globaldist}(a-d) show the global maps of MP materials without considering the atomic types, the following three figures show the distribution of samples in the physical feature spaces. In Figure \ref{fig:globaldist}(e), the MP materials are mapped into the DeeperGATGNN latent feature space. This latent feature extraction model was trained for predicting formation energy, which includes both the geometric and the atomic-type information. This process may have mixed the differences of material families leading to fewer distinct density clusters compared to Figure \ref{fig:globaldist}(a) or (b). In Figure \ref{fig:globaldist}(f) which shows the map in the XRD feature space, we also observe fewer distinct density clusters. However, we can still find island-shaped small clusters, showing the tinkering process of a related set of materials. Figure \ref{fig:globaldist}(g) shows the MP materials in the composition space without considering the material structures. The figure shows clear material family clusters. Overall, we find that the global maps of MP materials show different patterns based on the feature space that incorporate different sources of information on the crystal materials.

\subsection{Global mapping of material families }
\label{subsec:global_mapping}
\begin{figure*}
    \centering
    \begin{subfigure}[b]{0.24\columnwidth}
        \includegraphics[width=1\columnwidth]{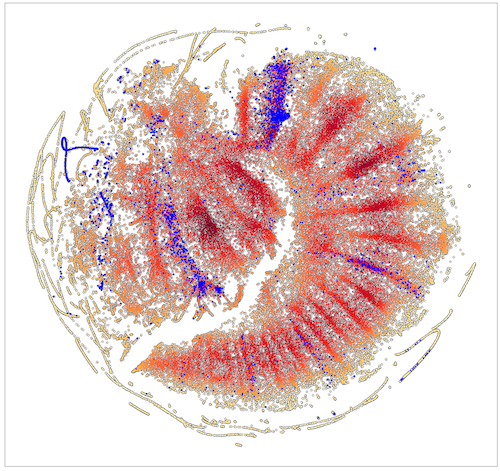}  
        \caption{}
        \label{fig:ccod0_abc3_dis}
    \end{subfigure}%
    \centering
    \begin{subfigure}[b]{0.24\columnwidth}
\includegraphics[width=1\columnwidth]{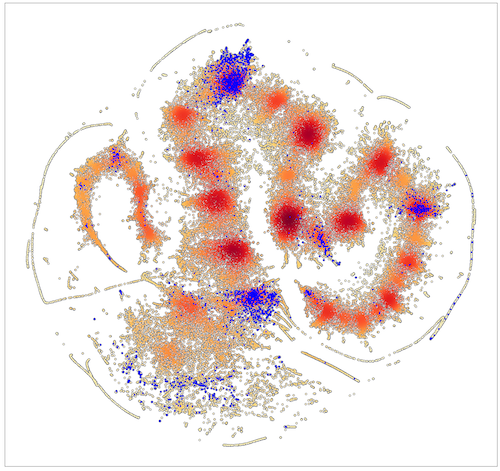}  
        \caption{}
        \label{fig:fcod0_abc3_dis}
    \end{subfigure}%
    \centering
    \begin{subfigure}[b]{0.24\columnwidth}
\includegraphics[width=1\columnwidth]{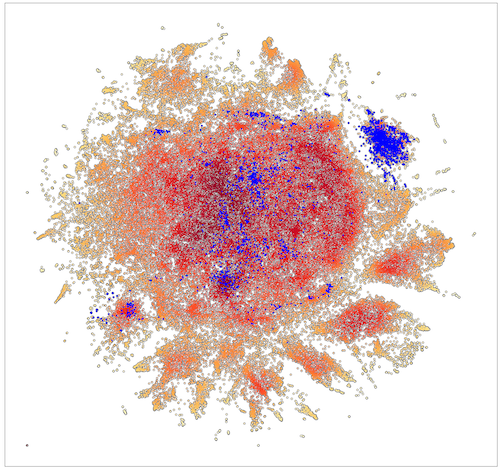}  
        \caption{}
        \label{fig:comp_abc3_dist}
    \end{subfigure}%
    \centering
    \begin{subfigure}[b]{0.24\columnwidth}
\includegraphics[width=1\columnwidth]{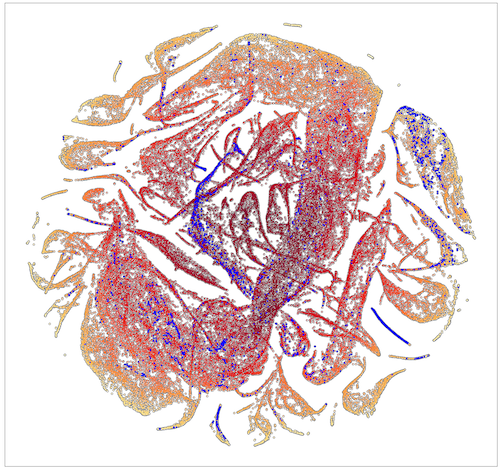}  
        \caption{}
        \label{fig:topo_abc3_dist}
    \end{subfigure}%
    \centering

    \begin{subfigure}[b]{0.24\columnwidth}
\includegraphics[width=1\columnwidth]{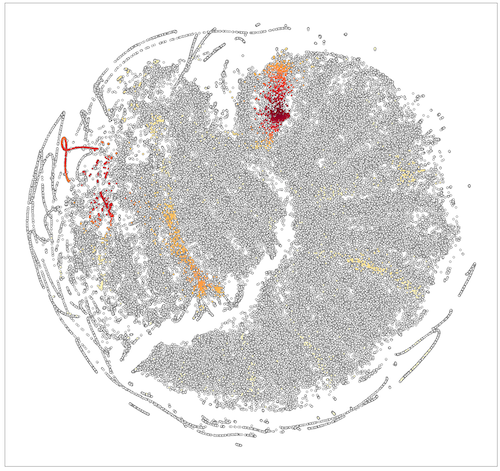}  
        \caption{}
        \label{fig:ccod0_abc3_dens}
    \end{subfigure}%
    \centering
    \begin{subfigure}[b]{0.24\columnwidth}
\includegraphics[width=1\columnwidth]{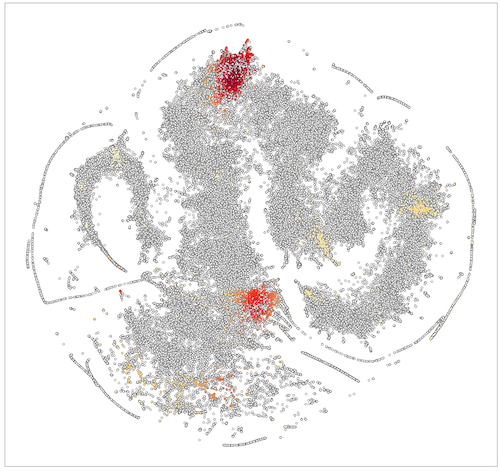}  
        \caption{}
        \label{fig:fcod0_abc3_den}
    \end{subfigure}%
    \centering
    \begin{subfigure}[b]{0.24\columnwidth}
\includegraphics[width=1\columnwidth]{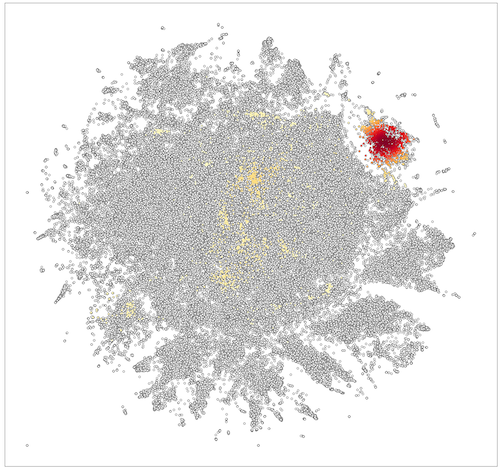}  
        \caption{}
        \label{fig:comp_abc3_den}
    \end{subfigure}%
    \centering
    \begin{subfigure}[b]{0.24\columnwidth}
\includegraphics[width=1\columnwidth]{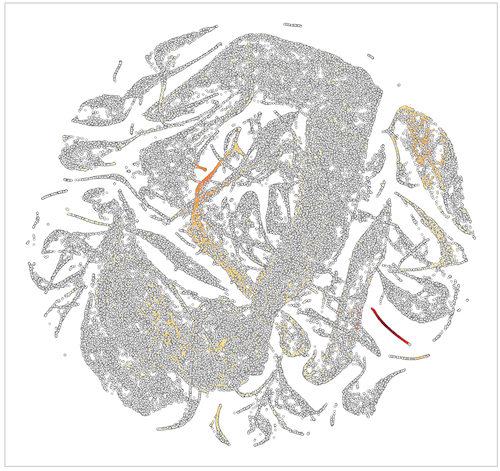}  
        \caption{}
        \label{fig:topo_abc3_dens}
    \end{subfigure}%

\caption{ ABC$_3$ prototype materials with respect to global density distributions of all MP datasets.  
In (a)-(d), ABC$_3$ materials distribution is represented in blue scatter with a background of global distribution density. 
In (e)-(h), ABC$_3$ materials distribution density is represented with global distributions in the background (gray circles). 
The density representation used a yellow-red color scheme, where red indicates a high density.
Mapping descriptors:
(a) \& (e) :  Cartesian Coordinates,  
(b) \& (f) :  Fractional Coordinates, 
(c) \& (g) :  Composition,   
(d) \& (h) :  Topology. 
 }     
\label{fig:ABC3MP}     
\end{figure*}

\paragraph{Mapping of ABC$_3$ Materials.} 
ABC$_3$ prototype is one of the most common composition types of crystal materials. Here we study the distribution of 4,358 crystals with the ABC$_3$ composition prototype within the global maps in Figure \ref{fig:globaldist}, which allows us to understand how the material family changes with respect to the features. 
In Figure \ref{fig:ABC3MP}(a) to (d), the ABC$_3$ materials are marked with blue color dots which scatter over the global MP distribution map. Figure \ref{fig:ABC3MP}(e) to (h) show the density of ABC$_3$ compounds of corresponding distributions in Figure \ref{fig:ABC3MP}(a) to (d), which shows the MP material distributions along with their densities represented as yellow-red heat maps. 
We notice that for ABC$_3$ materials, there are four distinct major clusters formed in both the coordinate map Figure \ref{fig:ABC3MP}(a) and (b). Their corresponding density maps in terms of ABC$_3$ materials in Figure \ref{fig:ABC3MP}(e) and (f) also show the clusters clearly. In the composition map of Figure \ref{fig:ABC3MP}(c), the ABC$_3$ materials also show up as a few large clusters with the largest cluster in the top right corner. Comparing these three maps, we can conclude that ABC$_3$ materials tend to form structures with a certain similarity in terms of Cartesian and fractional coordinates. In the topology map Figure \ref{fig:ABC3MP}(d), the ABC$_3$ prototypes also show up as a few clusters, but with different cluster shapes compared to those in those coordinate maps. In comparison, we have not observed clear clusters for the ABC$_3$ materials in all three other feature maps including atomic pairwise distance, DeeperGATGNN latent features, and XRD features (See Supplementary Figure S1).

In addition to the material distribution, we are also able to uncover the connection between different mapping descriptors. As shown in Figure \ref{fig:ABC3MP}, the top right major cluster on the composition map Figure \ref{fig:ABC3MP}(g) and the lower right island-shaped cluster on the topology map Figure \ref{fig:ABC3MP}(b) contains exactly the same group of crystal materials.

\paragraph{Mapping of material properties.}

\begin{figure*}[th]
    \centering
    \begin{subfigure}[b]{0.33\textwidth}
        \centering
        \includegraphics[width=1.0\textwidth]{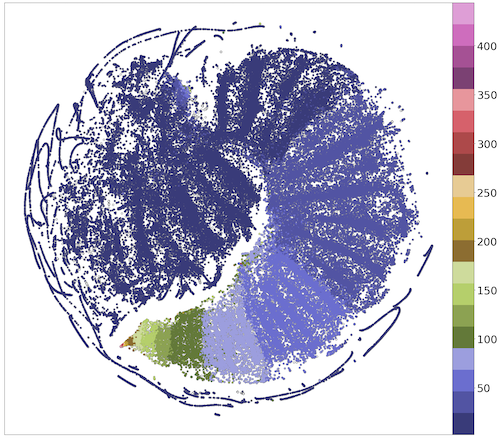}  
        \caption{}
        \label{fig:globalnsitescod0}
    \end{subfigure}%
    \hfill
    \begin{subfigure}[b]{0.33\textwidth}
        \centering
        \includegraphics[width=1.0\textwidth]{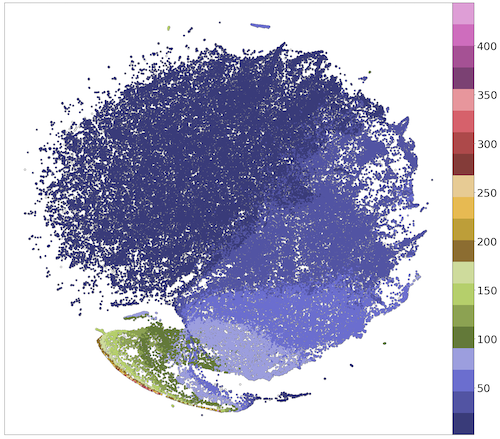}  
        \caption{}
        \label{fig:globalnsitespwdm}
    \end{subfigure}%
    \hfill
    \begin{subfigure}[b]{0.33\textwidth}
        \centering
        \includegraphics[width=1.0\textwidth]{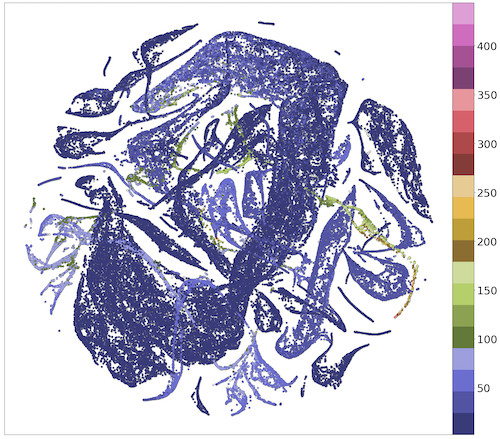}  
        \caption{}
        \label{fig:globalnsitesTopog}
    \end{subfigure}%
    \hfill

\caption{
Distribution of materials in terms of atomic site numbers with different descriptors
(a) Cartesian coordinates 0-padding,
(b) Atomic pairwise distance,
(c) Topology.
}     
\label{fig:globalnsites}

\end{figure*}

With a global map of all MP materials, it is interesting to check how a given set of materials with specific physicochemical (e.g. piezoelectric) or geometric (e.g. cubic) properties are distributed. To explore these patterns, we first check how the materials are distributed in terms of their number of atomic sites within the unit cell with different mapping descriptors. Figure \ref{fig:globalnsites} shows the mapping result. Please note that to reduce the issue of too many different values (no. of atomic sites), we divide the whole atom site number range into five classes (indicated as five different colors) and each class is further divided into four different levels (indicated as the lightness of a specific color). In the MP dataset, most crystal structures have fewer than 100 atomic sites. This is why all the plots in this figure have the majority of the crystal materials colored blue.

Figure \ref{fig:globalnsites}(a) shows the distribution of materials in terms of their number of atomic sites over a global map of MP materials in the space of Cartesian coordinates with zero-padding. 
We can see that there exists a smooth transition between different numbers of sites. Because we add zero padding for dimension alignment, the materials with a similar number of atomic sites tend to have similar descriptor features, which lead to stripe-shaped clusters after t-SNE mapping. The proportion of the area for each color region is consistent with the distribution of the MP data set where almost 67\% of the materials have fewer than 25 atoms within their unit cells. 
Figure \ref{fig:globalnsites}(b) shows the distribution of materials in terms of their number of atomic sites over the global pairwise distance map of MP materials. 
We find that distribution has smooth transitions between areas with a different number of sites, without distinct clusters as shown in Figure \ref{fig:globalnsites}(a). 
Figure \ref{fig:globalnsites}(c) shows the distribution of materials in terms of their number of atomic sites over the global persistence homology feature map of MP materials. It is found that the map has many connected island-shaped clusters as mentioned before. 
In the center of the map, there is a giant clamp-shaped cluster with all the materials in denim (dark blue) and other small strip-shaped clusters in color navy(light blue). Comparing Figure \ref{fig:globalnsites}(b) and (c), it is found that the latter map has more scattered points with different colors while the former one has smooth gradient areas. This is due to the fact that the persistence topology feature is a more generalized descriptor that tends to have more even distribution across the entire domain. 
Here we have shown that Figure \ref{fig:globalnsites}(c) manages to show the material clusters while other descriptors do not as shown in Supplementary file Figure S2.

\begin{figure*}[th]
    \centering
    \begin{subfigure}[b]{0.315\textwidth}
        \centering
        \includegraphics[width=1.0\textwidth]{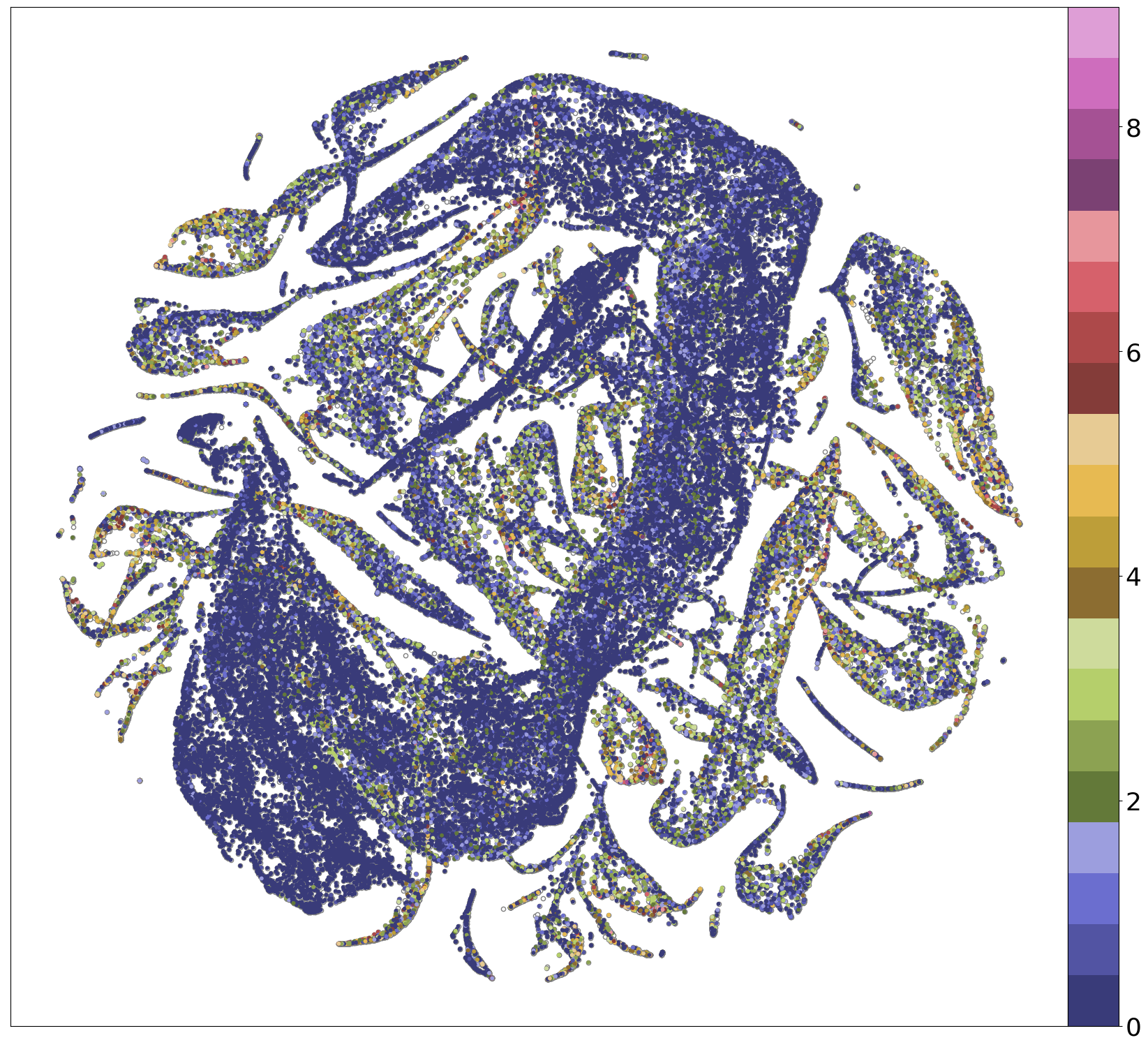}
        \caption{}
        \label{fig:pwdm_bg}
    \end{subfigure}%
    \hfill
    \begin{subfigure}[b]{0.33\textwidth}
        \centering
        \includegraphics[width=1.0\textwidth]{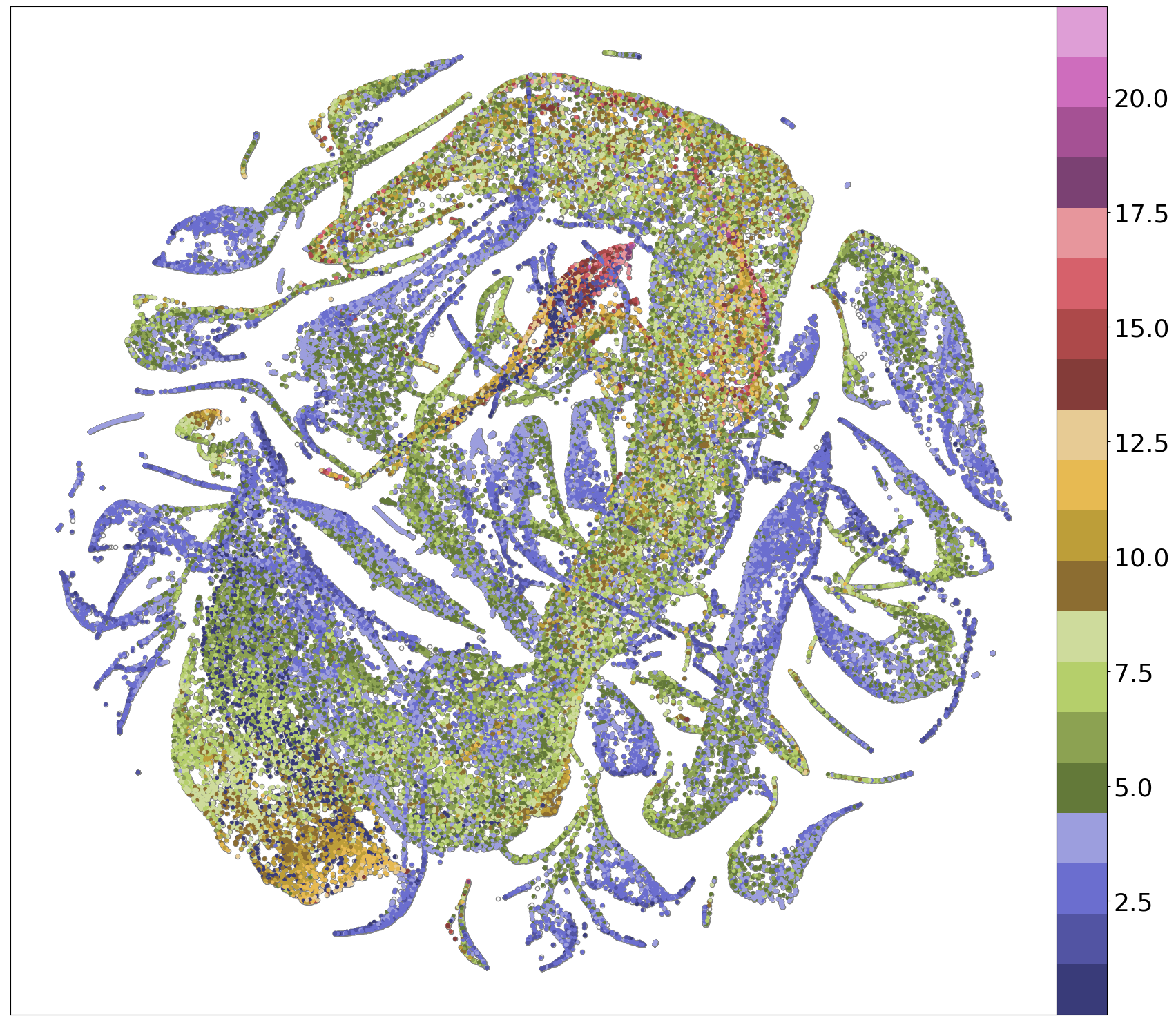}  
        \caption{}
        \label{fig:pwdm_den}
    \end{subfigure}%
    \hfill
    \begin{subfigure}[b]{0.33\textwidth}
        \centering
        \includegraphics[width=1.0\textwidth]{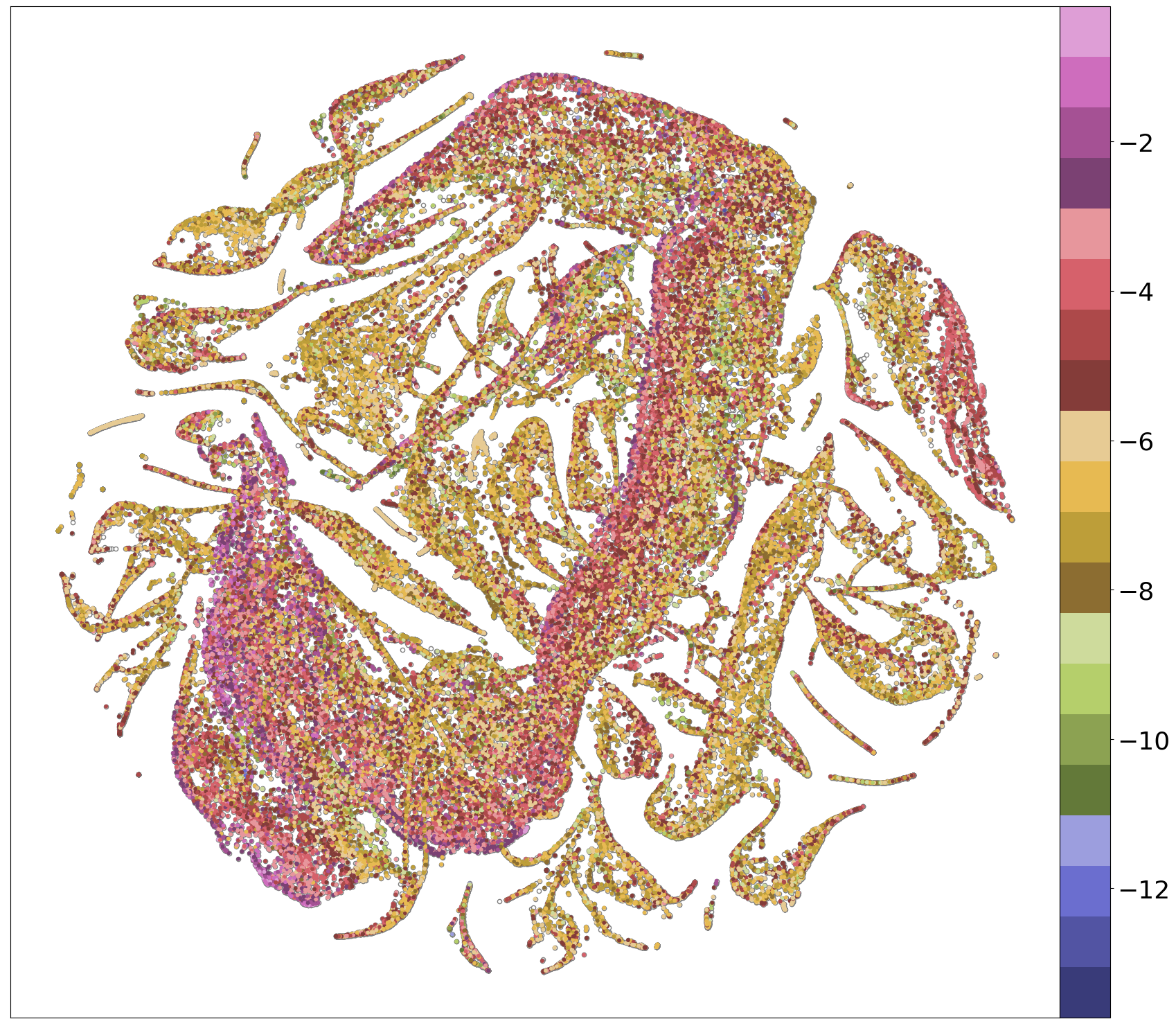}  
        \caption{}
        \label{fig:pwdm_fepa}
    \end{subfigure}%
    \hfill

\caption{ Distributions of materials with different properties including 
(a) Band gap (eV),
(b) Density (g/cm$^3$),
(c) formation energy per atom (ev/atom).
}     
\label{fig:pwdmprop}

\end{figure*}

To further understand how different material properties are distributed in the global material space, we plot the global map of three different material properties including band gap, density, and formation energy per atom in Figure \ref{fig:pwdmprop} where the colors and their lightness represent the relative magnitudes of the properties and the map is the persistence homology topology space as we described previously.

Figure \ref{fig:pwdmprop}(a) shows the distribution of all MP materials with different band gaps over the global map of persistence homology topology feature space. An outstanding pattern is the major C-shaped cluster, in which the materials all share a band gap value below 2 eV. Figure \ref{fig:pwdmprop}(b) shows the distribution of the densities over the map in the persistence homology topology feature space. It shows a high-density region of a strip shape in the top middle area with densities between 16 g/cm$^3$ to 10 g/cm$^3$. 
On the bottom left, there is a large cluster with a smooth transition in density values. This is interesting because only a small portion of MP crystal materials are complex structures with large density values. In this density map, we observe three large clusters with a density larger than 10 g/cm$^3$. 
Figure \ref{fig:pwdmprop}(c) shows the distribution of materials with different formation energy per atom over the global map of MP materials in the space of persistence homology topology feature space. First, we find that in general, the materials of most of the topological clusters tend to have similar formation energy per atom values. 
And within most of these individual topological clusters, we notice a smooth transition of the colors from one edge to another. We also find that the materials with the lowest formation energy per atom tend to be located in the region of the big C-shaped stripe. Some of the patterns may be due to the materials families in the MP database being derived due to the tinkering discovery process of materials over time.

\FloatBarrier

\begin{figure*}
    \centering
    \begin{subfigure}[b]{0.32\columnwidth}
        \includegraphics[width=1\columnwidth]{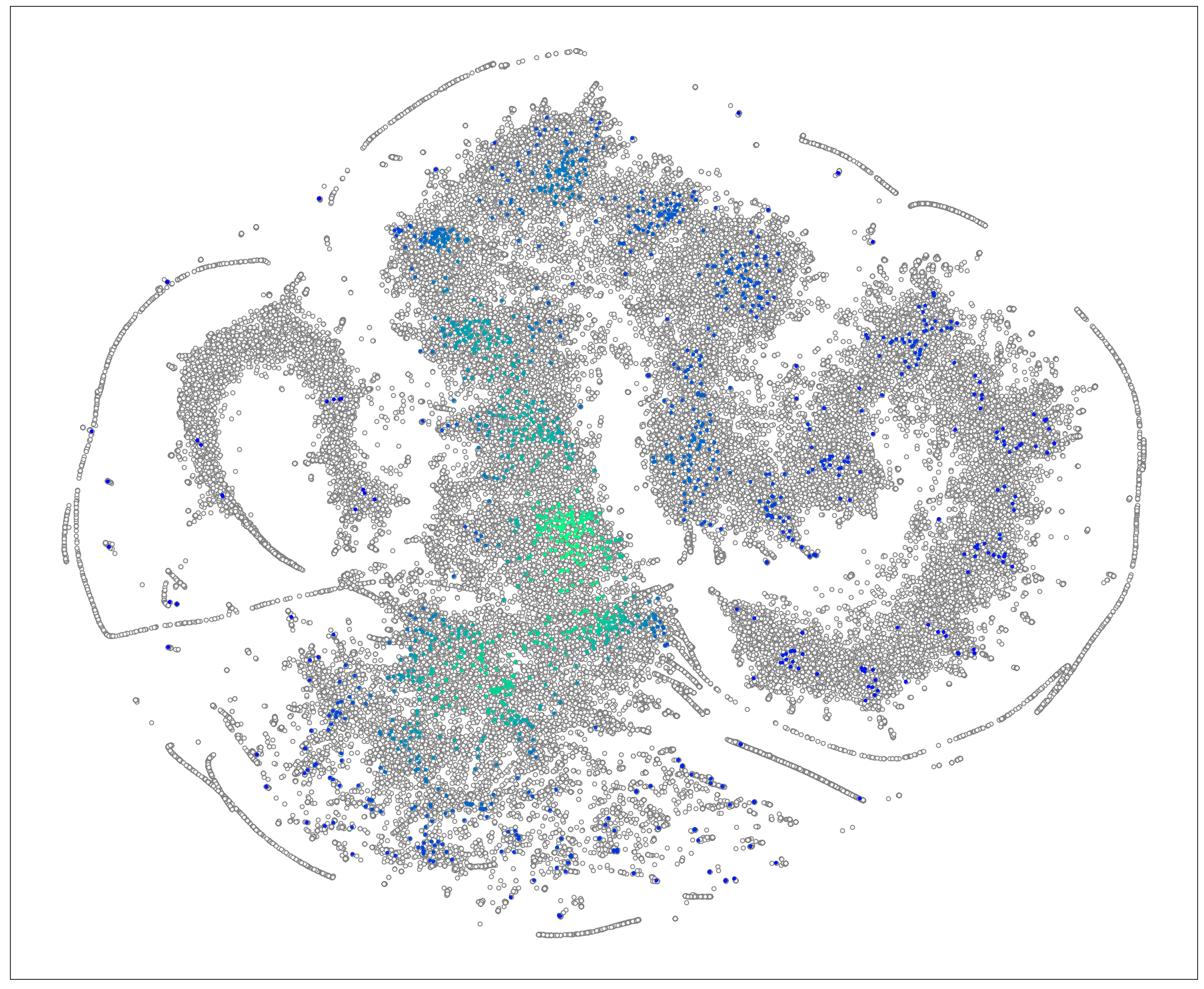}  
        \caption{}
        \label{fig:gpb_fcod}
    \end{subfigure}%
    \hfill
    \begin{subfigure}[b]{0.32\columnwidth}
        \includegraphics[width=1\columnwidth]{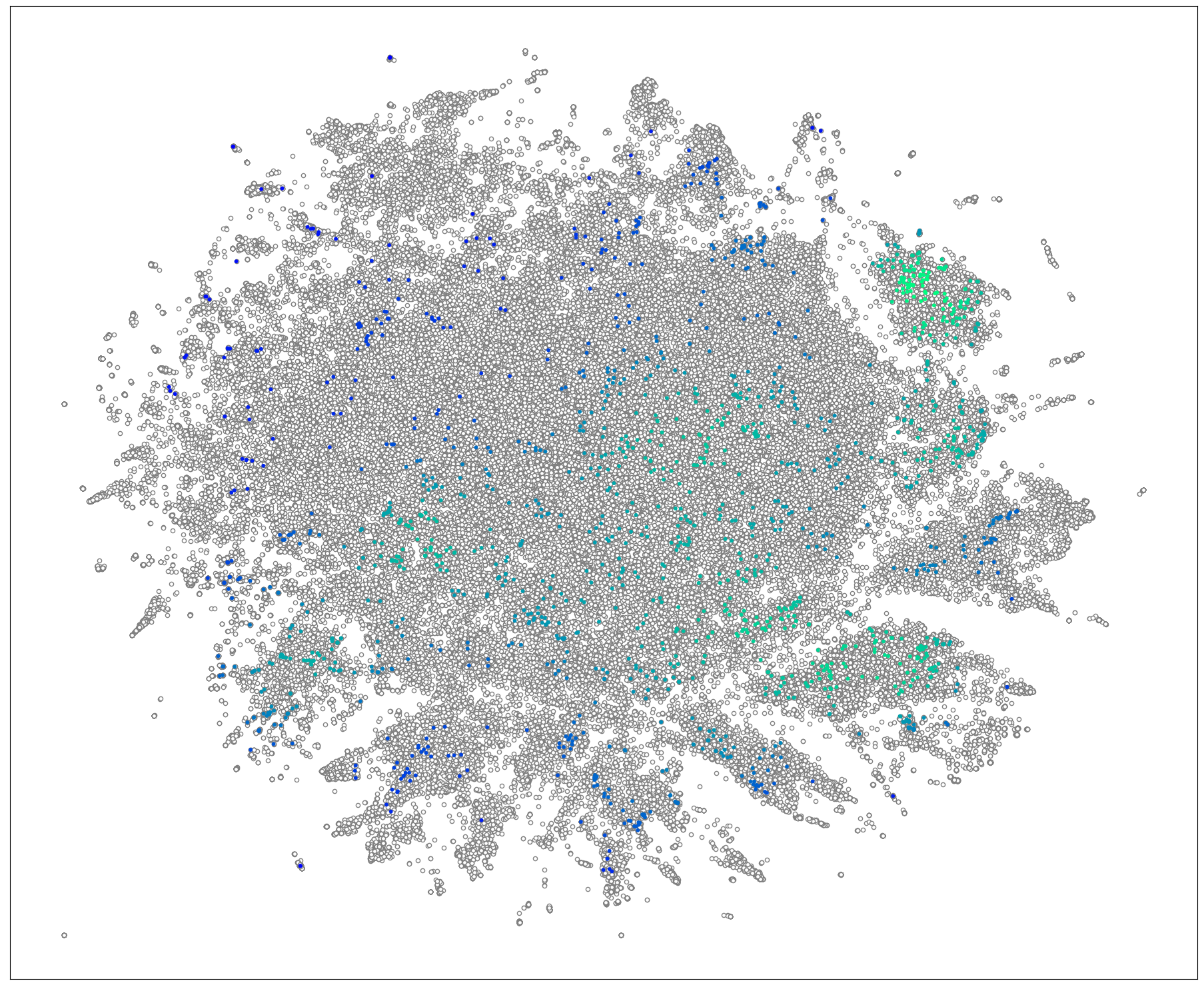}  
        \caption{}
        \label{fig:gpb_comp}
    \end{subfigure}%
    \hfill
    \begin{subfigure}[b]{0.32\columnwidth}
        \includegraphics[width=1\columnwidth]{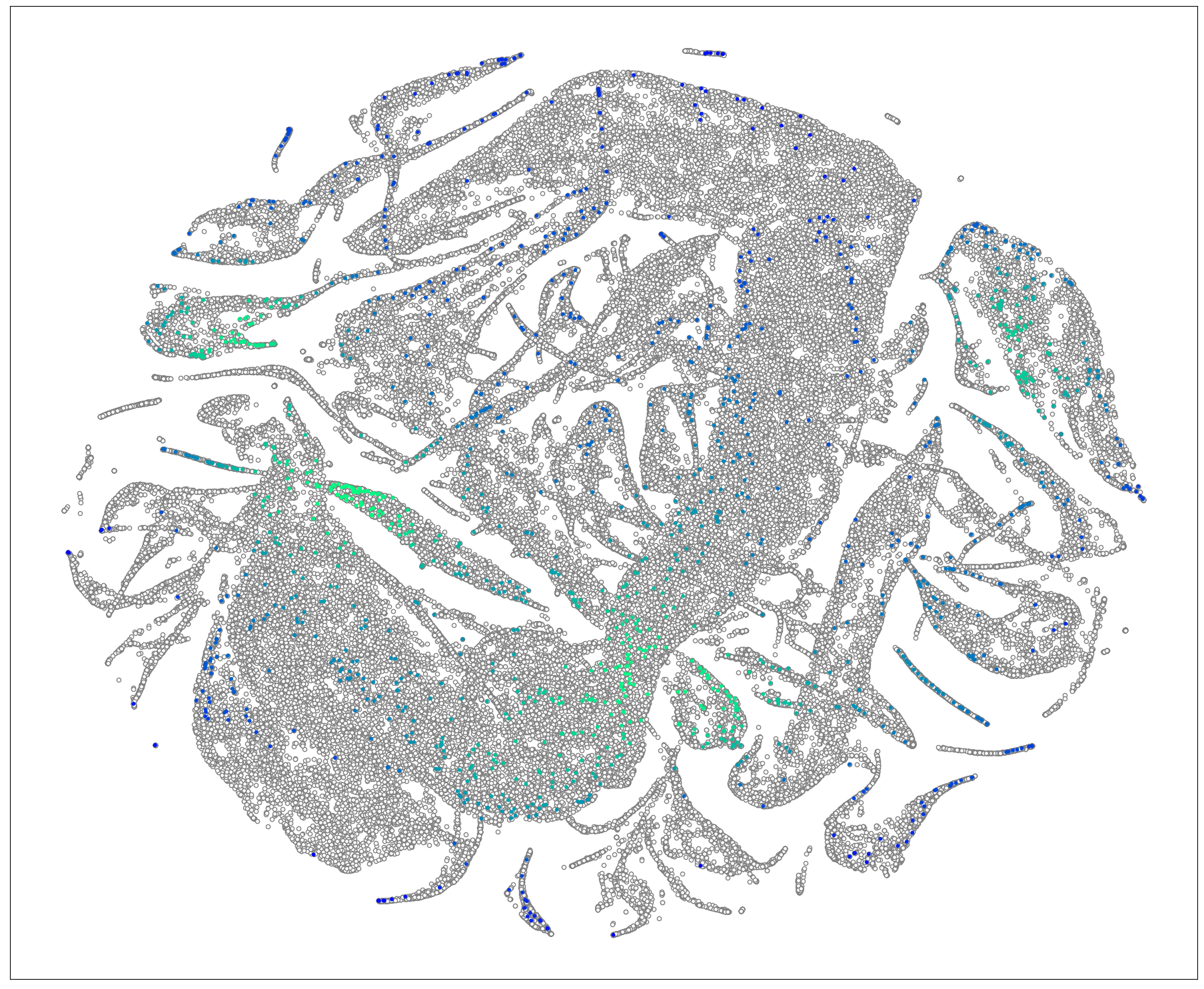}  
        \caption{} 
        \label{fig:gpb_topo}
    \end{subfigure}%

    \caption{ Piezoelectric property distribution with respect to global distributions of all MP materials in the space of 
(a) Fractional Coordinates,
(b) Composition, 
(c) Topology.      
}   
\label{fig:globalBinay}
\end{figure*}

Since different material maps reflect different structural clusters, it is interesting to check how a given material property is distributed over different feature maps. To investigate the issue, we use the piezoelectric property as an example, which is one of the most interesting properties of crystal materials. However, there are not many materials annotated with piezoelectric properties compared to the size of the MP data set. So here we only use two classes to show the distributions of piezoelectric crystals over three different maps in Figure \ref{fig:globalBinay}. 
Here piezoelectric materials are marked in blue-green points with grey points indicating the background of the whole MP data set. The blue-green color indicates the density of the data points and the green color indicates high-density areas.

Figure \ref{fig:globalBinay}(a) shows the density heat map of the piezoelectric materials over the global map of MP materials in the space of Fractional Coordinates with 0-padding. We find that the piezoelectric materials are evenly distributed across the whole domain of the MP map. Figure \ref{fig:globalBinay}(b) shows the density heat map of the piezoelectric materials over the global map of MP materials in the space of composition features.
On the top right region of this map, the background MP materials form an island cluster clearly separated from the main cluster. This island cluster tends to have many piezoelectric materials within it with a high density. 
Figure \ref{fig:globalBinay}(c) shows the density heat map of the piezoelectric materials over the MP map in the space of persistence homology topology feature. Although there is not many connections between the piezoelectric materials with Figure \ref{fig:pwdmprop} can be found, we did notice that there are overlapping data points in the cluster mentioned in Figure \ref{fig:ABC3MP}(g) which are materials of the ABC$_3$ prototype. 
\FloatBarrier

\subsection{Local mapping of ABC$_3$ materials}

While global mapping of all MP materials gives us a whole picture of structural and property distributions, the relationships can be too complex to show up as distinct patterns due to the high diversity of materials in the whole MP dataset. It is thus desirable to study the material property distribution within the local mapping of a subset of the whole MP dataset. 

\paragraph {Structural subset.} In section \ref{subsec:global_mapping}, we discovered a string cluster (805 materials) located at the bottom left of the topology map. We decide to further inspect this group of clusters within the scope of 4,358) ABC$_3$ prototype materials. 

\begin{figure*}[thb]

    \begin{subfigure}[t]{0.32\textwidth}
         \centering
         \includegraphics[width=\textwidth]{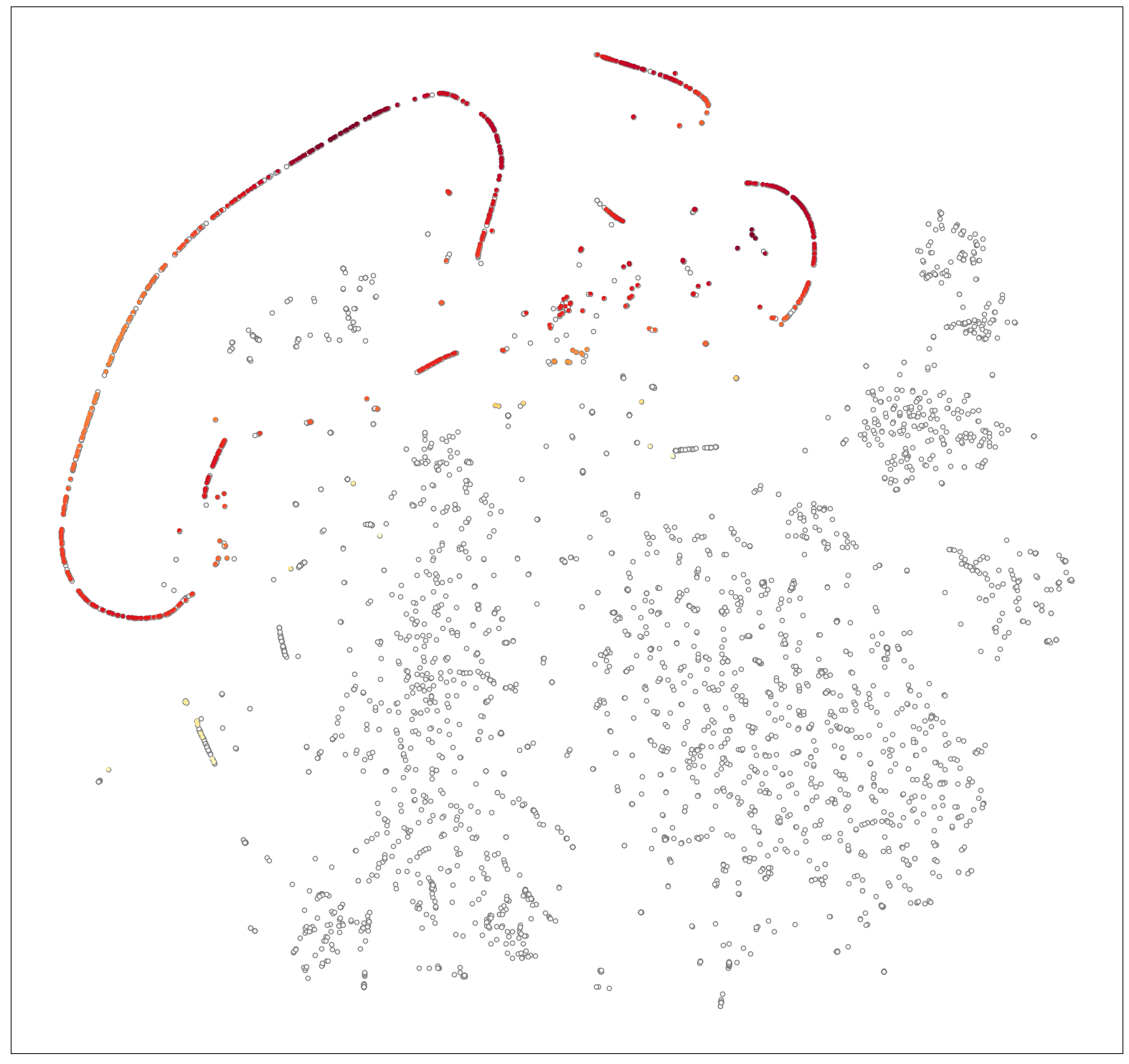}
         \caption{}
     \end{subfigure}
    \hfill
    \begin{subfigure}[t]{0.32\textwidth}
         \centering
         \includegraphics[width=\textwidth]{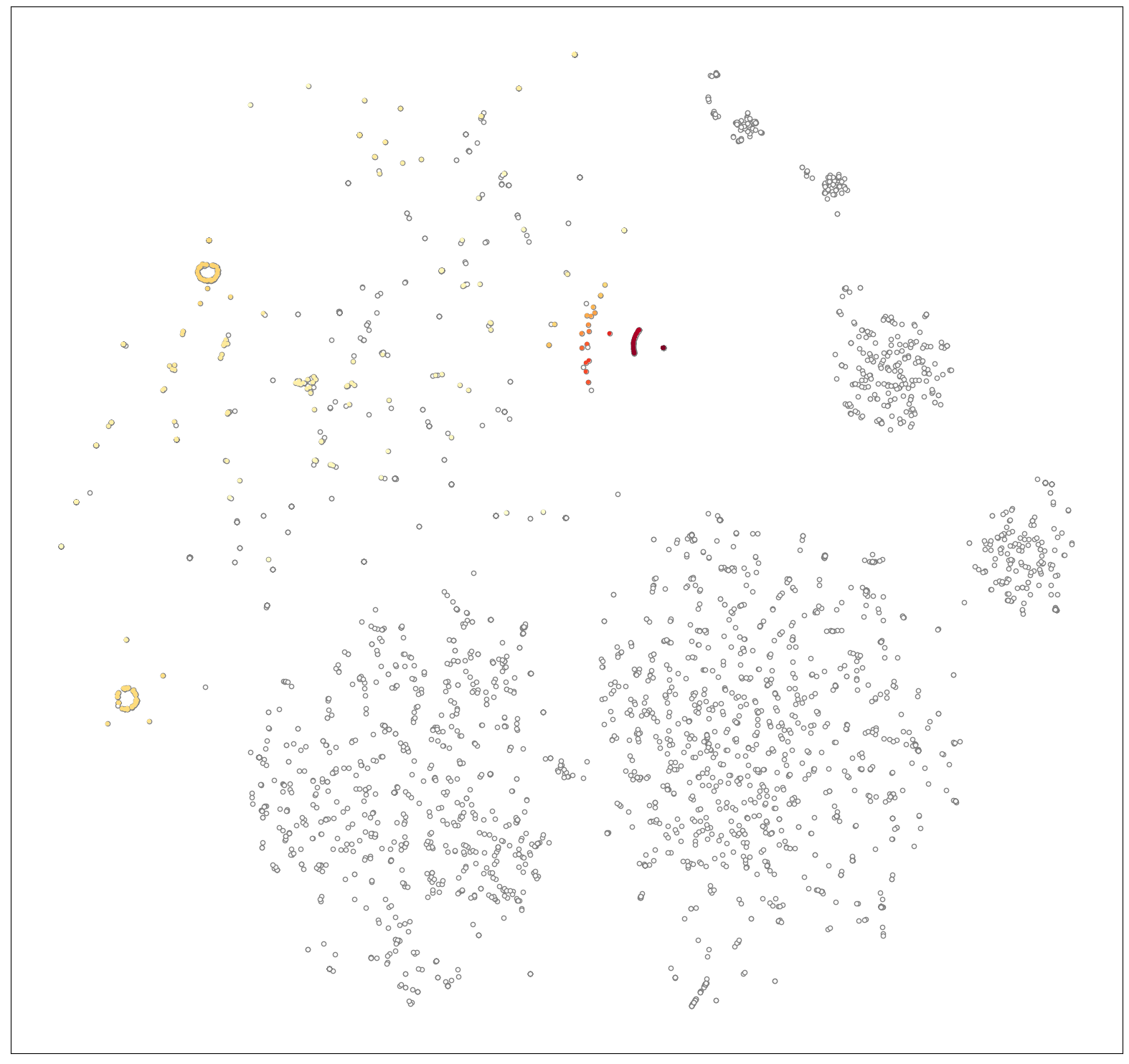}
         \caption{}
     \end{subfigure}
    \hfill
    \begin{subfigure}[t]{0.32\textwidth}
         \centering
         \includegraphics[width=\textwidth]{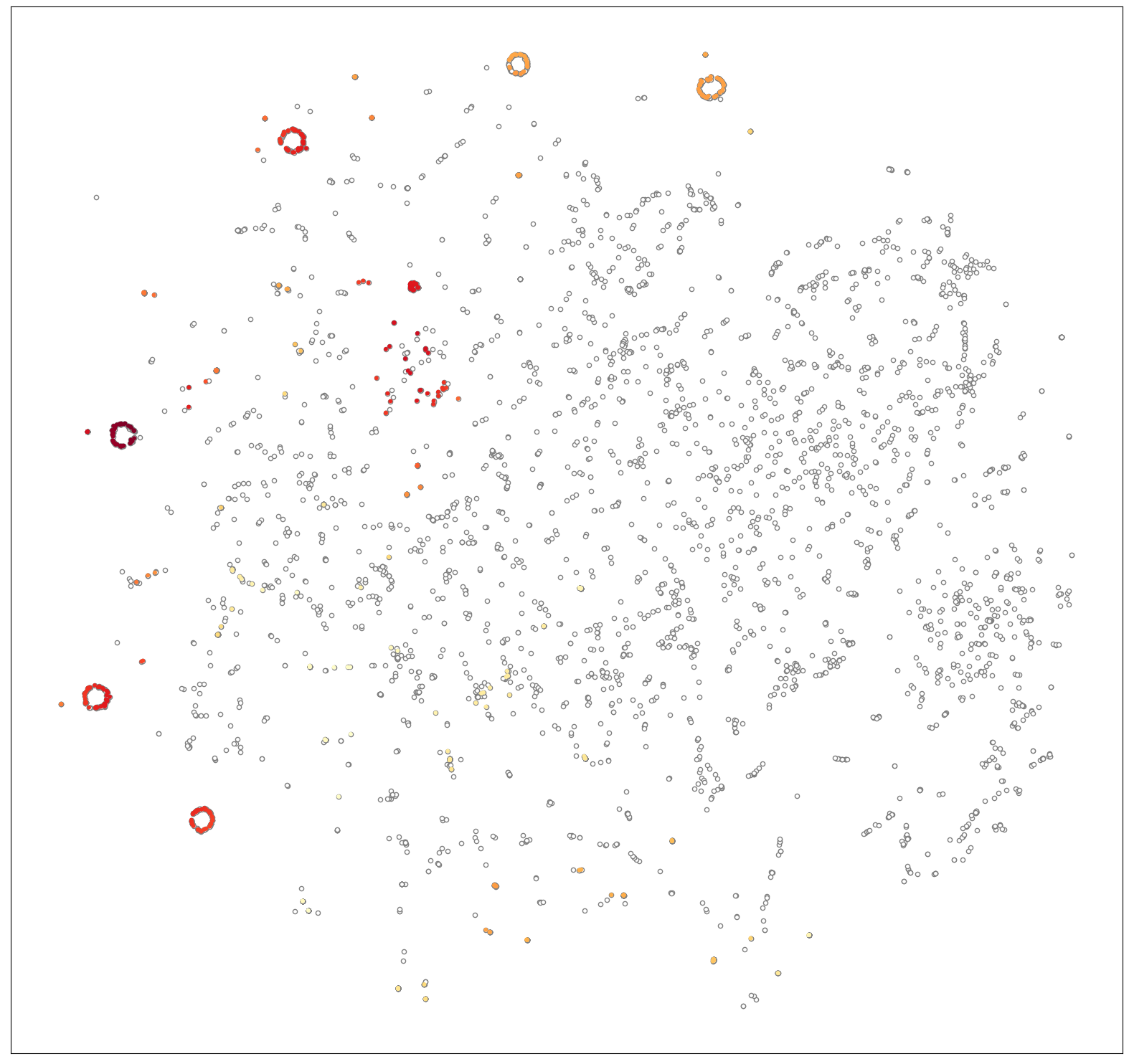}
         \caption{}
     \end{subfigure}
    \hfill

    \begin{subfigure}[t]{0.32\textwidth}
         \centering
         \includegraphics[width=\textwidth]{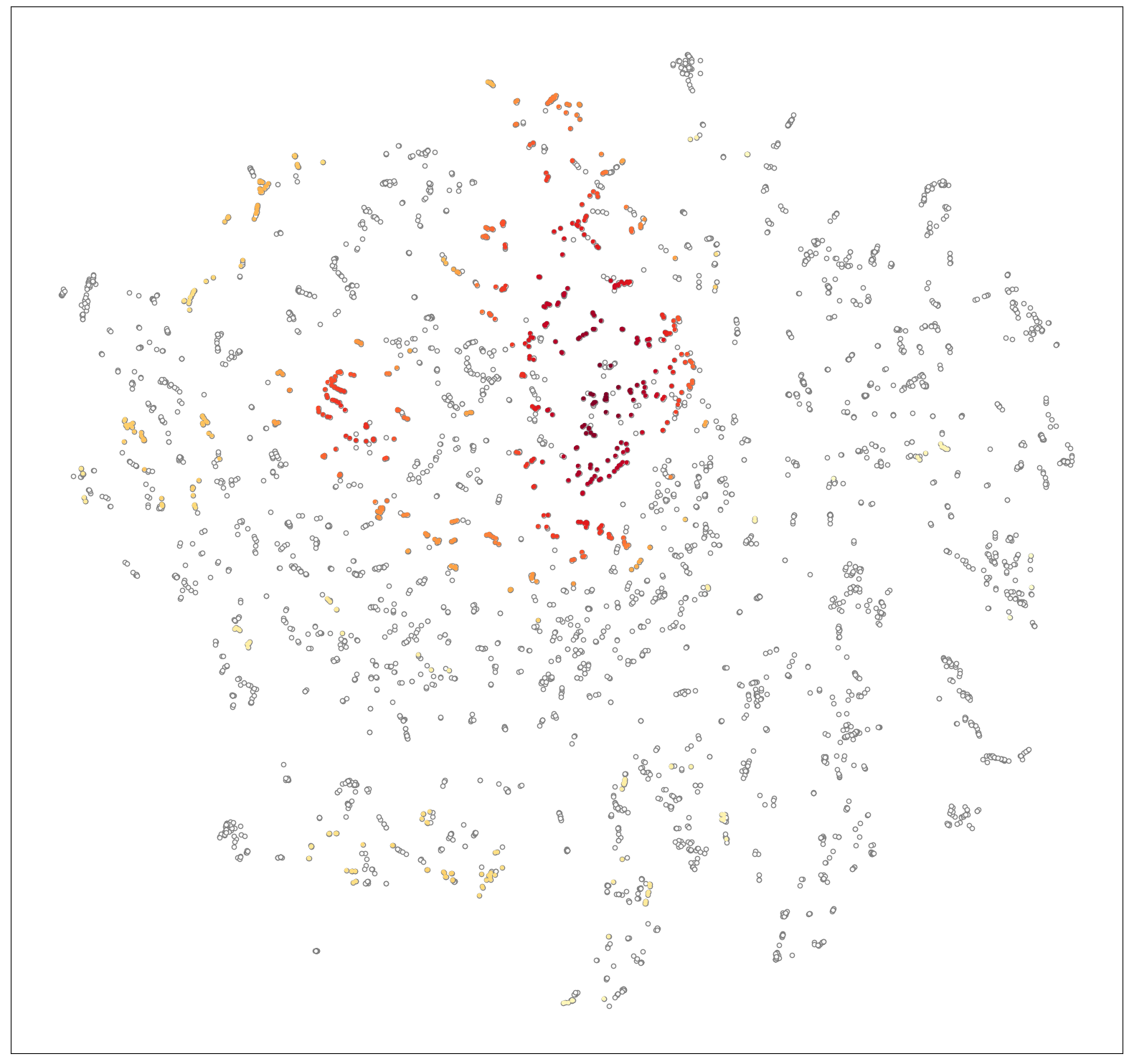}
         \caption{}
     \end{subfigure}
     \hfill
    \begin{subfigure}[t]{0.32\textwidth}
         \centering
         \includegraphics[width=\textwidth]{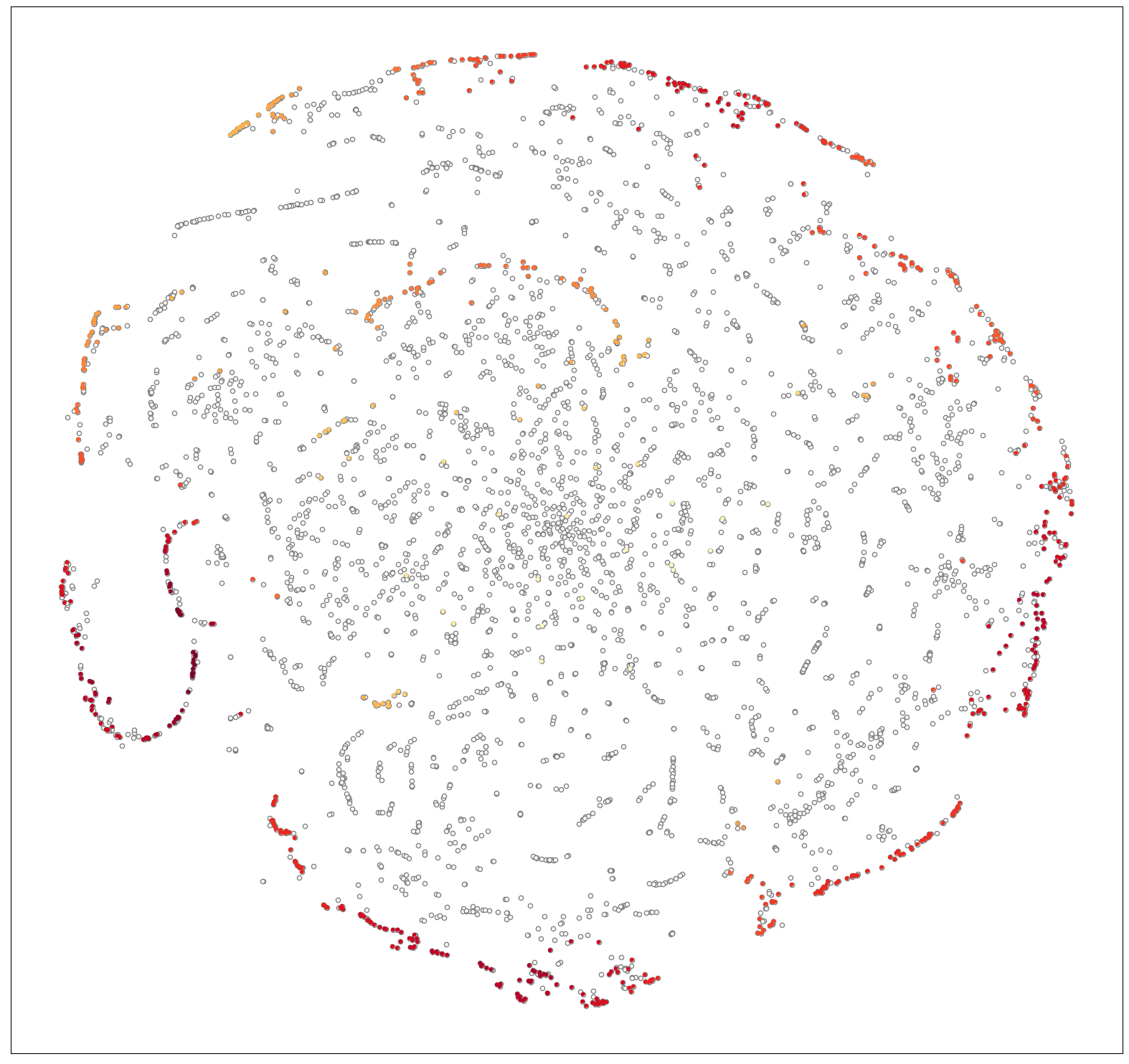}
         \caption{}
     \end{subfigure}
     \hfill
    \begin{subfigure}[t]{0.32\textwidth}
         \centering
         \includegraphics[width=\textwidth]{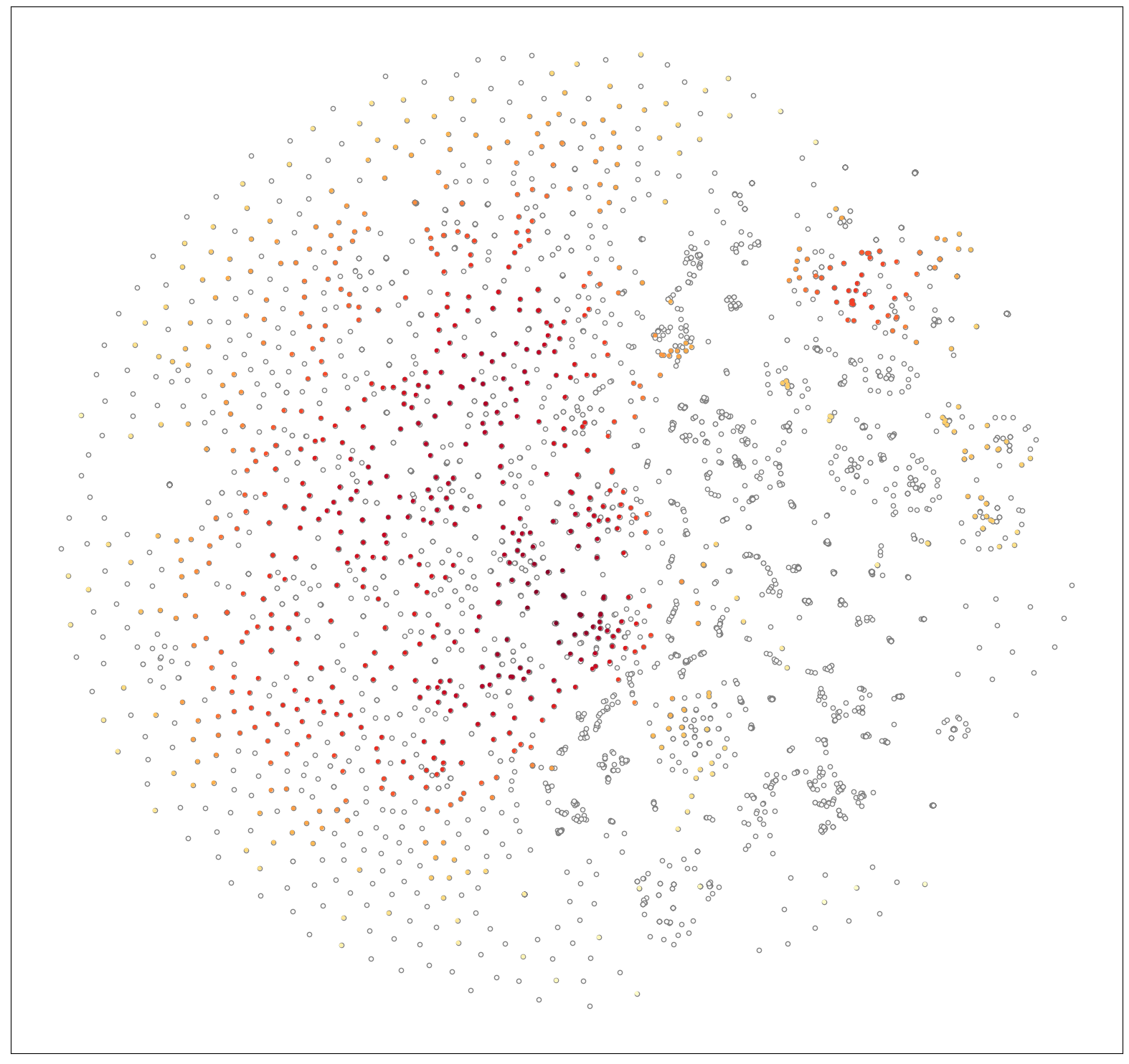}
         \caption{}
     \end{subfigure}
     \hfill

\caption{ 
Distribution density of target prototype materials over the map of all ABC$_3$ prototype materials(gray circles). 
The density representation uses a yellow-red color scheme, where red indicates a high density.
Mapping descriptors:
(a) Cartesian Coordinates 0-padding, 
(b) Fractional Coordinates 0-padding, 
(c) Atomic pairwise distance,  
(d) DeeperGATGNN latent feature,  
(e) XRD feature, 
(f) Composition. 
}   
\label{fig:localabc3}
     
\end{figure*}

Figure \ref{fig:localabc3} shows the distribution densities of the local cluster with 805 materials marked with yellow-red heat maps scatters over the global ABC$_3$ prototype distribution map in grey small circle background. Compared with global mapping, a smaller data set displays the characteristic difference between descriptors. Figures \ref{fig:localabc3}(a), (b), and (c) is the mapping of Cartesian coordinates, fractional coordinates, and pairwise distance. All three descriptors capture the structure information, while Figure \ref{fig:localabc3}(a) shows the experimental tinkering process leading to the discovery of new materials throughout history which leads to a string of materials with similar structures. At the same time, Figures \ref{fig:localabc3}(b) and (c) both show the structure similarity of this group of materials with similar crystal lattice cells and atomic pairwise distance. 
All the circle clusters distributed throughout the entire map in Figure \ref{fig:localabc3}(c) are cubic crystal systems. Such structural similarity is also generating clusters over the XRD descriptor map, one of the most chaotic mapping descriptors that hardly show any pattern over the global MP distribution. In Figure \ref{fig:localabc3}(e), the distribution of the 805 clusters over ABC$_3$ XRD descriptor shows clear string-shaped clusters. Finally, Figure \ref{fig:localabc3}(d) and (f) are DeepGATGNN and composition descriptor maps, both of which show data points evenly scattered over the entire map. The backgroud of Figure \ref{fig:localabc3}(f) is also a scattered point cloud indicating there is no special group of materials in ABC$_3$ prototypes that has unique dominating compositions.

\FloatBarrier

\paragraph{Properties subset.}  
Now, from the material property perspective, we use different mapping to study the distribution of piezoelectric materials from the global composition map to the oxide materials map. %

\begin{figure*}[ht]
    \centering
    \begin{subfigure}[t]{0.49\textwidth}
         \centering
         \includegraphics[width=\textwidth]{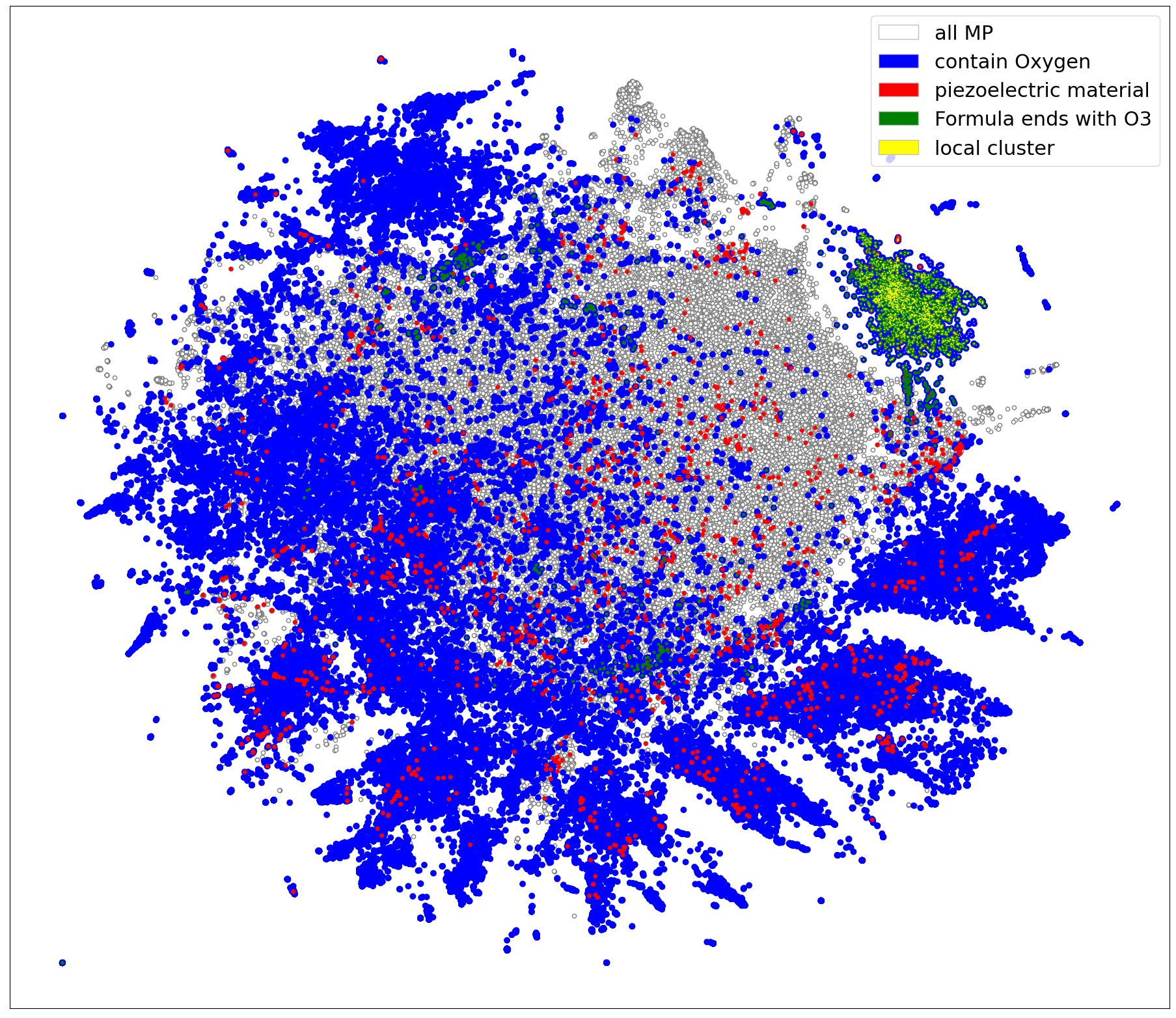}
         \caption{}
         \label{fig:pizeozoom}
    \end{subfigure}
    \hfill
    \centering
    \begin{subfigure}[t]{0.49\textwidth}
         \centering
         \includegraphics[width=\textwidth]{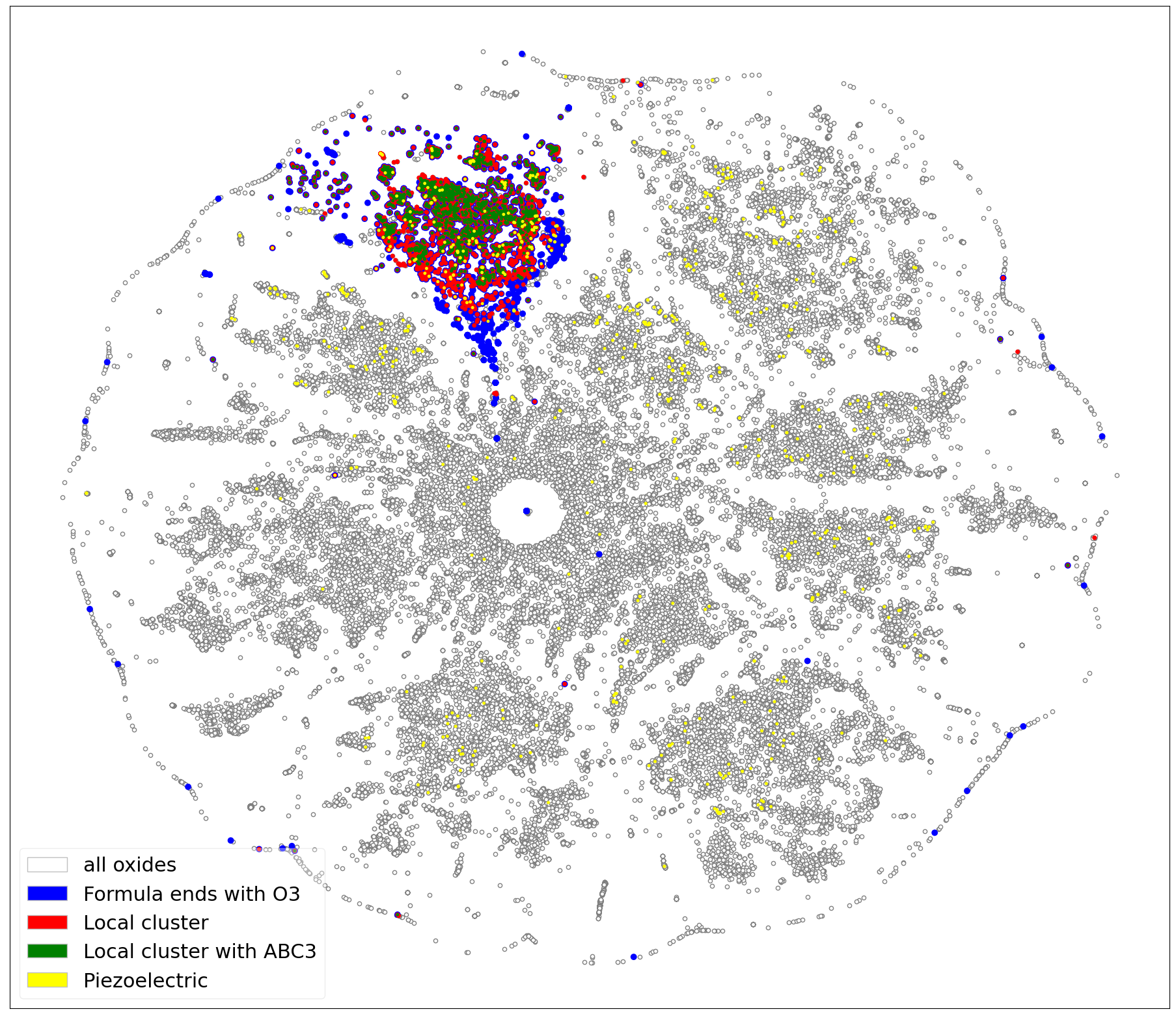}
         \caption{}
         \label{fig:oxidecomp}
    \end{subfigure}
    \hfill
    
    \begin{subfigure}[t]{0.49\textwidth}
         \centering
         \includegraphics[width=\textwidth]{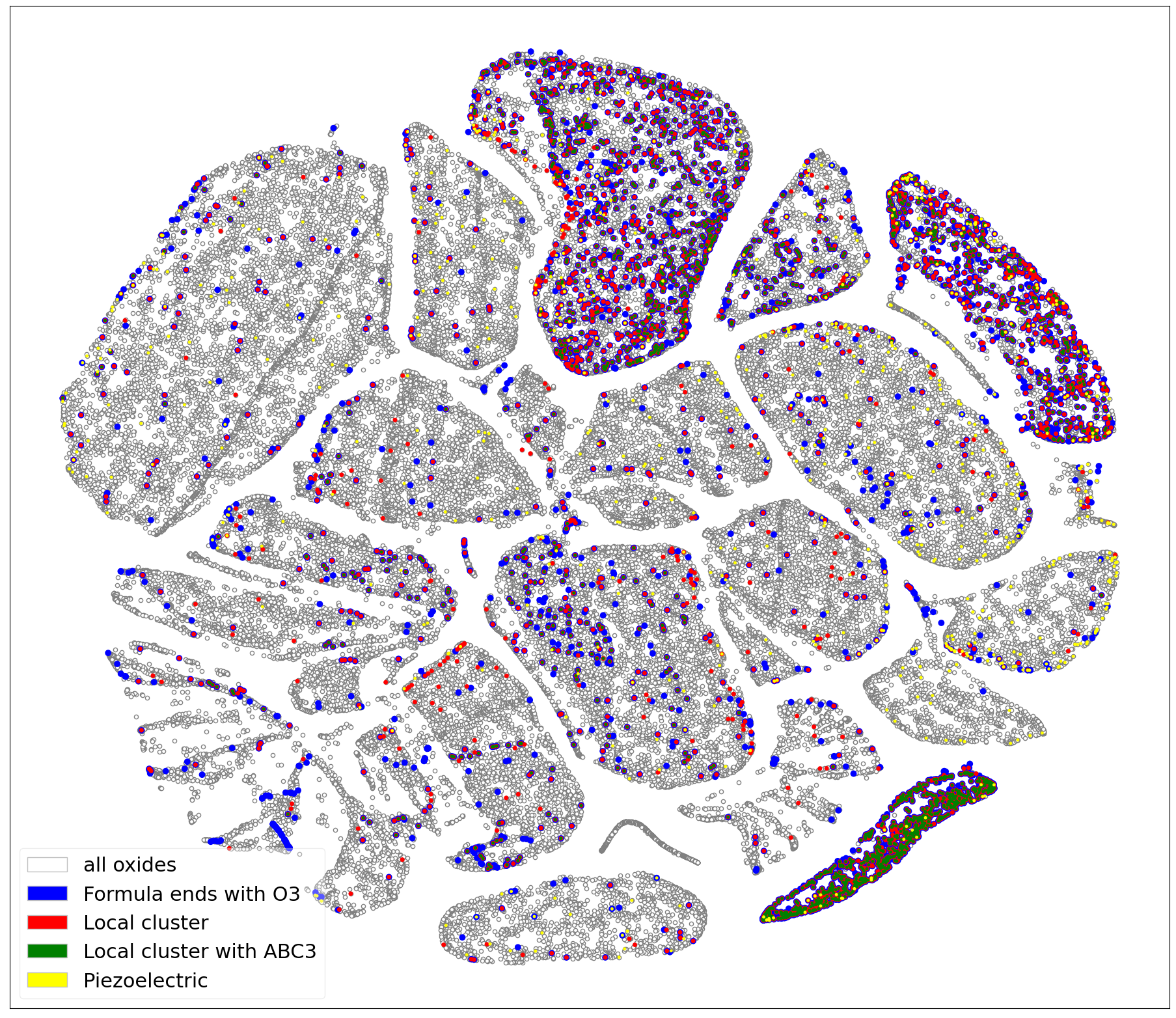}
         \caption{}
         \label{fig:oxidetopo}
    \end{subfigure}
    \hfill
    \begin{subfigure}[t]{0.45\textwidth}
         \centering
         \includegraphics[width=\textwidth]{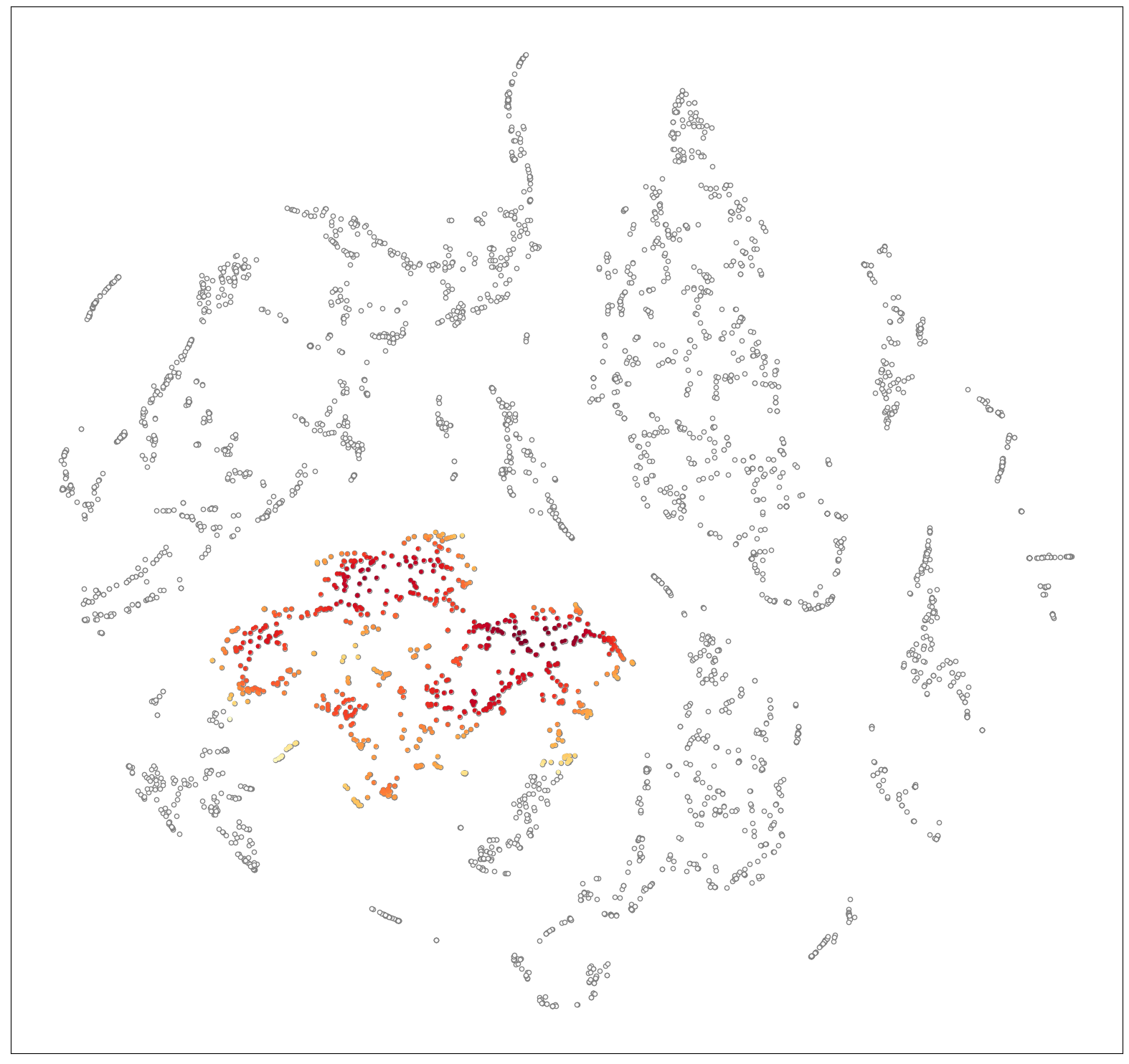}
         \caption{}
         \label{fig:abc3805topo}
    \end{subfigure}
    \hfill

\caption{ 
(a) Oxides, piezoelectric materials, ABO$_3$ prototypes with respect to the global map of all MP materials in the composition space.
(b) Piezoelectric materials, prototypes with formulas ending with O$_3$, local cluster materials from (a), and ABC$_3$ prototypes within the local cluster over the composition map of all oxides materials(gray circles).
(c) Piezoelectric materials, prototypes with formulas ending with O$_3$, local cluster materials from (a), and ABC$_3$ prototypes within the local cluster over the  topology map of all oxides materials(gray circles).
(d) Distribution density of 805 ABC$_3$ prototype materials over the topology map of all ABC$_3$ prototype materials(gray circles).
}
\label{fig:localprops}
\end{figure*}

Figure \ref{fig:localprops}(a) shows the distribution of oxides, piezoelectric materials, and ABO$_3$ prototype materials on top of the global composition map of all MP materials. Here we mark all oxides as blue dots, piezoelectric materials as red dots, ABO$_3$ materials as green dots, and all MP materials as grey circles. We can see that oxide materials are distributed almost everywhere in many of the composition clusters. We notice a local cluster formed on the top right of the t-SNE distribution map, in which  many of the which are piezoelectric oxide materials labeled as yellow points.

To obtain better visualization of how the piezoelectric materials are distributed within the oxides, we plot their distribution map as shown in Figure \ref{fig:localprops}(b). The background of this map is all oxide materials in the composition space. We then mark all formulas ending with O$_3$ as blue, piezoelectric materials as yellow, and ABC$_3$ materials as red. It is interesting to see that O$_3$ prototype materials form a unifying cluster on the top of the map Figure \ref{fig:localprops}(b). Since the composition feature is generated using one-hot encoding, it is no surprise that O$_3$ prototypes and ABC$_3$ prototypes are located in the same region. 
We also find that the local cluster observed in Figure \ref{fig:localprops}(a) still exists as an individual cluster in this map shown in red points, which suggests that we can further reduce the scope of the material mapping to all O$_3$ prototypes.

Figure \ref{fig:localprops}(c) shows the distribution of piezoelectric materials and O$_3$ prototype materials over the global map of MP oxide materials in the space of persistence homology topology features. 
There are over 20 background clusters formed indicating different structure topologies among all oxide materials. Among them, the O$_3$ prototype materials are spread over 3 major clusters (dark color). The most interesting cluster is the one located at the bottom right of the figure.

\FloatBarrier

\subsection{Case study of local clusters of material families}

\begin{figure}[ht]
    \begin{subfigure}[t]{0.46\textwidth}
         \centering
         \includegraphics[width=\textwidth]{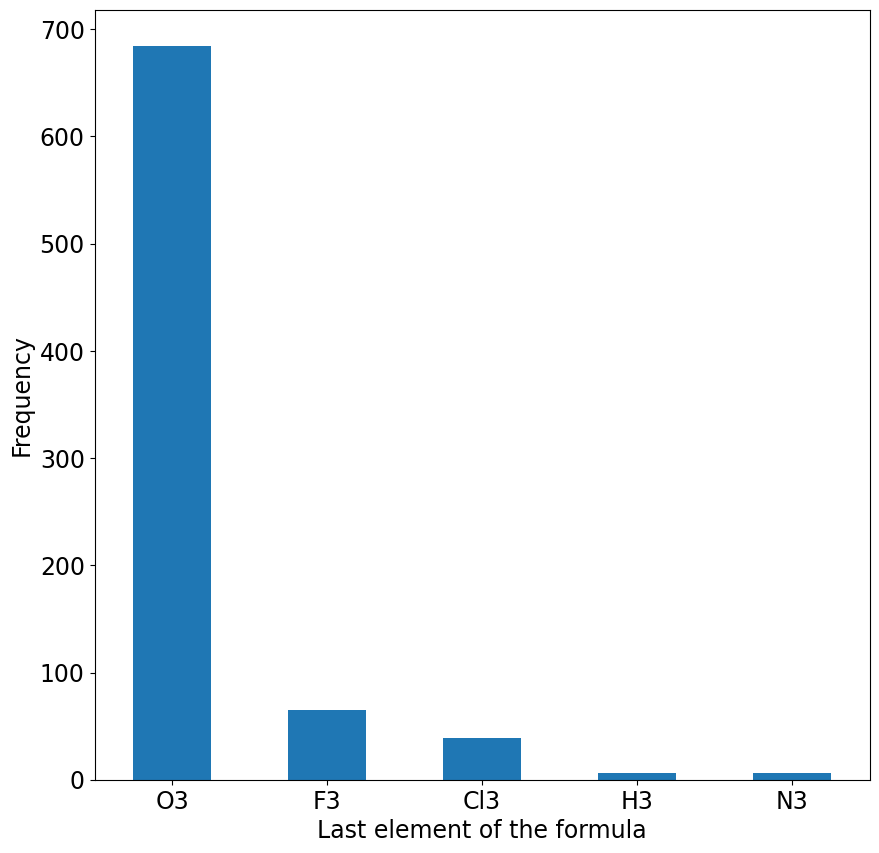}
         \caption{}
    \end{subfigure}
    \hfill
    \begin{subfigure}[t]{0.45\textwidth}
         \centering
         \includegraphics[width=\textwidth]{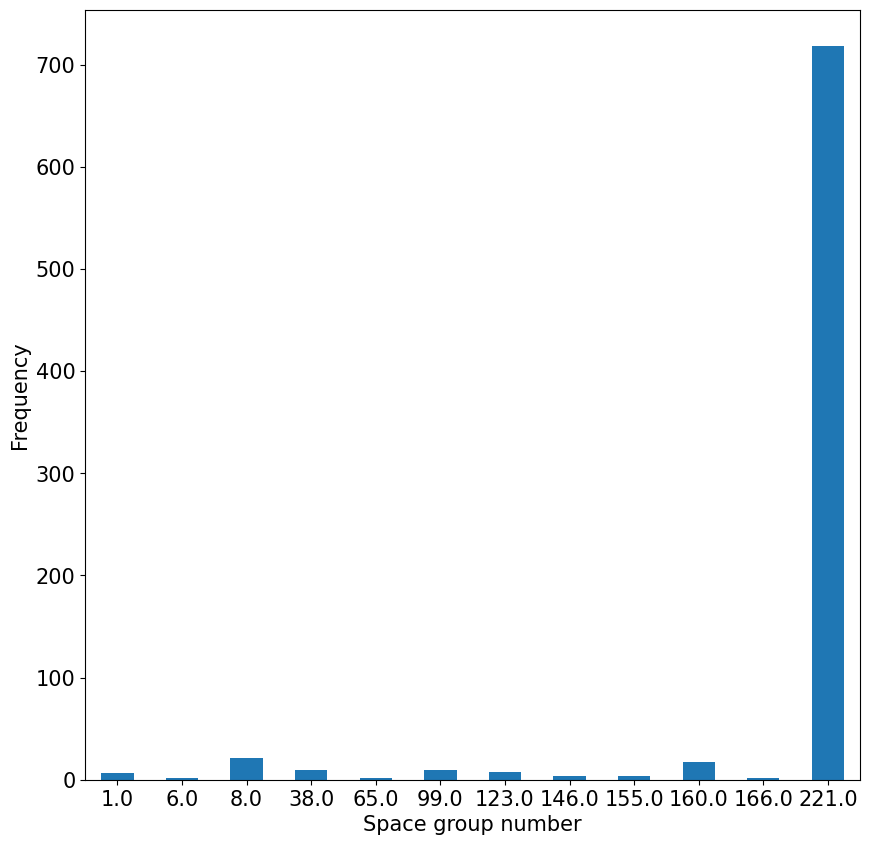}
         \caption{}
    \end{subfigure}

    \begin{subfigure}[t]{0.32\textwidth}
         \centering
         \includegraphics[width=\textwidth]{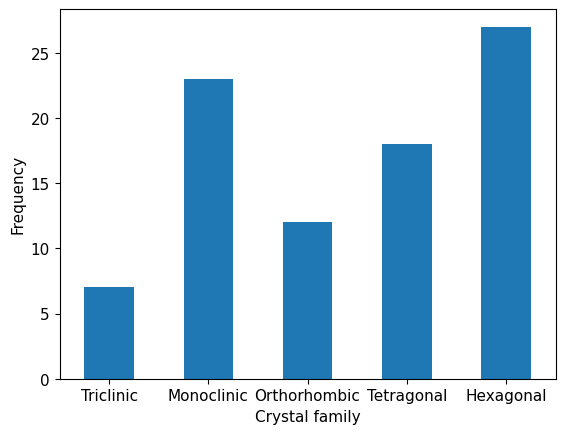}
         \caption{}
    \end{subfigure}
    \hfill
    \begin{subfigure}[t]{0.32\textwidth}
         \centering
         \includegraphics[width=\textwidth]{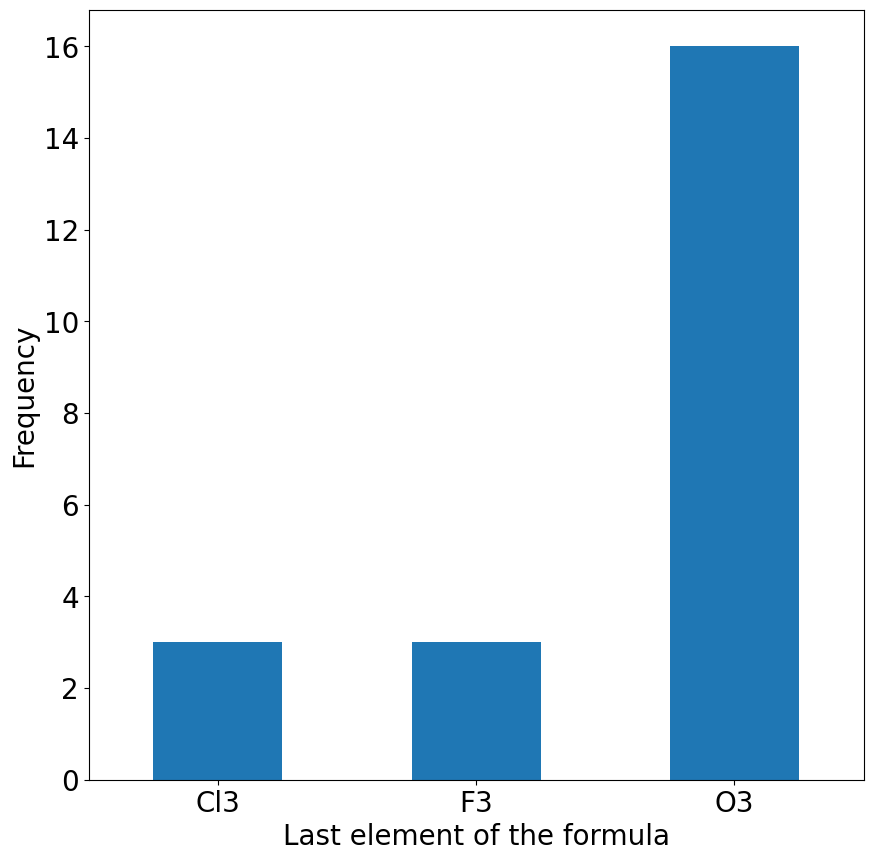}
         \caption{}
    \end{subfigure} 
    \begin{subfigure}[t]{0.32\textwidth}
         \centering
         \includegraphics[width=\textwidth]{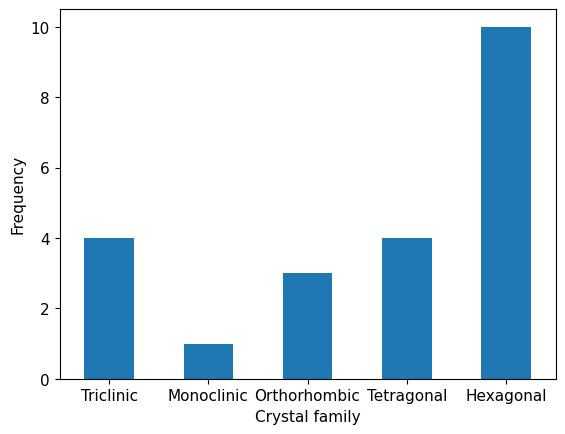}
         \caption{}
    \end{subfigure} 

  \caption{Case study of local clusters:
  (a) the frequency of C$_3$ in the formulas of 805 cluster materials,
  (b) the frequency of space group number of 805 cluster materials,
  (c) the frequency of crystal system in the local non-cubic materials,
  (d) the frequency of C$_3$ in the formulas of piezoelectric materials,
  (e) the frequency of crystal system in piezoelectric materials.
  }
  \label{fig:Casesstudy}
\end{figure}

In the previous section, we studied a local cluster with 805 materials within the scope of ABC$_3$ materials. Figure \ref{fig:localprops}(d) shows the local 805-cluster over the topology map of all ABC$_3$ materials. We found this group of materials is mainly distributed at the bottom right of Figure \ref{fig:localprops}(c), as mentioned above, which is also part of the major cluster in Figure \ref{fig:localprops}(b). This is a small group of ABC$_3$ materials sharing the same clustering tendency using different mapping methods. We acquire the formulas and MP ids of this group of materials (See Supplementary Table 1) and analyze their compositions. We inspect the formulas and found that 84.9\% of them belong to the ABO$_3$ prototypes (684) as illustrated in Figure \ref{fig:Casesstudy}(a) such as AcBO$_3$, ScPdO$_3$, and YTiO$_3$, etc. While other C$_3$ includes elements such as F, Cl, H, and N, with examples including CuAgF$_3$, RbAgCl$_3$, CsYbH$_3$, and LaMoN$_3$. From the structural perspective, 89\% (718) of the materials from this group have the space group number of 221 which makes them all belong to the cubic system as shown in Figure \ref{fig:Casesstudy}(b). From the properties perspective, in Figure \ref{fig:localprops}(a) and (b), we start from all materials and reduce the scope to oxides material. The group of clusters mentioned above still falls into the cluster in the color-marked area in Figure \ref{fig:localprops}(b). At the same time, in Figure \ref{fig:localprops}(c), this same group of materials reaches a cluster from structural scope reduction. This is interesting because not only the global distribution cluster indicates a uniqueness of composition of this group for material, but also from Figure \ref{fig:globalBinay}(b) and (c) the piezoelectric materials have distribution over this region(See Supplementary Figure S3). 
To our surprise, there are no piezoelectric materials within the group of 718 cubic materials. There are 22 piezoelectric materials shown in Table \ref{tab:805pizeo}. They are all from other crystal families which contain 87 non-cubic crystals. Figure \ref{fig:Casesstudy}(c) and (e) shows that almost half of the hexagonal materials are piezoelectric and Figure \ref{fig:Casesstudy}(d) shows the formula composition of this group of piezoelectric materials.

\begin{table}[h]
\begin{center}
\caption{List of piezoelectric material formulas and MP indexes}
\label{tab:805pizeo}

\begin{tabular}{llllllll}
\toprule
    mp id & formula & mp id & formula & mp id & formula & mp id & formula\\
\midrule
  mp-5777 &  BaTiO3  & mp-998712 &  TlGeF3 &    mp-998560 & InSnCl3  &   mp-4342 &   KNbO3 \\
  mp-5020 &  BaTiO3  &  mp-13108 &  SrHfO3 &     mp-22988 & CsGeCl3  &   mp-5246 &   KNbO3 \\
  mp-5986 &  BaTiO3  &  mp-22981 &   TlIO3 &    mp-998298 &  CsSnF3  &  mp-20459 &  TiPbO3 \\
  mp-4559 &   BaCO3  &  mp-27193 &   RbIO3 &    mp-562987 &   VCuO3  & mp-755018 &  RbTaO3 \\
 mp-29798 &  TlBrO3  & mp-552729 &    KIO3 &    mp-998193 &  RbSnF3  &  mp-20337 &  ZrPbO3 \\
mp-675022 & CsPbCl3  &   mp-7375 &   KNbO3 \\
\bottomrule
\end{tabular}

\end{center}
\end{table}

\FloatBarrier
\subsection{Quality evaluation of clusters in mapping }

\begin{figure*}[th]
    \centering 
    \includegraphics[width=0.7\columnwidth]{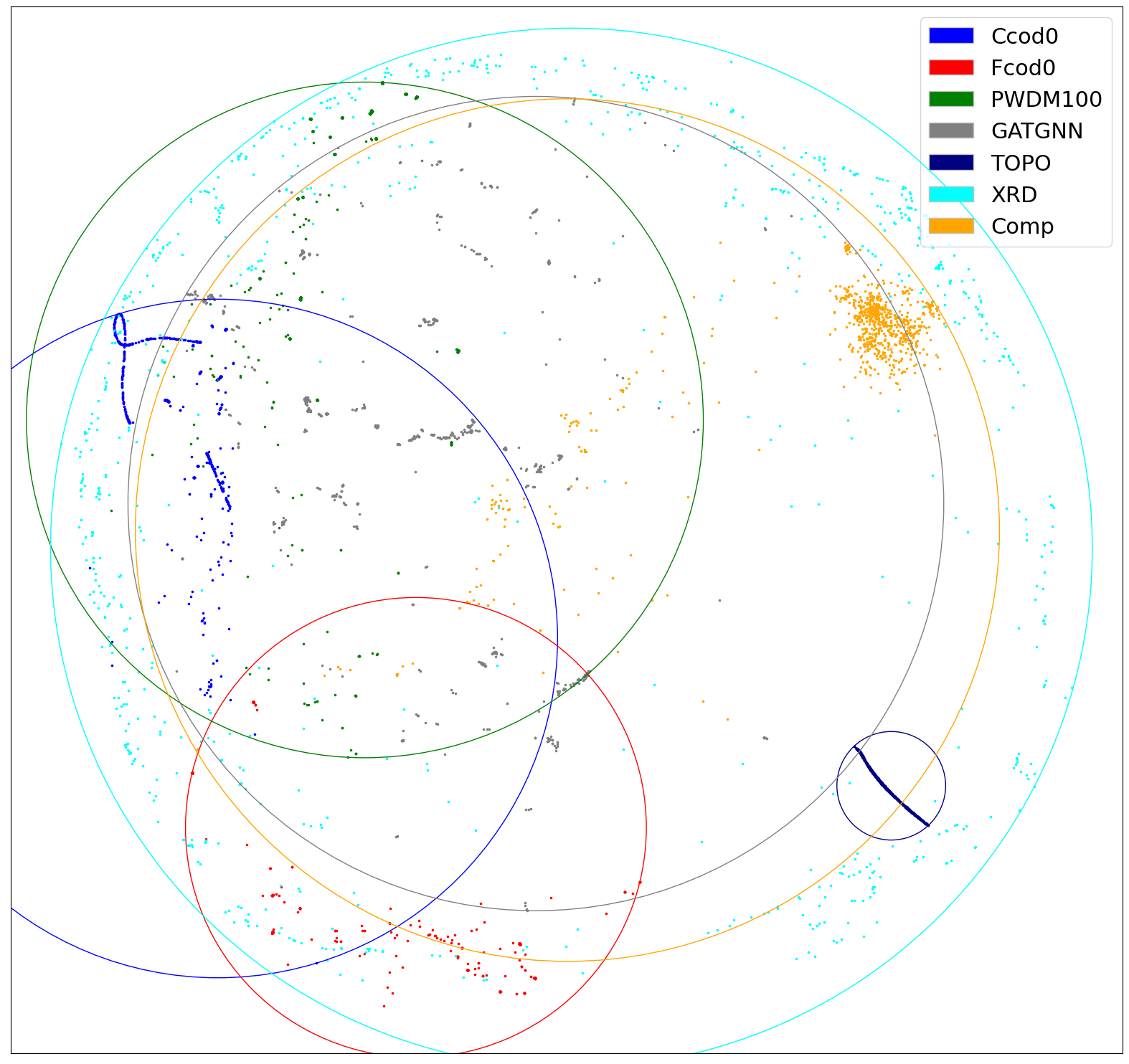}  
    
\caption{ 
Smallest enclosing circles for local ABC$_3$ clusters over different global MP maps with all descriptor features.
The color of the cluster and circle indicates their descriptor as shown in the top right corner and also Table \ref{tab:radius}
}   
\label{fig:circles} 
\end{figure*}

For a given set of materials with common properties (e.g piezoelectric materials), we can generate a variety of maps using different background global or local feature maps. Here we propose a measure to quantitatively evaluate the quality of the mapping of materials clusters over the background map: given a material background map generated using t-SNE dimension reduction, the clustering quality of a subset of material is evaluated by the radius of the smallest circle that encircles all the materials of the subset over the map. Figure \ref{fig:circles} shows the circles that encircle the cluster materials mentioned in section 3.3 local mapping of ABC$_3$ materials (805) on the global MP maps with different feature spaces with their radius shown in Table \ref{tab:radius}. It is found that the persistence homology feature map gives the smallest radius indicating the most compact clusters for the local 805 in the ABC$_3$ material family. This is the reason why we studied this group of materials.

\begin{table}[h]
\begin{center}
\caption{
Radius of the smallest enclosing circles over different MP maps.}
\label{tab:radius}

\definecolor{navy}{HTML}{000080}
\definecolor{blue}{HTML}{0000FF}
\definecolor{cyan}{HTML}{00FFFF}
\definecolor{green}{HTML}{15B01A}
\definecolor{olive}{HTML}{808000}
\definecolor{red}{HTML}{FF0000}
\definecolor{purp}{HTML}{800080}
\definecolor{grey}{HTML}{808080}
\definecolor{purp}{HTML}{800080}
\definecolor{brown}{HTML}{A52A2a}
\definecolor{orange}{HTML}{FFA500}
\definecolor{black}{HTML}{000000}

\begin{tabular}{|l|l|l|}
\hline

\textbf{Cartesian coordinates with 0 padding(Ccod0)}         &  \textbf{0.317} &  \cellcolor{blue}          \\ \hline
\textbf{Fractional coordinates with 0 padding(Fcod0) }      &  \textbf{0.215} &  \cellcolor{red}        \\ \hline
\textbf{Atomic pair wise distance matrix (PWDM100)   }       &  \textbf{0.316} &  \cellcolor{green}           \\ \hline
DeeperGATGNN network (GATGNN)                               &  0.381 &  \cellcolor{grey}         \\ \hline
\textbf{Topology (TOPO) }                                    &  \textbf{0.050} &  \cellcolor{navy}         \\ \hline
XRD  (XRD)                                                  &  0.487 &  \cellcolor{cyan}         \\ \hline         
Atomic elemental composition (Comp)                         &  0.404 &  \cellcolor{orange}         \\ \hline

\end{tabular}
\end{center}
\end{table}

\FloatBarrier

\section{Conclusion}

Crystal structure and functional properties have sophisticated relationships. This work proposes the global mapping approaches to map crystal materials of the Materials Project database into structural, physical, and latent feature space using t-SNE dimension reduction and visualization. These maps have shown interesting patterns such as the island clusters that reflect the tinkering design history of existing materials. The global maps of whole MP materials allow us to examine the distribution of different material families including e.g. ABC$_3$ prototype materials and materials with known properties such as band gaps, density, and formation energy. Different kinds of clusters with various shapes have been found that demonstrates the close relationship between structures and function/properties. 

Our studies also show the advantage of using local mapping to examine the distribution of a group of materials with similar structural, compositional, or physical properties. We find that the ABC$_3$ materials form different types of clusters in the composition, topology, and neural latent feature space, each showing distinct but interesting patterns. We also find that using the local mapping approach, the piezoelectric materials of the ABO$_3$ prototype can show clear clusters, which can be used to find potential piezoelectric materials inside those clusters. Finally, we evaluate a procedure to check the most appropriate global mapping space for a given materials family. Overall, our study shows the strong relationships between the structure and functions of crystal materials and global mapping is an effective approach for materials distribution investigation.

\section{Code and Data Availability}

All the raw datasets are downloaded from the Material projects database at \href{http://www.materialspoject.org}{http://www.materialspoject.org}. Our source code and the coordinates of the material maps after t-SNE processing can be accessed freely at \href{https://github.com/usccolumbia/matglobalmapping}{https://github.com/usccolumbia/matglobalmapping}. 
Due to the size of the generated features for the whole MP dataset, they can be downloaded from  
\href{https://figshare.com/articles/dataset/7_generated_mp_dataset_136k_features/21980081}{figshare.com}

\section{Contribution}
Conceptualization, J.H.; methodology,Q.L.,S.S, R.D., N.F. ; software, Q.L., N.F., S.S., R.D.; resources, J.H.; writing--original draft preparation, Q.L.,J.H.; writing--review and editing, Q.L., R.D., N.F., L.W., S.S., J.H.; visualization,Q.L., N.F.; supervision, J.H.;  funding acquisition, J.H.

\section*{Acknowledgement}
The research reported in this work was supported in part by National Science Foundation under the grant 2110033 and 10009560. The views, perspectives, and content do not necessarily represent the official views of the NSF.

\bibliographystyle{unsrt}

\bibliography{references}

\end{document}


\maketitle
\begin{figure*}[th]
    \centering
    \begin{subfigure}[b]{0.32\columnwidth}
        \includegraphics[width=1\columnwidth]{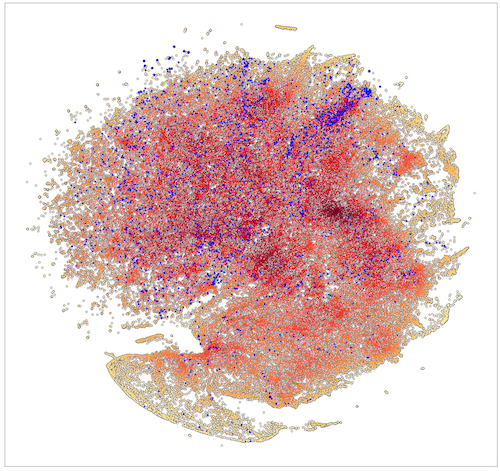}  
        \caption{}
        \label{fig:pwdm100_abc3_dist}
    \end{subfigure}%
    \centering
    \begin{subfigure}[b]{0.32\columnwidth}
        \includegraphics[width=1\columnwidth]{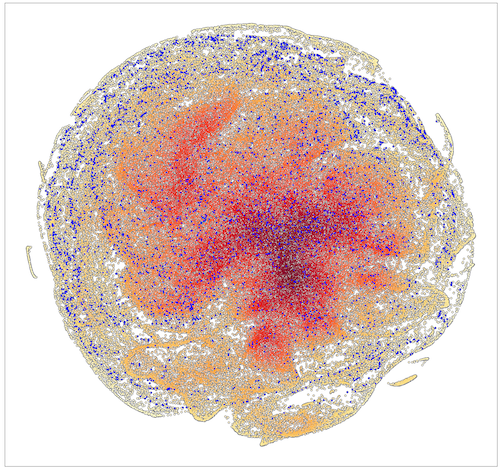}  
        \caption{}
        \label{fig:xrd_abc3_dist}
    \end{subfigure}%
    \centering
    \begin{subfigure}[b]{0.32\columnwidth}
        \includegraphics[width=1\columnwidth]{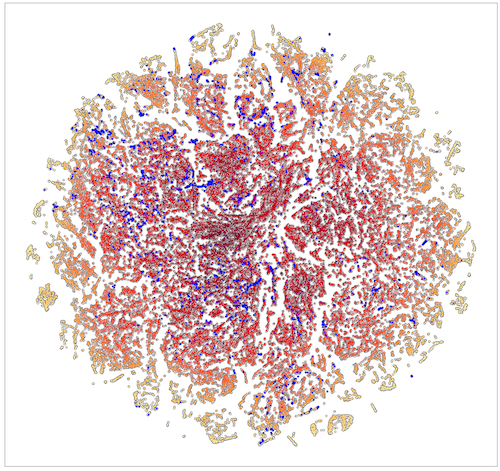}  
        \caption{}
        \label{fig:gat_abc3_dist}
    \end{subfigure}%

    \centering
    \begin{subfigure}[b]{0.32\columnwidth}
        \includegraphics[width=1\columnwidth]{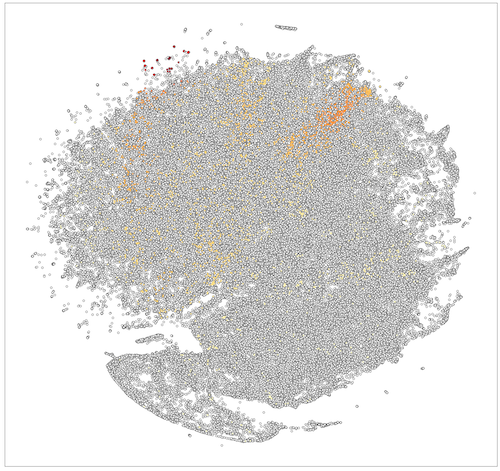}  
        \caption{}
        \label{fig:pwdm100_abc3_dens}
    \end{subfigure}%
    \centering
    \begin{subfigure}[b]{0.32\columnwidth}
        \includegraphics[width=1\columnwidth]{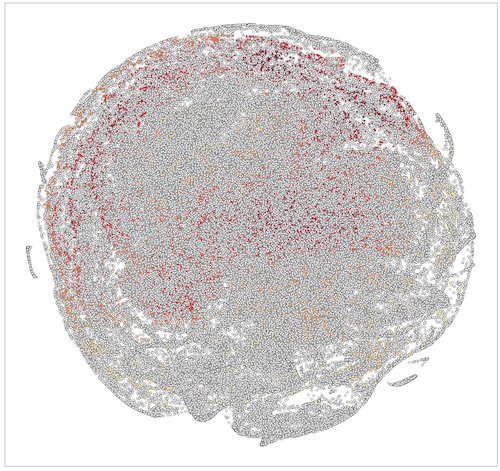}  
        \caption{}
        \label{fig:xrd_abc3_dens}
    \end{subfigure}%
    \centering
    \begin{subfigure}[b]{0.32\columnwidth}
        \includegraphics[width=1\columnwidth]{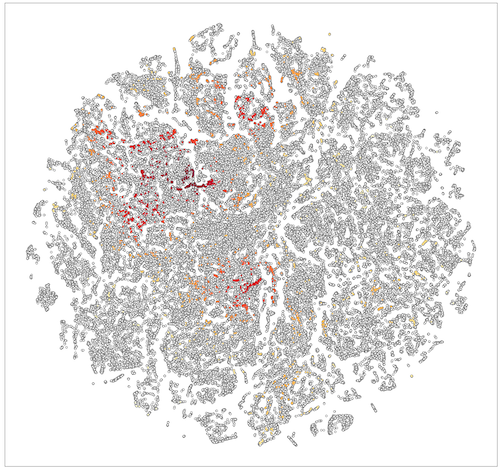}  
        \caption{}
        \label{fig:gat_abc3_den}
    \end{subfigure}%
    \centering

\caption{ ABC3 materials with respect to global density distributions of all MP data-set. 
In figs. (a)-(c), ABC3 materials distribution is represented in blue scatter with a background of global distribution density. 
In figs. (d)–(f), ABC3 materials distribution density is represented with global distributions in the background (grey circles). 
The density representation used an orange-red color scheme, where red indicates a high density.
Mapping descriptors:
Fig (a)\&(d)   :  Atomic pairwise distance   
Fig (b)\&(e)   :  XRD feature 
Fig (c)\&(f)   :  Deep GATGNN
}     
\label{fig:ABC3MP}     
\end{figure*}

\begin{figure*}[th]
    \centering
    \begin{subfigure}[b]{0.49\textwidth}
        \centering
        \includegraphics[width=1.0\textwidth]{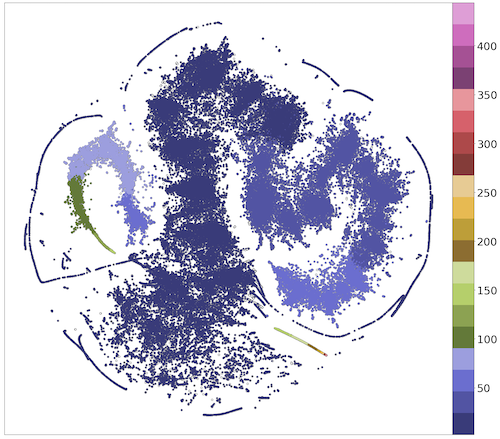}  
        \caption{}
        \label{fig:globalnsitesfodvg}
    \end{subfigure}%
    \hfill  
    \begin{subfigure}[b]{0.49\textwidth}
        \centering
        \includegraphics[width=1.0\textwidth]{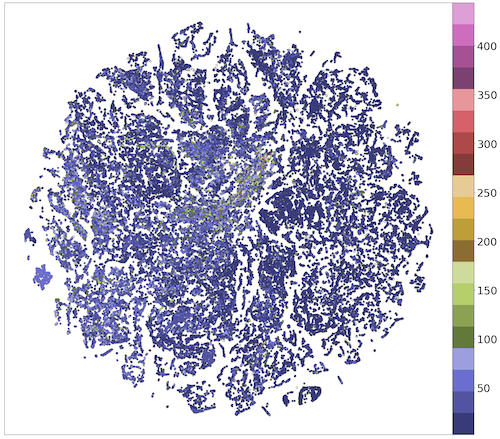}
        \caption{}
        \label{fig:globalnsitesGATGNNg}
    \end{subfigure}%
    \hfill
    \begin{subfigure}[b]{0.49\textwidth}
        \centering
        \includegraphics[width=1.0\textwidth]{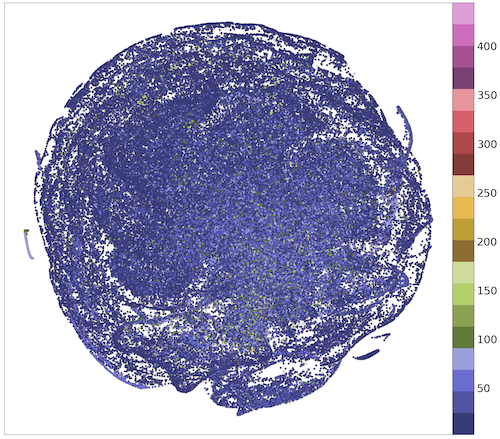}  
        \caption{}
        \label{fig:globalnsitesxrdg}
    \end{subfigure}%
    \hfill
    \begin{subfigure}[b]{0.49\textwidth}
        \centering
        \includegraphics[width=1.0\textwidth]{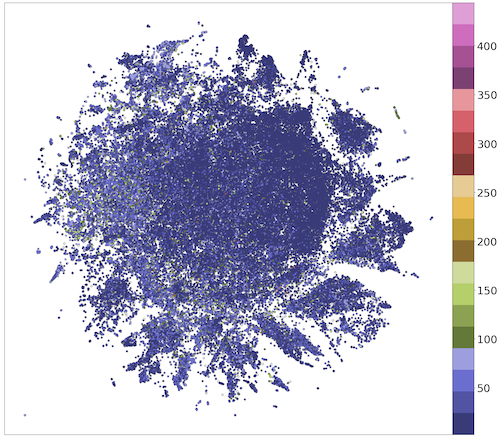}  
        \caption{ }
        \label{fig:globalnsitescompg}
    \end{subfigure}%
    \hfill

\caption{Number of atomic sites distributions over descriptor spaces.  
(a) Fractional Coordinates 0-padding
(b) DeeperGATGNN Latent    
(c) XRD feature
(d) Composition.  We find that the distribution of atomic sites have no clear clusters in physical and neural fingerprint space including DeeperGATGNN latent space, XRD, and composition space. 
}     
\label{fig:globalnsites}

\end{figure*}

\begin{figure*}[ht]
    \centering
    \begin{subfigure}[t]{0.49\textwidth}
         \centering
         \includegraphics[width=\textwidth]{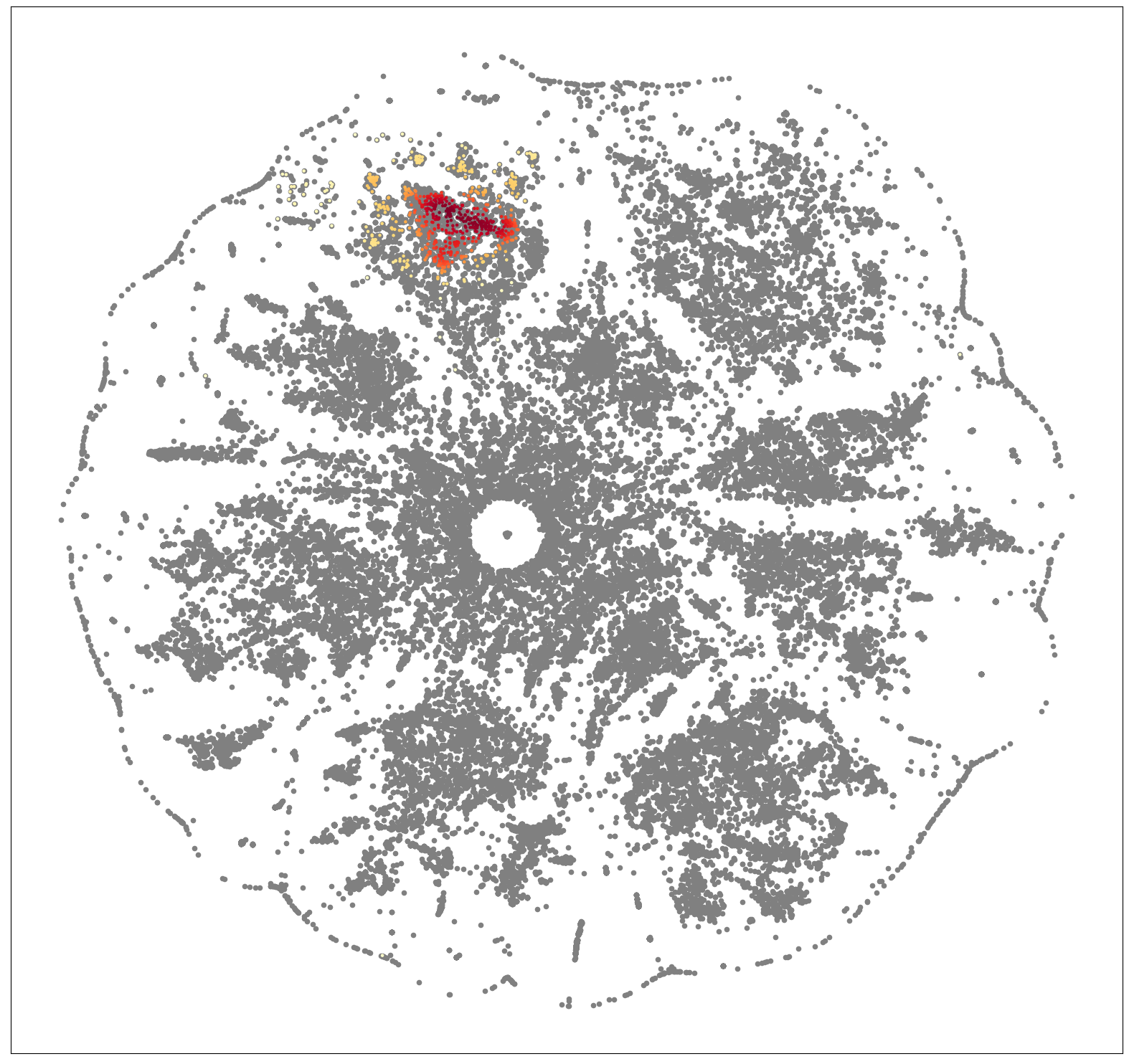}
         \caption{}
         \label{fig:pizeozoom}
    \end{subfigure}
    \hfill
    \centering
    \begin{subfigure}[t]{0.49\textwidth}
         \centering
         \includegraphics[width=\textwidth]{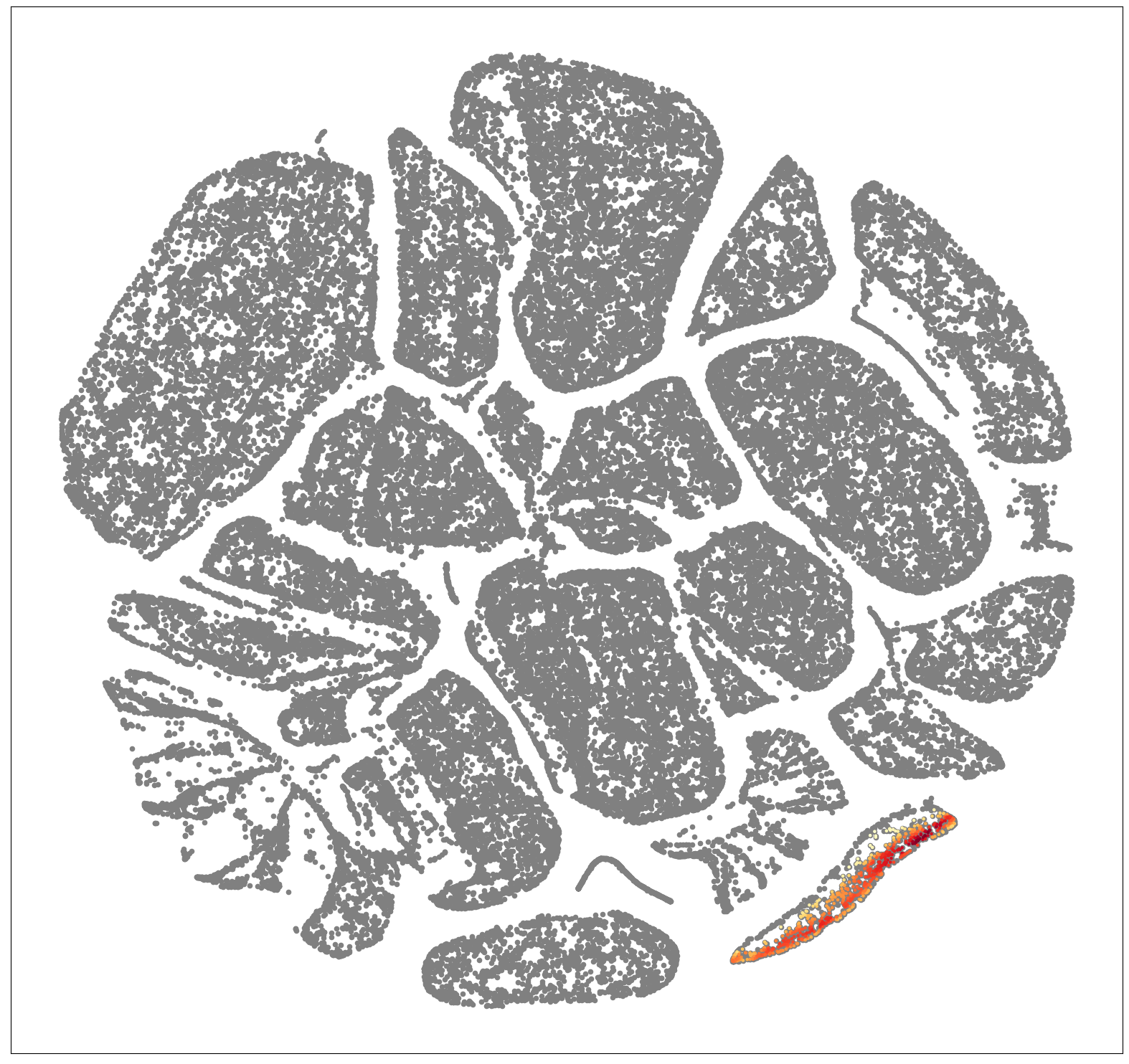}
         \caption{}
         \label{fig:oxidecomp}
    \end{subfigure}
    \hfill

\caption{ 
(a) Distribution density of 805 ABC$_3$ prototype materials over the composition map of all oxides materials(gray circles).
(b) Distribution density of 805 ABC$_3$ prototype materials over the topology map of all oxides materials(gray circles).
}
\label{fig:localprops}
\end{figure*}

\FloatBarrier

\setlength{\hoffset}{-1.5cm}

\begin{center}
\begin{longtable}  {llllllllllll}
\label{tab:mpid805}
\caption{List of local cluster materials' formula and MP index}
\hline
\toprule
 &       mp\_id &  formula &    &       mp\_id &  formula &    &       mp\_id &  formula &    &       mp\_id &  formula  \\
\midrule
\endhead
\midrule
\multicolumn{3}{r}{{Continued on next page}} \\
\midrule
\endfoot

\bottomrule
\endlastfoot
0   &  mp-1183052 &    AcBO3    &    50  &   mvc-11224 &    AlVO3 & 100 &   mp-973877 &    PaBO3 &  150 &  mp-1183443 &   BeSbO3       \\
1   &  mp-1183053 &   AcGaO3    &    51  &   mvc-14600 &   AlSbO3 & 101 &  mp-1183242 &    BPdO3 &  151 &  mp-1183449 &    BeVO3       \\
2   &  mp-1183115 &   AcAlO3    &    52  &  mp-1183160 &   AlInO3 & 102 &  mp-1185996 &    MnBO3 &  152 &  mp-1184148 &   ErBeO3       \\
3   &  mp-1183139 &   AcNiO3    &    53  &  mp-1185485 &   LuAlO3 & 103 &  mp-1187416 &    ThBO3 &  153 &  mp-1184610 &   HoBeO3       \\
4   &  mp-1183142 &   AcMgO3    &    54  &     mp-5323 &   CeAlO3 & 104 &   mp-971704 &    ZnBO3 &  154 &  mp-1186392 &   PaBeO3       \\
5   &  mp-1183145 &   AcPdO3    &    55  &   mvc-10882 &   AlMoO3 & 105 &  mp-1017465 &    BaVO3 &  155 &  mp-1187310 &   TbBeO3       \\
6   &  mp-1183150 &   AcSiO3    &    56  &  mp-1183154 &   AlPbO3 & 106 &  mp-1076782 &   BaCoO3 &  156 &  mp-1187421 &   ThBeO3       \\
7   &  mp-1183152 &   AcScO3    &    57  &     mp-5304 &   LaAlO3 & 107 &  mp-1183211 &    BSbO3 &  157 &  mp-1227955 &   BaBiO3       \\
8   &  mp-1183168 &    AcVO3    &    58  &   mp-972066 &   ZrAlO3 & 108 &  mp-1183291 &   BaErO3 &  158 &     mp-4559 &    BaCO3       \\
9   &  mp-1183417 &   BeAgO3    &    59  &   mp-973892 &   EuAlO3 & 109 &  mp-1183310 &   BaEuO3 &  159 &     mp-4783 &   BaPrO3       \\
10  &  mp-1185297 &   LiAcO3    &    60  &   mvc-11155 &   AlNiO3 & 110 &    mp-19322 &   BaMoO3 &  160 &     mp-8037 &   BaThO3       \\
11  &   mp-861502 &   AcFeO3    &    61  &   mvc-11225 &   AlFeO3 & 111 &     mp-3834 &   BaZrO3 &  161 &   mp-978503 &   SmBeO3       \\
12  &   mp-864606 &   AcCuO3    &    62  &   mvc-11226 &   AlCoO3 & 112 &   mp-545783 &   BaBiO3 &  162 &   mp-981934 &   LaBeO3       \\
13  &   mp-864911 &   AcMnO3    &    63  &   mvc-15476 &   AlBiO3 & 113 &     mp-5777 &   BaTiO3 &  163 &   mp-984552 &   BeZnO3       \\
14  &   mp-865927 &   AcTiO3    &    64  &    mp-14254 &   NdAlO3 & 114 &  mp-1016852 &   BaMnO3 &  164 &  mp-1183403 &   BeBiO3       \\
15  &   mp-866101 &   AcCrO3    &    65  &     mp-8218 &   PrAlO3 & 115 &  mp-1017591 &   BaRuO3 &  165 &  mp-1183413 &   BeHgO3       \\
16  &   mp-975463 &   RbAcO3    &    66  &   mvc-11052 &   AlSnO3 & 116 &  mp-1076932 &   BaTiO3 &  166 &  mp-1183415 &   BeGeO3       \\
17  &   mp-979294 &   SrAgO3    &    67  &   mvc-11693 &   TiAlO3 & 117 &  mp-1120765 &   BaNiO3 &  167 &  mp-1183418 &   BeCdO3       \\
18  &   mp-985292 &    AgBO3    &    68  &   mvc-15239 &   AlCuO3 & 118 &  mp-1183288 &   BaInO3 &  168 &  mp-1183426 &   BeCuO3       \\
19  &   mp-998201 &  RbAgCl3    &    69  &   mvc-15604 &   AlCrO3 & 119 &  mp-1183393 &   BaTlO3 &  169 &  mp-1183436 &   BeInO3       \\
20  &   mp-998414 &   CuAgF3    &    70  &  mp-1178268 &   GdAlO3 & 120 &     mp-2929 &   BaTbO3 &  170 &  mp-1183447 &   BePdO3       \\
21  &   mvc-15475 &   AlAgO3    &    71  &  mp-1183134 &   AlSbO3 & 121 &     mp-2998 &   BaTiO3 &  171 &  mp-1183499 &   BePbO3       \\
22  &  mp-1178562 &   AgBrO3    &    72  &   mp-758371 &   YbAlO3 & 122 &     mp-3163 &   BaSnO3 &  172 &  mp-1183685 &   CeBeO3       \\
23  &  mp-1184234 &   FeAgO3    &    73  &  mp-1183128 &    AlVO3 & 123 &   mp-504715 &   BaTiO3 &  173 &  mp-1184325 &   EuBeO3       \\
24  &  mp-1186101 &   NaAgO3    &    74  &   mp-984723 &   DyAlO3 & 124 &   mp-754678 &   BaTaO3 &  174 &  mp-1184468 &   GdBeO3       \\
25  &  mp-1206915 &   CoAgF3    &    75  &   mvc-15603 &   MnAlO3 & 125 &   mp-755877 &   BaNdO3 &  175 &  mp-1184563 &   HfBeO3       \\
26  &  mp-1209766 &   NiAgF3    &    76  &  mp-1184150 &   ErAlO3 & 126 &     mp-8557 &   BaPaO3 &  176 &  mp-1186216 &   NbBiO3       \\
27  &   mp-867844 &   RbAgF3    &    77  &  mp-1183911 &   CsAsO3 & 127 &   mp-998552 &   BaHfO3 &  177 &  mp-1186533 &   PmBeO3       \\
28  &   mp-975421 &   RbAgO3    &    78  &  mp-1184937 &    KAsO3 & 128 &  mp-1016821 &   BaSiO3 &  178 &  mp-1186742 &   PrBeO3       \\
29  &   mp-976229 &    KAgO3    &    79  &   mp-976133 &   NaAsO3 & 129 &  mp-1016823 &   BaGeO3 &  179 &  mp-1187465 &   TiBeO3       \\
30  &   mp-998610 &  TlAgCl3    &    80  &   mp-976763 &   NiAsO3 & 130 &  mp-1016850 &   BaRhO3 &  180 &  mp-1187905 &   YbBeO3       \\
31  &  mp-1075933 &   SmAgO3    &    81  &   mp-983564 &   EuAsO3 & 131 &  mp-1076800 &   BaCuO3 &  181 &  mp-1188011 &   ZrBeO3       \\
32  &  mp-1077622 &   EuAgO3    &    82  &  mp-1184913 &   InAsO3 & 132 &  mp-1178510 &   BaPdO3 &  182 &  mp-1207000 &   DyBiO3       \\
33  &  mp-1094009 &    YAgO3    &    83  &   mp-975425 &   RbAsO3 & 133 &  mp-1227982 &    BaCO3 &  183 &  mp-1239170 &   ZrBeO3       \\
34  &  mp-1183169 &   AlAgO3    &    84  &  mp-1184215 &   FeAsO3 & 134 &    mp-19990 &   BaTiO3 &  184 &   mp-550008 &   ScBiO3       \\
35  &  mp-1186999 &   SiAgO3    &    85  &   mp-973156 &   ScAsO3 & 135 &     mp-5020 &   BaTiO3 &  185 &   mp-972222 &   TaBeO3       \\
36  &  mp-1210575 &   MgAgF3    &    86  &   mp-978851 &   SrAsO3 & 136 &   mp-561598 &   BaPbO3 &  186 &   mp-972395 &   TmBeO3       \\
37  &    mp-12725 &   NbAgO3    &    87  &  mp-1183397 &    CoBO3 & 137 &     mp-5663 &   BaCeO3 &  187 &   mp-975579 &   NdBeO3       \\
38  &    mp-13819 &    KAgF3    &    88  &  mp-1183631 &    CdBO3 & 138 &   mp-644497 &   BaTiO3 &  188 &   mp-976727 &   PrBiO3       \\
39  &   mp-998358 &   MgAgF3    &    89  &  mp-1184546 &    HoBO3 & 139 &   mp-995191 &   BaTiO3 &  189 &   mp-984722 &   BeGaO3       \\
40  &   mp-998365 &   NiAgF3    &    90  &  mp-1187494 &    TlBO3 & 140 &  mp-1183289 &   BaScO3 &  190 &   mvc-13598 &    YBiO3       \\
41  &  mp-1076000 &   LaAgO3    &    91  &   mp-972442 &    SnBO3 & 141 &  mp-1183292 &   BaCdO3 &  191 &  mp-1184051 &   CuBiO3       \\
42  &  mp-1206344 &   MnAgF3    &    92  &  mp-1184042 &    CuBO3 & 142 &  mp-1183394 &   BaTmO3 &  192 &    mp-22979 &   GaBiO3       \\
43  &  mp-1206910 &   CoAgF3    &    93  &  mp-1188026 &    ZrBO3 & 143 &  mp-1183406 &   BaYbO3 &  193 &   mp-545379 &   InBiO3       \\
44  &    mp-14099 &   ZnAgF3    &    94  &  mp-1214407 &    BaBO3 & 144 &    mp-19035 &   BaFeO3 &  194 &   mp-551451 &   FeBiO3       \\
45  &   mp-998152 &   TlAgF3    &    95  &   mp-973738 &    GeBO3 & 145 &    mp-21280 &   BaPbO3 &  195 &   mp-971706 &   ZnBiO3       \\
46  &   mp-998554 &   KAgCl3    &    96  &   mp-984554 &    BPbO3 & 146 &     mp-3020 &   BaNbO3 &  196 &   mp-975384 &   RbBiO3       \\
47  &   mvc-15521 &    YAgO3    &    97  &  mp-1183194 &    BTeO3 & 147 &     mp-5986 &   BaTiO3 &  197 &  mp-1184562 &   GeBiO3       \\
48  &  mp-1183176 &   AlNiO3    &    98  &  mp-1184023 &    GaBO3 & 148 &  mp-1183286 &   BaDyO3 &  198 &  mp-1185847 &   MgBiO3       \\
49  &    mp-23080 &   AlBiO3    &    99  &  mp-1184625 &    HfBO3 & 149 &  mp-1183405 &   BaZnO3 &  199 &  mp-1187744 &    VBiO3       \\
\\
\\
\\
\\
200 &   mp-972827 &   SiBiO3  &  250 &   mp-975380 &   RbCdO3  &   300 &   mp-998195 &  RbInCl3  &     350 &  mp-1183976 &   CsTeO3 \\
201 &   mp-974740 &   NdBiO3  &  251 &   mp-976293 &   NaCdO3  &   301 &   mp-998230 &  LiSnCl3  &     351 &  mp-1183997 &   CsThO3 \\
202 &   mp-982040 &    KBiO3  &  252 &   mp-998623 &   TlCdF3  &   302 &   mp-998233 &  InGeCl3  &     352 &  mp-1185545 &   CsPbO3 \\
203 &   mp-983578 &   CsBiO3  &  253 &  mp-1016845 &   ZrCdO3  &   303 &   mp-998535 &  InPbCl3  &     353 &  mp-1185551 &   CsTbO3 \\
204 &   mp-998363 &   NiBiO3  &  254 &  mp-1016846 &   TiCdO3  &   304 &   mp-998560 &  InSnCl3  &     354 &  mp-1185552 &   CsTaO3 \\
205 &  mp-1183479 &   BiMoO3  &  255 &  mp-1016849 &   CdRuO3  &   305 &   mp-998732 &  TlInCl3  &     355 &  mp-1185564 &   CsYbO3 \\
206 &   mp-655631 &   FeBiO3  &  256 &  mp-1016875 &   CdRhO3  &   306 &  mp-1206518 &  RbFeCl3  &     356 &  mp-1187572 &   YbCrO3 \\
207 &    mp-29798 &   TlBrO3  &  257 &  mp-1016881 &   CdSnO3  &   307 &    mp-22988 &  CsGeCl3  &     357 &    mp-19257 &   SmCrO3 \\
208 &    mp-27249 &    CuCO3  &  258 &  mp-1016903 &   CdGeO3  &   308 &   mp-998420 &   KCrCl3  &     358 &    mp-22364 &   CrPbO3 \\
209 &   mp-547864 &    CdCO3  &  259 &  mp-1016904 &    VCdO3  &   309 &  mp-1183692 &   CoSnO3  &     359 &   mp-561947 &   CsHgF3 \\
210 &  mp-1179357 &   SnCCl3  &  260 &  mp-1017446 &   HfCdO3  &   310 &  mp-1183715 &   CoRhO3  &     360 &     mp-5811 &   CsPbF3 \\
211 &  mp-1218370 &    SrCO3  &  261 &  mp-1207995 &   TlCdN3  &   311 &   mp-573180 &   LaCoO3  &     361 &   mp-613384 &    CsIO3 \\
212 &  mp-1016853 &    CaVO3  &  262 &  mp-1209203 &   RbCdN3  &   312 &   mp-998160 &   TlCoF3  &     362 &     mp-8398 &   CsYbF3 \\
213 &  mp-1068389 &   SnCCl3  &  263 &    mp-22345 &   TiCdO3  &   313 &  mp-1183681 &   CoSbO3  &     363 &   mp-984051 &   CsGeO3 \\
214 &  mp-1068985 &   SnCCl3  &  264 &     mp-8399 &   CsCdF3  &   314 &  mp-1186980 &   ScCoO3  &     364 &   mp-998235 &   CsTlF3 \\
215 &  mp-1096983 &     H3CI  &  265 &   mp-973869 &   LaCdO3  &   315 &    mp-20427 &   PrCoO3  &     365 &   mp-998298 &   CsSnF3 \\
216 &   mp-654812 &    CdCO3  &  266 &  mp-1211584 &    KCdN3  &   316 &    mp-22549 &   SmCoO3  &     366 &  mp-1076070 &   LaCuO3 \\
217 &  mp-1001571 &   CaFeO3  &  267 &    mp-22593 &    CeVO3  &   317 &   mp-985545 &   CoTeO3  &     367 &  mp-1096944 &   CsNbO3 \\
218 &  mp-1016884 &   CaGeO3  &  268 &     mp-6951 &   RbCdF3  &   318 &  mp-1184947 &    KCoO3  &     368 &  mp-1183913 &   CsPdO3 \\
219 &  mp-1099873 &   CaCuO3  &  269 &   mp-754524 &   CeTiO3  &   319 &  mp-1187746 &    VCoO3  &     369 &  mp-1183917 &   CsMoO3 \\
220 &  mp-1183538 &   CaTmO3  &  270 &   mp-973613 &   LiCeO3  &   320 &   mp-510611 &   LaCoO3  &     370 &  mp-1183918 &   CsLaO3 \\
221 &  mp-1183540 &   CaZnO3  &  271 &    mp-22540 &   CeGaO3  &   321 &   mvc-11732 &    YCoO3  &     371 &  mp-1183919 &   CsPrO3 \\
222 &  mp-1183543 &   CaInO3  &  272 &    mp-33365 &   CeGaO3  &   322 &  mp-1075975 &   EuCoO3  &     372 &  mp-1183920 &   CsPaO3 \\
223 &  mp-1183570 &   CaHgO3  &  273 &   mp-977389 &   CeCuO3  &   323 &  mp-1183772 &   CoMoO3  &     373 &  mp-1183971 &   CsRuO3 \\
224 &   mp-546910 &   CaSnO3  &  274 &  mp-1023129 &   KCuCl3  &   324 &  mp-1076820 &   MgCoO3  &     374 &  mp-1183977 &   CsTiO3 \\
225 &     mp-7986 &   CaSnO3  &  275 &  mp-1183706 &   CeMnO3  &   325 &  mp-1186126 &   NaCoO3  &     375 &  mp-1184022 &   CsSmO3 \\
226 &   mp-998432 &   CaTcO3  &  276 &  mp-1228834 &  CsHgCl3  &   326 &    mp-20031 &   NdCoO3  &     376 &  mp-1185544 &   CsTlO3 \\
227 &  mp-1016835 &   CaRuO3  &  277 &    mp-20530 &   CeCrO3  &   327 &    mp-20951 &   GdCoO3  &     377 &  mp-1185550 &   CsEuO3 \\
228 &  mp-1017467 &   CaMnO3  &  278 &   mp-570591 &  CsHgCl3  &   328 &   mp-559946 &    KCoF3  &     378 &  mp-1185561 &   CsSbO3 \\
229 &  mp-1097061 &   CaTlF3  &  279 &   mp-864636 &   CeFeO3  &   329 &  mp-1187577 &   TmCrO3  &     379 &  mp-1187483 &   TlCuO3 \\
230 &   mp-998150 &   CaTlF3  &  280 &   mp-866095 &   CeNiO3  &   330 &  mp-1226313 &   CrMoO3  &     380 &  mp-1206323 &   CsYbH3 \\
231 &   mvc-14243 &   CaCoO3  &  281 &   mp-998147 &  TlCuCl3  &   331 &    mp-18770 &    YCrO3  &     381 &  mp-1207081 &   CsTmH3 \\
232 &  mp-1016873 &   CaHfO3  &  282 &  mp-1068377 &  CsEuCl3  &   332 &    mp-19062 &   NdCrO3  &     382 &  mp-1207302 &   CsPdF3 \\
233 &  mp-1183539 &   CaHoO3  &  283 &  mp-1070375 &  CsSnCl3  &   333 &  mp-1076360 &   RbCrO3  &     383 &   mp-558694 &   CsFeF3 \\
234 &  mp-1183562 &   CaLaO3  &  284 &  mp-1147710 &  CsPdCl3  &   334 &  mp-1147577 &   GdCrO3  &     384 &     mp-8396 &   CsEuF3 \\
235 &  mp-1207169 &   CaH3Pd  &  285 &  mp-1206801 &  RbEuCl3  &   335 &  mp-1183712 &   CrGeO3  &     385 &   mp-984082 &   CsGdO3 \\
236 &     mp-5827 &   CaTiO3  &  286 &    mp-23037 &  CsPbCl3  &   336 &  mp-1183719 &   CrRhO3  &     386 &   mp-984745 &   CsSnO3 \\
237 &  mp-1016833 &   CaRhO3  &  287 &   mp-570290 &  MnTlCl3  &   337 &  mp-1185311 &   LiCrO3  &     387 &   mp-998234 &   CsTlF3 \\
238 &  mp-1099934 &   CaCoO3  &  288 &   mp-675022 &  CsPbCl3  &   338 &    mp-18841 &   LaCrO3  &     388 &   mp-998322 &   CsInF3 \\
239 &  mp-1183544 &    CaYO3  &  289 &   mp-998156 &  RbSnCl3  &   339 &   mp-555036 &    KCrF3  &     389 &  mp-1147748 &   TiCuO3 \\
240 &  mp-1207078 &   CaNiH3  &  290 &   mp-998231 &   KGeCl3  &   340 &   mp-566131 &    KCrF3  &     390 &  mp-1184073 &   CuPdO3 \\
241 &   mp-542112 &   CaZrO3  &  291 &   mp-998605 &  RbPbCl3  &   341 &   mp-754941 &   EuCrO3  &     391 &  mp-1186515 &   PrCuO3 \\
242 &   mp-998151 &  CaTlCl3  &  292 &   mp-998744 &  TlGeCl3  &   342 &  mp-1076732 &    KCrO3  &     392 &  mp-1187321 &   TbCuO3 \\
243 &  mp-1016854 &   MnCdO3  &  293 &  mp-1070599 &  CsYbCl3  &   343 &  mp-1183695 &   CrTcO3  &     393 &  mp-1187469 &   TiCuO3 \\
244 &  mp-1016879 &   CdSiO3  &  294 &  mp-1080231 &  CsTlCl3  &   344 &  mp-1183696 &   CrCuO3  &     394 &   mp-554973 &   TlCuF3 \\
245 &    mp-10175 &    KCdF3  &  295 &  mp-1209191 &  RbPbCl3  &   345 &  mp-1206325 &   RbCrF3  &     395 &   mp-975443 &   InCuO3 \\
246 &  mp-1183925 &   CsCdO3  &  296 &   mp-567939 &  CsSmCl3  &   346 &    mp-19353 &   PrCrO3  &     396 &   mvc-11697 &    YCuO3 \\
247 &  mp-1184992 &    KCdO3  &  297 &   mp-579768 &  CsTmCl3  &   347 &    mp-20029 &   SrCrO3  &     397 &  mp-1076317 &   MgCuO3 \\
248 &  mp-1185203 &   LiCdO3  &  298 &   mp-675460 &  LiTiCl3  &   348 &  mp-1076642 &   NaCrO3  &     398 &  mp-1076711 &   SrCuO3 \\
249 &   mp-568544 &  CsCdCl3  &  299 &   mp-998173 &  RbPdCl3  &   349 &  mp-1183922 &   CsNdO3  &     399 &  mp-1100778 &   ZnCuO3 \\
\\
\\
\\
\\
400 &  mp-1183771 &   DyCuO3    &   450 &   mp-771055 &   EuZrO3    &   500 &   mp-510624 &   SrFeO3    &   550 &  mp-1016819 &   HfZnO3 \\
401 &  mp-1187562 &   YbCuO3    &   451 &   mp-862606 &   EuSnO3    &   501 &  mp-1076368 &   MgFeO3    &   551 &  mp-1016825 &   HfMgO3 \\
402 &   mp-554601 &   RbCuF3    &   452 &   mp-865181 &   EuGeO3    &   502 &  mp-1184032 &   GaFeO3    &   552 &  mp-1018647 &   YbNiH3 \\
403 &   mp-562987 &    VCuO3    &   453 &   mp-866030 &   EuSiO3    &   503 &  mp-1186134 &   NaFeO3    &   553 &  mp-1186747 &   PrHfO3 \\
404 &   mp-973068 &   ScCuO3    &   454 &   mp-866036 &   EuRuO3    &   504 &  mp-1186220 &   NbFeO3    &   554 &    mp-13108 &   SrHfO3 \\
405 &  mp-1076706 &   SmCuO3    &   455 &   mp-972931 &   EuMgO3    &   505 &  mp-1207083 &   FeSnO3    &   555 &   mp-998227 &    KHfO3 \\
406 &  mp-1184044 &   CuHgO3    &   456 &   mp-976284 &   EuSbO3    &   506 &   mvc-11229 &    YFeO3    &   556 &  mp-1016837 &   SnHgO3 \\
407 &  mp-1184117 &   CuPbO3    &   457 &   mp-977412 &   EuNiO3    &   507 &  mp-1184206 &   GaSiO3    &   557 &  mp-1016842 &   TiHgO3 \\
408 &  mp-1184157 &   ErCuO3    &   458 &  mp-1099572 &   MnInF3    &   508 &  mp-1184243 &   GaSnO3    &   558 &  mp-1016874 &   SiHgO3 \\
409 &  mp-1224030 &   InCuO3    &   459 &  mp-1184359 &   EuHgO3    &   509 &     mp-9831 &   LaGaO3    &   559 &  mp-1016938 &    VHgO3 \\
410 &     mp-5566 &    KCuF3    &   460 &  mp-1184411 &   EuMoO3    &   510 &  mp-1097026 &   LaGaO3    &   560 &  mp-1184597 &   HfThO3 \\
411 &  mp-1075902 &   EuCuO3    &   461 &  mp-1185344 &   LiEuO3    &   511 &  mp-1183980 &   GaGeO3    &   561 &  mp-1185882 &   MgHgO3 \\
412 &  mp-1184481 &   GdCuO3    &   462 &   mp-555123 &    KMnF3    &   512 &  mp-1183983 &   GaNiO3    &   562 &  mp-1186886 &   RbHgO3 \\
413 &  mp-1184736 &   HoCuO3    &   463 &     mp-7482 &   RbHgF3    &   513 &  mp-1184220 &   GaTeO3    &   563 &  mp-1187530 &   TlHgO3 \\
414 &  mp-1187487 &   ThCuO3    &   464 &   mp-753781 &   EuHfO3    &   514 &  mp-1185488 &   LuGaO3    &   564 &     mp-4551 &   SrHfO3 \\
415 &   mp-975364 &   RbCuO3    &   465 &   mp-756315 &    EuVO3    &   515 &  mp-1187561 &   YbGaO3    &   565 &  mp-1016871 &   ZrHgO3 \\
416 &  mp-1184153 &   DySnO3    &   466 &     mp-9060 &   RbPdF3    &   516 &   mp-971733 &   ZnGaO3    &   566 &  mp-1080187 &   MnHgO3 \\
417 &  mp-1183821 &   DyRhO3    &   467 &  mp-1120777 &   NaTiF3    &   517 &   mp-978504 &   SmGaO3    &   567 &  mp-1184813 &   HoSiO3 \\
418 &  mp-1183823 &   DyInO3    &   468 &   mp-556424 &    RbVF3    &   518 &     mp-9833 &   PrGaO3    &   568 &  mp-1184944 &    KHoO3 \\
419 &  mp-1184778 &   HoErO3    &   469 &     mp-5878 &    KZnF3    &   519 &  mp-1184484 &   GdSiO3    &   569 &  mp-1207180 &   HoSbO3 \\
420 &  mp-1184954 &    KDyO3    &   470 &   mp-998193 &   RbSnF3    &   520 &  mp-1184491 &   GdThO3    &   570 &   mp-971975 &   SrHgO3 \\
421 &  mp-1185308 &   LiDyO3    &   471 &   mp-998761 &   TlNiF3    &   521 &  mp-1184503 &   GdRhO3    &   571 &   mp-971996 &   SrHoO3 \\
422 &  mp-1186557 &   PmErO3    &   472 &   mp-998710 &   TlHgF3    &   522 &  mp-1184599 &   GdZrO3    &   572 &   mp-973818 &   LiHoO3 \\
423 &  mp-1187145 &   SrDyO3    &   473 &   mp-998712 &   TlGeF3    &   523 &  mp-1184987 &    KGdO3    &   573 &   mp-975296 &   RbHoO3 \\
424 &  mp-1187147 &   SrErO3    &   474 &   mp-554465 &   RbFeF3    &   524 &  mp-1185401 &   LiGdO3    &   574 &  mp-1184833 &   HoRhO3 \\
425 &  mp-1207184 &   DySbO3    &   475 &   mp-556891 &    KFeF3    &   525 &  mp-1187481 &   TlGaO3    &   575 &    mp-22981 &    TlIO3 \\
426 &   mp-975386 &   RbDyO3    &   476 &     mp-6952 &   RbYbF3    &   526 &   mp-975308 &   RbGdO3    &   576 &    mp-27193 &    RbIO3 \\
427 &  mp-1183905 &   EuErO3    &   477 &   mp-998762 &   MnTlF3    &   527 &     mp-9834 &   NdGaO3    &   577 &   mp-545825 &    NaIO3 \\
428 &  mp-1184164 &   ErGaO3    &   478 &    mp-20572 &    NaVF3    &   528 &  mp-1017447 &   HgGeO3    &   578 &   mp-546200 &    TlIO3 \\
429 &  mp-1184207 &   ErZnO3    &   479 &    mp-21043 &   RbPbF3    &   529 &  mp-1187065 &   SnGeO3    &   579 &   mp-552729 &     KIO3 \\
430 &  mp-1184219 &   ErTlO3    &   480 &   mp-557257 &     KVF3    &   530 &   mp-975347 &   NdGeO3    &   580 &   mp-558843 &     KIO3 \\
431 &  mp-1184486 &   GdErO3    &   481 &   mp-558749 &   RbMnF3    &   531 &   mp-978926 &   SmGeO3    &   581 &   mp-973977 &   HoSnO3 \\
432 &  mp-1185310 &   LiErO3    &   482 &   mp-560976 &    KNiF3    &   532 &  mp-1016930 &   ZnGeO3    &   582 &  mp-1184751 &   InSiO3 \\
433 &  mp-1187342 &   TbErO3    &   483 &   mp-675295 &    InPF3    &   533 &  mp-1016999 &   SrGeO3    &   583 &  mp-1184765 &   InSbO3 \\
434 &   mp-973830 &    KErO3    &   484 &     mp-7483 &    KHgF3    &   534 &  mp-1017566 &   GePbO3    &   584 &  mp-1186316 &   NdInO3 \\
435 &   mp-975361 &   RbErO3    &   485 &   mp-998159 &   TlFeF3    &   535 &  mp-1184557 &   GePdO3    &   585 &  mp-1186597 &   PmInO3 \\
436 &  mp-1075904 &   EuFeO3    &   486 &   mp-998607 &   RbTlF3    &   536 &  mp-1185514 &   LuGeO3    &   586 &  mp-1223446 &     KIO3 \\
437 &  mp-1099619 &   EuMnO3    &   487 &   mp-998739 &   MgTlF3    &   537 &  mp-1187533 &   TlGeO3    &   587 &   mp-974734 &    KInO3 \\
438 &  mp-1100775 &   SrEuO3    &   488 &   mp-998745 &   TlZnF3    &   538 &   mp-865758 &   YbGeO3    &   588 &  mp-1185889 &   MgInO3 \\
439 &  mp-1184304 &   EuInO3    &   489 &   mp-998786 &   TlPdF3    &   539 &   mp-982558 &   HoGeO3    &   589 &  mp-1186138 &   NaInO3 \\
440 &  mp-1184309 &   EuLuO3    &   490 &   mp-998241 &   MnInF3    &   540 &  mp-1016827 &   MgGeO3    &   590 &  mp-1187569 &   YbInO3 \\
441 &  mp-1184348 &   EuPbO3    &   491 &  mp-1184349 &   FeRhO3    &   541 &  mp-1184641 &   GeTeO3    &   591 &   mp-972021 &   SrInO3 \\
442 &  mp-1184353 &   EuTmO3    &   492 &   mp-552676 &   LaFeO3    &   542 &  mp-1185303 &   LiGeO3    &   592 &   mp-975299 &   RbInO3 \\
443 &  mp-1184369 &   EuPdO3    &   493 &     mp-9061 &    KPdF3    &   543 &  mp-1186133 &   NaGeO3    &   593 &  mp-1040469 &    KMoO3 \\
444 &  mp-1184447 &   EuTcO3    &   494 &   mp-973579 &   FePbO3    &   544 &  mp-1186885 &   RbGeO3    &   594 &  mp-1076633 &     KVO3 \\
445 &  mp-1186571 &   PmEuO3    &   495 &  mp-1099588 &   SmFeO3    &   545 &  mp-1187655 &   TmGeO3    &   595 &  mp-1184945 &     KYO3 \\
446 &  mp-1187715 &   YbEuO3    &   496 &  mp-1184033 &   GaFeO3    &   546 &   mp-981132 &   TbGeO3    &   596 &  mp-1184946 &    KZnO3 \\
447 &   mp-541365 &   LiEuH3    &   497 &   mp-973648 &   FeRhO3    &   547 &  mp-1017629 &   MgNiH3    &   597 &  mp-1184948 &    KMnO3 \\
448 &   mp-644246 &   EuH3Pd    &   498 &  mp-1185320 &   LiFeO3    &   548 &  mp-1018646 &   YbH3Pd    &   598 &  mp-1184950 &    KZrO3 \\
449 &   mp-755572 &   EuNbO3    &   499 &   mp-975322 &   RbFeO3    &   549 &  mp-1207079 &   SrH3Pd    &   599 &  mp-1206970 &    KPaO3 \\

\\
\\
\\
\\
600 &     mp-3614 &    KTaO3    &       650 &  mp-1016929 &   MgRhO3    &   700 &  mp-1099668 &   SmNiO3    &   750 &  mp-1186757 &   SrZnO3 \\
601 &     mp-7375 &    KNbO3    &       651 &  mp-1185612 &   MgZnO3    &   701 &  mp-1186113 &   NiSbO3    &   751 &  mp-1186758 &   SrTmO3 \\
602 &   mp-760376 &    KTiO3    &       652 &  mp-1185613 &   MgTlO3    &   702 &  mp-1187595 &   YbNiO3    &   752 &  mp-1186927 &   SbPdO3 \\
603 &   mp-935811 &    KNbO3    &       653 &  mp-1187347 &   TbMgO3    &   703 &  mp-1188025 &   ZrNiO3    &   753 &  mp-1187190 &   SrPdO3 \\
604 &   mp-972985 &    KPdO3    &       654 &   mp-972160 &   YbMgO3    &   704 &   mp-974108 &   NiPbO3    &   754 &  mp-1187389 &    TbVO3 \\
605 &   mp-973267 &    KTcO3    &       655 &   mp-975521 &   NdMgO3    &   705 &   mp-975373 &   RbNiO3    &   755 &  mp-1187400 &   TcSnO3 \\
606 &   mp-973958 &    KRuO3    &       656 &   mp-976446 &   MgMnO3    &   706 &   mp-976848 &   NiRuO3    &   756 &  mp-1187515 &   YbSnO3 \\
607 &   mp-975134 &    KRhO3    &       657 &   mp-976722 &   MgPbO3    &   707 &   mp-998781 &   TlNiO3    &   757 &  mp-1207179 &   SmSbO3 \\
608 &   mp-984454 &    KTlO3    &       658 &  mp-1075973 &   SmMnO3    &   708 &  mp-1070440 &    VPbO3    &   758 &    mp-22534 &   SrRuO3 \\
609 &  mp-1075921 &   LaNiO3    &       659 &  mp-1185896 &   MgMoO3    &   709 &  mp-1183043 &   ZrSiO3    &   759 &     mp-3323 &   SrZrO3 \\
610 &  mp-1096800 &   LaScO3    &       660 &   mp-975346 &   RbMnO3    &   710 &  mp-1186893 &   RbSnO3    &   760 &     mp-4346 &   SrRuO3 \\
611 &  mp-1184935 &    KPbO3    &       661 &  mp-1016932 &   MnZnO3    &   711 &  mp-1186948 &   SbPbO3    &   761 &     mp-5229 &   SrTiO3 \\
612 &  mp-1184940 &    KTeO3    &       662 &    mp-37214 &   MnPbO3    &   712 &  mp-1187364 &   TbSiO3    &   762 &   mp-973269 &   SbTeO3 \\
613 &  mp-1184941 &    KLuO3    &       663 &   mvc-10883 &    YMnO3    &   713 &  mp-1187493 &   YbScO3    &   763 &   mp-975207 &   RbSmO3 \\
614 &  mp-1184949 &    KTmO3    &       664 &   mvc-15452 &    YMoO3    &   714 &  mp-1187557 &   TmScO3    &   764 &   mp-975255 &   RbPdO3 \\
615 &  mp-1184953 &    KYbO3    &       665 &  mp-1040471 &   NaMoO3    &   715 &  mp-1187797 &   YbZnO3    &   765 &   mp-978490 &   SiPdO3 \\
616 &  mp-1185357 &   LiLaO3    &       666 &  mp-1186106 &   MoPbO3    &   716 &  mp-1187862 &    YSiO3    &   766 &   mp-979272 &   TePdO3 \\
617 &    mp-19025 &   LaMnO3    &       667 &  mp-1186997 &   SiMoO3    &   717 &  mp-1206844 &   RbPaO3    &   767 &   mvc-15474 &    YSnO3 \\
618 &     mp-4342 &    KNbO3    &       668 &  mp-1187460 &   ThMoO3    &   718 &    mp-20459 &   TiPbO3    &   768 &  mp-1068577 &   ZrPbO3 \\
619 &     mp-5246 &    KNbO3    &       669 &    mp-18747 &   SrMoO3    &   719 &   mp-755018 &   RbTaO3    &   769 &  mp-1068742 &   SrZrO3 \\
620 &   mp-981313 &    KTbO3    &       670 &   mp-975292 &   RbMoO3    &   720 &   mp-865218 &   YbTiO3    &   770 &  mp-1183045 &   ZrTiO3 \\
621 &   mp-989515 &   LaMoN3    &       671 &   mp-977100 &   NiMoO3    &   721 &   mp-975159 &   RbZnO3    &   771 &  mp-1186471 &   PrPbO3 \\
622 &  mp-1184964 &   LaMgO3    &       672 &   mp-989557 &    YTcN3    &   722 &   mp-975173 &    RbYO3    &   772 &  mp-1186897 &   RbRhO3 \\
623 &  mp-1184965 &   LaZnO3    &       673 &   mp-989613 &    TiPN3    &   723 &   mp-975190 &   RbTeO3    &   773 &  mp-1187192 &   SrSbO3 \\
624 &  mp-1185065 &   LaZrO3    &       674 &  mp-1099591 &    NaVO3    &   724 &   mp-978489 &   SiPbO3    &   774 &  mp-1187442 &   ThPdO3 \\
625 &  mp-1185122 &   LaSnO3    &       675 &  mp-1186154 &   NaNiO3    &   725 &   mp-978506 &   SmSiO3    &   775 &  mp-1187516 &   YbSiO3 \\
626 &    mp-19053 &    LaVO3    &       676 &  mp-1186199 &   NaYbO3    &   726 &   mp-978952 &   SnPbO3    &   776 &  mp-1187516 &   YbSiO3 \\
627 &     mp-8020 &   LaTiO3    &       677 &     mp-4170 &   NaTaO3    &   727 &   mp-980058 &   RbTbO3    &   777 &  mp-1187516 &   YbSiO3 \\
628 &  mp-1185263 &   LiPaO3    &       678 &   mp-976968 &   NaPdO3    &   728 &  mp-1183044 &   ZrTlO3    &   778 &  mp-1187516 &   YbSiO3 \\
629 &  mp-1185420 &   LiRuO3    &       679 &   mp-977107 &   NaRuO3    &   729 &  mp-1187425 &   ThSiO3    &   779 &  mp-1187518 &   YbTcO3 \\
630 &   mp-976331 &   LiLuO3    &       680 &   mp-977117 &   NaTeO3    &   730 &  mp-1187488 &   TiTeO3    &   780 &  mp-1187570 &   YbPdO3 \\
631 &  mp-1185417 &   LiYbO3    &       681 &   mp-977123 &   NaTiO3    &   731 &  mp-1187529 &   TlSiO3    &   781 &  mp-1187623 &   TlTcO3 \\
632 &   mp-973136 &   LiSmO3    &       682 &   mp-977176 &   NaTcO3    &   732 &    mp-12631 &   SrTcO3    &   782 &    mp-20337 &   ZrPbO3 \\
633 &  mp-1185236 &   LiTmO3    &       683 &  mp-1186153 &   NaPbO3    &   733 &    mp-18717 &    SrVO3    &   783 &   mp-972058 &   YbTmO3 \\
634 &  mp-1185261 &   LiNiO3    &       684 &  mp-1186198 &   NaZnO3    &   734 &    mp-19845 &   TiPbO3    &   784 &   mp-972978 &    ScVO3 \\
635 &  mp-1185388 &   LiPmO3    &       685 &     mp-3136 &   NaNbO3    &   735 &    mp-20702 &    SmVO3    &   785 &   mp-974337 &   RuPbO3 \\
636 &  mp-1185404 &   LiPrO3    &       686 &   mp-865120 &   NaPaO3    &   736 &   mp-754696 &   PrTiO3    &   786 &   mp-975206 &   RbSbO3 \\
637 &   mp-862870 &   LiTcO3    &       687 &  mp-1075911 &   RbNbO3    &   737 &   mp-981376 &   ScSnO3    &   787 &   mp-978493 &   SiSnO3 \\
638 &   mp-976283 &    LiYO3    &       688 &  mp-1187578 &   YbNbO3    &   738 &   mp-981550 &   RbTmO3    &   788 &   mp-980537 &   RbYbO3 \\
639 &  mp-1185387 &   LiNdO3    &       689 &  mp-1206341 &   NdSbO3    &   739 &   mp-983513 &   PmRuO3    &   789 &  mp-1016902 &   ZnSnO3 \\
640 &   mp-973486 &    LuVO3    &       690 &    mp-19253 &    NdVO3    &   740 &   mvc-11152 &    YSbO3    &   790 &  mp-1016927 &   TiZnO3 \\
641 &   mp-973681 &   LuRhO3    &       691 &     mp-7006 &   SrNbO3    &   741 &   mvc-11730 &     YVO3    &   791 &  mp-1018639 &   TiSnO3 \\
642 &  mp-1016820 &   MgSnO3    &       692 &   mp-975913 &   PrNbO3    &   742 &  mp-1016876 &   ZnSiO3    &   792 &  mp-1069346 &    PrVO3 \\
643 &  mp-1017000 &   MgZrO3    &       693 &   mp-977408 &   NbTlO3    &   743 &  mp-1016880 &   ZnRhO3    &   793 &  mp-1076534 &   RbTaO3 \\
644 &  mp-1185490 &   LuPdO3    &       694 &  mp-1101333 &   SrNdO3    &   744 &  mp-1016883 &   ZrZnO3    &   794 &  mp-1100777 &    VTeO3 \\
645 &  mp-1186600 &   PmMgO3    &       695 &  mp-1186345 &   NiPdO3    &   745 &  mp-1016931 &    VZnO3    &   795 &  mp-1186401 &   PaSiO3 \\
646 &   mp-973655 &   LuSiO3    &       696 &  mp-1187352 &   TbNiO3    &   746 &  mp-1017441 &   SrRhO3    &   796 &  mp-1187377 &   TbTeO3 \\
647 &  mp-1016829 &   MgRuO3    &       697 &   mvc-15448 &    YNiO3    &   747 &  mp-1076638 &    RbVO3    &   797 &   mp-546973 &   SrSnO3 \\
648 &  mp-1016830 &   MgTiO3    &       698 &  mp-1097917 &   TlNiO3    &   748 &  mp-1186755 &   SrTaO3    &   798 &   mp-613402 &   SrZrO3 \\
649 &  mp-1016886 &    MgVO3    &       699 &  mp-1099583 &   SmTiO3    &   749 &  mp-1186756 &   SrYbO3    &   799 &   mp-754046 &   PrTlO3 \\      
800 &   mp-973123 &   ScPdO3    &       801 &   mp-975179 &   RbTlO3    &   802 &   mp-975208 &   RbRuO3    &   802 &   mp-975208 &   RbRuO3 \\
803 &   mp-981367 &   SrTlO3    &       804 &   mvc-11198 &    YTiO3    &       &             &             &       &             &         \\
\end{longtable}

\end{center}